\documentclass[12pt]{article}

\usepackage{comment}
\usepackage{amsmath}
\usepackage{mathrsfs}
\usepackage{graphicx}
\usepackage{algorithm}
\usepackage{algpseudocode}
\algrenewcommand\algorithmicrequire{\textbf{Input:}}
\algrenewcommand\algorithmicensure{\textbf{Output:}}

\usepackage[utf8]{inputenc}
\usepackage{comment}
\usepackage{bbm}
\usepackage{natbib}
\usepackage{url}
\usepackage{latexsym, epsfig, amssymb, amsmath, booktabs}
\usepackage{rotating, tabularx, paralist}
\usepackage{color}
\usepackage{graphicx}
\usepackage[OT1]{fontenc}
\usepackage{lscape}
\usepackage{multirow,multicol}
\usepackage{subcaption}
\usepackage{amsthm}
\usepackage{appendix}

\usepackage[left=1in, right=1in, top=1.5in, bottom=1.5in]{geometry} 

\newtheorem{condition}{Condition}
\newtheorem{lemma}{Lemma}

\newcommand{\e}{\mathbb{E}}
\newcommand{\R}{\mathbb{R}}

\newcommand{\Var}{\mbox{Var}}
\newcommand{\bx}{\mbox{\bf x}}
\newcommand{\bX}{\mbox{\bf X}}

\newcommand{\bZ}{\mbox{\bf Z}}

\newcommand{\bW}{\mbox{\bf W}}

\newcommand{\bR}{\mbox{\bf R}}

\newcommand{\by}{\mbox{\bf y}}

\def\X{\mbox{\bf X}}

\def\calS{\mathcal{S}}

\def\wh{\widehat}
\def\wt{\widetilde}

\newcommand{\E}{\mathbb{E}} 

\newcommand{\bOmega}{\boldsymbol{\Omega}}
\newcommand{\bSigma}{\boldsymbol{\Sigma}}
\newcommand{\br}{\boldsymbol{r}}

\newcommand{\argmin}{\mbox{arg\,min}}
\newcommand{\argmax}{\mbox{arg\,max}}

\newcommand{\beps}{\mbox{\boldmath $\epsilon$}}
\newcommand{\bbeta}{\mbox{\boldmath $\beta$}}
\newcommand{\bSig}{\mbox{\boldmath $\Sigma$}}

\newcommand\independent{\protect\mathpalette{\protect\independenT}{\perp}}
\def\independenT#1#2{\mathrel{\setbox0\hbox{$#1#2$}%
		\copy0\kern-\wd0\mkern4mu\box0}}

\DeclareMathOperator*{\FDR}{FDR}
\DeclareMathOperator*{\FDP}{FDP}

\DeclareMathOperator{\avg}{avg}
\DeclareMathOperator{\diag}{diag}

\DeclareMathOperator{\Lap}{Lap}

\newtheorem{theorem}{Theorem}
\newtheorem{proposition}[theorem]{Proposition}
\newtheorem{example}{Example}
\newtheorem{definition}{Definition}
\newtheorem{remark}{Remark}

\newcommand{\ignore}[1]{}{}

\begin{document}

	\def\spacingset#1{\renewcommand{\baselinestretch}%
		{#1}\small\normalsize} \spacingset{1}
	
		\title{\bf Knockoff Inference under Privacy Constraints}
		\author{
			Zhanrui Cai$^1$ \quad Yingying Fan$^2$ \quad Lan Gao$^3$ \thanks{Authors are listed alphabetically.} \\
			\\
			University of Hong Kong$^1$ \\
			University of Southern California$^2$ \\
			The University of Tennessee$^3$
		}
		\date{}
		\maketitle	
	\bigskip
	\begin{abstract}
		Model-X knockoff framework \citep{candes2018panning} offers a model-free variable selection method that ensures finite-sample false discovery rate (FDR) control. However, the complexity of generating knockoff variables, coupled with the model-free assumption, presents significant challenges for protecting data privacy in this context. We propose a comprehensive framework for knockoff inference within the differential privacy paradigm. Our proposed method guarantees robust privacy protection while preserving the exact FDR control entailed by the original model-X knockoff procedure. We further conduct power analysis and establish sufficient conditions under which the noise added for privacy preservation does not asymptotically compromise power. Through various applications, we demonstrate that the differential privacy knockoff 
method can be effectively utilized to safeguard privacy during variable selection with FDR control in both low and high dimensional settings.
	\end{abstract}
	
	\noindent%
	\vfill
	
	\newpage
	\spacingset{1.8}

\section{Introduction}
	
	In the era of big data, selecting important variables is of fundamental importance across a range of fields, including medical research, social sciences, and finance. Identifying important variables not only provides meaningful insights into the data but also enables informed decision-making.  Various variable selection methods have been proposed in the literature, such as the Lasso \citep{tibshirani1996regression}, SCAD \citep{fan2001variable}, and feature screening \citep{fan2008sure, li2012feature}, among others. Typically, given the response variable $Y$ and the covariates $X = (X_1, X_2, \dots, X_p)^T$, the goal is to identify the smallest subset $\calS_0\subset\{1, 2, \dots, p\}$, such that $Y$ is independent of covariates outside of $\calS_0$ conditional on $\{X_j\}_{j\in\calS_0}$. Here, we assume the existence and uniqueness of such $\calS_0$; see \cite{candes2018panning} for discussions on when such assumptions can be justified. More recently, it is increasingly popular to develop methods that control the FDR of the selected variables. Let $S$ denote the selected subset of variables based on the sample data. The FDR is defined as the expectation of the false discovery proportion (FDP). Specifically,
    \begin{equation}\label{def:DFR}
		\mbox{FDR}:=\E[\mbox{FDP}(S)], \quad \mbox{FDP}(S) = \frac{ |S \cap  \calS_0^c|}{1\vee|S|},
	\end{equation}
where $|\cdot|$ stands for the cardinality of a set. 
	The model-X knockoff framework \citep{candes2018panning} provides a powerful tool to enable model-free FDR control even in finite samples. The key idea is to construct knockoff variables that are exchangeable in distribution with the original variables $X$, but are independent of $Y$ conditional on $X$.  Since its first proposal in \cite{barber2015controlling}, the knockoff method has become increasingly popular and has been extended and studied under various cases, such as the power analysis \citep{fan2020rank} and robust inference \citep{barber2020robust, fan2023ark, fan2025asymptotic}. 
	
	As data volumes and complexities increase, the risk of privacy violations also becomes a critical concern. Privacy-preserving data analysis not only builds trust between individuals and organizations that collect and store personal data but also complies with legal requirements, such as the European General Data Protection Regulation \citep{voigt2017eu}. Differential privacy (DP), proposed by \cite{dwork2006calibrating}, offers a rigorous mathematical foundation for privacy preservation in practice. After nearly two decades of development, it has become the standard approach for ensuring data privacy. For example, the 2020 Census conducted by the U.S. Census Bureau published DP summary statistics and spatial locations to ensure privacy for respondents \citep{wines20222020}. The core principle of DP lies in adding properly scaled random noise, thereby limiting the extent to which an individual’s information can be inferred from the output. In general, DP methods offer strong privacy guarantees without compromising too much data utility. 
	
	FDR control with privacy protection was first studied in \cite{dwork2021differentially} based on the Benjamini-Hochberg procedure. The paper provides conservative bounds for the FDR,  which are unavoidable because the privacy noises are directly added to the $p$-values. \cite{xia2023adaptive} proposed adding noises to the transformed $p$-values, and achieved exact finite-sample FDR control under the privacy constraint. However, those two methods are only applicable when the $p$-values for each hypothesis are available. Under the high-dimensional linear model, \cite{cai2023private} proposed the DP-FDR control approach based on sample splitting and mirror statistics. How to control the FDR for high-dimensional data in general nonlinear models while ensuring privacy protection remains an open question in the literature.

	In this paper, we analyze the model-X knockoff framework under the differential privacy constraint. By integrating the mirror peeling algorithm into the model-X knockoff framework, we ensure the differential privacy requirement while preserving the FDR control property of the knockoff procedure. Since the noise added for privacy preservation can compromise power, 
    we further conduct the power analysis for the proposed DP model-X knockoff. Our theory shows that under mild sensitivity constraints of the knockoff statistics,  DP model-X knockoff does not suffer power loss compared with its non-DP counterpart asymptotically. 
    Moreover, according to the sensitivity level, commonly encountered knockoff statistics can be broadly classified to have dimension-free sensitivity or dimension-dependent sensitivity, where the former is based on marginal dependence statistics and the latter depends on joint dependence measures. The dimension-free ones can accommodate higher dimensional data in controlling FDR under DP, while the dimension-dependent ones have more restrictive constraints on dimensionality. Yet, the joint-dependence-measure based knockoff statistics generally have higher power because they can be more accurate in measuring conditional variable importance. Motivated by this, we also propose a general two-step strategy based on sample splitting under the high-dimensional setting and prove the unconditional FDR control under DP. 
	
	This paper is organized as follows. Section \ref{sec:preliminaries} introduces the background of differential privacy and the knockoff method. Section \ref{Sec:DPknockoff_theory} proposes the general framework of DP-knockoff, and Section \ref{Sec6.Sensitivity} discusses the sensitivity of knockoff statistics through four examples. 
    Section \ref{Sec:highdim-knockoff}  analyzes the DP-knockoff under the high-dimensional setting. Section \ref{sec:simu} provides numerical evidence in simulation  analysis. We conclude the paper in Section \ref{sec:discussion} with discussion. 
	
    \subsection{Notation}
    Throughout the paper, $C$ denotes a positive constant, whose value may vary across lines unless otherwise specified. 
    $| \mathcal{H} |$ denotes the cardinality of a set $\mathcal{H}$.
    For any positive integer $n$, denote by  $[n] := \{1, 2, \ldots, n\}$. For $a,b \in \R$, let  $a\vee b :=  \max\{a,b\}$ and $a\wedge b:= \min\{a,b\}$. 
    For a vector $x$, $\|x\|_{k}$ denotes its $l_k$-norm.
    For a matrix $A \in \mathbb{R}^{m \times n}$, denote by $A_{j}$ its $j$-th column and $A_{-j} $ the submatrix of $A$ excluding its $j$-th column. Let $\|A\|_2$ be its spectral norm and $\|A\|_{1,2} := \max_{j\in [n]} \| A_j\|_{2} $ be its $(1,2)$-th norm.  For two sequences $a_n$ and $b_n$, we write $a_n = o(b_n)$ or $ a_n \ll b_n $ if $a_n /b_n \to 0$, $ a_n \lesssim b_n $ if $a_n \leq C b_n$,  $ a_n \gtrsim b_n $ if $a_n \geq Cb_n$, and $a_n \asymp b_n$ if both $a_n \leq C b_n$ and $a_n \geq Cb_n$ hold. 
    For a random sequence $x_n$, $x_n = \mathfrak{o}_\mathbf{P}(1)$ denotes $x_n \to 0$ in probability. {$\Lap(\lambda)$ denotes the Laplace distribution with scale parameter $\lambda$ and density function $ \frac{1}{2 \lambda}\exp( - |x| / \lambda)  $.}
	\section{Preliminaries}\label{sec:preliminaries}
	\subsection{Differential Privacy}
	
    We start with briefly introducing the background of differential privacy (DP). First introduced by \cite{dwork2006calibrating}, DP provides a probabilistic framework for measuring the privacy loss incurred by a randomized algorithm. In general, Differential privacy (DP) guarantees that a randomized algorithm produces \emph{similar} output distributions on neighboring datasets $D$ and $D'$ that differ in only one data point. In this paper, we adopt the Gaussian-DP framework \citep{dong2019gaussian}, which quantifies privacy loss based on the statistical power of a hypothesis testing problem. Specifically, if the power is small when testing the two hypotheses 
	\begin{equation}
		\label{GDP_test}
		H_0:\text{ the underlying dataset is }D,\text{ versus } H_1:\text{ the underlying dataset is }D^{\prime},  
	\end{equation} 
    then the adversary cannot distinguish $D$ from $D'$ and the privacy is protected. 
	
	\begin{definition}[Gaussian DP, \cite{dong2019gaussian}]
		\begin{enumerate}
			\item \,	A mechanism $\mathcal{M}$ satisfies $f$-DP if any $\alpha$-level test of (\ref{GDP_test}) has power function $\beta(\alpha)\leq 1-f(\alpha)$ with $f(\cdot)$ a convex, continuous, non-increasing function satisfying $f(\alpha)\leq 1-\alpha$ for all $\alpha\in[0,1]$.
			\item \,  For some $\mu\geq 0$,	a mechanism $\mathcal{M}$ satisfies $\mu$-\textit{Gaussian Differential Privacy} ($\mu$-GDP) if $\mathcal{M}$ is $G_\mu$-DP, where $G_{\mu}(\alpha)=\Phi(\Phi^{-1}(1-\alpha)-\mu)$ and $\Phi(\cdot)$ is the cumulative distribution function of standard normal $N(0,1)$.
		\end{enumerate}
	\end{definition}
	If a mechanism satisfies $\mu$-GDP, then testing (\ref{GDP_test}) is at least as hard as testing $H_0: Z\sim N(0, 1)$ versus $H_1: Z\sim N(\mu, 1)$ based on a single observation from $Z$. Next, we introduce the {\it sensitivity}, which measures the change of the statistic or algorithm when only one data entry is altered. {\it Sensitivity} is closely related to the amount of noise required to guarantee privacy.
	
	\begin{definition}[Sensitivity]
		\label{def:sensitivity}
		For a vector-valued deterministic algorithm $T(\cdot):\mathcal{D}\to\mathbb{R}^{d}$, the $l_q$ sensitivity of $T(\cdot)$ is defined as $\Delta_q(T):=\sup_{D,D'\in\mathcal{D}}\|T(D)-T(D')\|_q$,		where $D$ and $D'$ only differ in one datum, and $\|\cdot\|_q$ denotes the $l_q$ norm of a vector.
	\end{definition}
	

    \begin{example}[Gaussian Mechanism]
        Let $T:\mathcal D \to \mathbb R^k$ be a statistic with $l_2$ sensitivity $\Delta_2(T)$. Let $\xi \sim \mathcal N\!\left(0, \frac{\Delta_2(T)^2}{\mu^2} I_k\right)$ for some $\mu>0$.
        Then the mechanism $M(D) = T(D) + \xi$ with $D\in \mathcal D$ satisfies $\mu$-GDP.
    \end{example}
    
	
	Gaussian DP possesses the following properties \citep{dong2019gaussian}. 
		\begin{itemize}
		\item {\it Post-processing}: Let $M(D)$ satisfy $\mu$-GDP, and $h(\cdot)$ be a deterministic function that maps $M(D)$ to real Euclidean space. Then $h\circ M(D)$ is also $\mu$-GDP.
		\item {\it Composition}: Let $M_1(\cdot)$ satisfy $\mu_1$-GDP and $M_2(\cdot)$ satisfy $\mu_2$-GDP. Then $M_1\circ M_2(\cdot)$ is $\sqrt{\mu_1^2+\mu_2^2}$-GDP.
	\end{itemize}

	\subsection{Model-X Knockoff Framework}	
     Given the response variable $Y$ and covariate vector $X = (X_1, \ldots, X_p)^T$,  the core idea in model-X knockoff  \citep{candes2018panning} is to generate the knockoff variables $\widetilde{X} = (\widetilde{X}_1, \ldots, \widetilde{X}_p)^T$ such that  
    \begin{equation} \label{knockoff-conditions}
       \widetilde{X} \independent Y | X \quad \mbox{and} \quad (X, \widetilde{X})_{\mbox{\tiny swap}(S)}   \stackrel{d}{ = }  (X, \widetilde{X}) ,
    \end{equation} 
    for any subset $S \subset [p]$, where $\mbox{swap}(S)$ is the operation of swapping the components $X_j$ and $\wt X_j$ in  $(X, \wt X)$ for each $j\in S$. 
    For Gaussian covariates $X \stackrel{d}{\sim} N(\mathbf{0},\bOmega^{-1})$, where $\bOmega$ is a positive definite precision matrix, we can construct $\wt{X}$ as
    \begin{equation} \label{exam-Gaussian-knockoff}
        \wt{X} =    (I_p - \diag (\boldsymbol{r}) \bOmega  ) X +  (2 \diag(\boldsymbol{r}) - \diag(\boldsymbol{r}) \bOmega  \diag(\boldsymbol{r})  )^{1/2} Z,
    \end{equation}
    where $\diag(\boldsymbol{r}) = \diag(r_1, \ldots, r_p)$ is a deterministic diagonal matrix ensuring that $2 \diag(\boldsymbol{r}) - \diag(\boldsymbol{r}) \bOmega  \diag(\boldsymbol{r})  $ is positive definite, and $Z \stackrel{d}{\sim} N({\bf 0}, I_p)$ is a standard Gaussian random vector and is independent of $X$. For instance, $\diag(\br)$ can be chosen as $\lambda_{\min}(\bOmega^{-1})I_p$; see \cite{candes2018panning} for additional ways to construct $\br$.  Note that for a given $X$, its knockoff $\widetilde X$ is random and nonunique. 
 
    Suppose we have $n$ i.i.d.  observations $\{(\bX_{i, \cdot}, Y_i) \}_{i=1}^n$ from the population $(X, Y)$. In matrix notation, let $\by=(Y_i)^{T}\in\mathbb{R}^n$ and $\bX=(\bX_{i,j})\in\mathbb{R}^{n\times p}$ be respectively the response vector and the design matrix.
    The knockoff procedure consists of three steps. 
    \begin{enumerate}
     	\item[1)]\, For $i \in [n]$ and the $i$th row $ \bX_{i, \cdot} $ in the design matrix $\bX$, generate the knockoff variables $\wt{\bX }_{i, \cdot} $ satisfying \eqref{knockoff-conditions} (e.g., Algorithm 1 in \cite{candes2018panning}). Let $\wt{\bX} =(\wt{\bX}_{i, j}) \in \mathbb{R}^{n \times p} $ denote the generated knockoff variable matrix. 
     	\item[2)]\, Construct the knockoff statistics $\{W_j\}_{j = 1}^p$ to evaluate the importance of the original variables. Specifically, $ {W}_j$ for the $j$th variable is defined as 
     	\begin{equation}  \label{knockoff-W}
     		 {W}_j = w_j( [\bX, \widetilde{\bX}], \by ),
     	\end{equation}
     where the function $w_j$ satisfies for any $S \subset [p]$, 
     
	\begin{equation*}
		 w_j([\X, \wt\X]_{\mbox{\tiny swap}(S)}, Y)=
		\begin{cases}
			w_j([\X, \wt\X], \by), & j\notin S; \\
			-w_j([\X, \wt\X], \by), & j\in S.
		\end{cases}
	\end{equation*}
	Intuitively, high quality knockoff statistics $W_j$'s should possess the properties that i) the important variables have large positive values for their $W_j$'s, while ii)  unimportant variables have small values (in magnitude)  symmetrically distributed around zero. For instance, $W_1, \dots, W_p$ can be constructed as the Lasso coefficient-difference statistics. 	 
    See \cite{barber2015controlling} for more examples on valid constructions of knockoff statistics.  
  	
 	\item[3)]\, For a prespecified level $q \in (0, 1)$ for FDR control,  select the set of important variables as $S^* := \{j \in [p]: W_j \geq T^* \}$ with the threshold defined by
 	\begin{equation} \label{knockoff-threshold}
           T^* := \inf \bigg\{t \in |{\mathcal{W}}|: \frac{1 + \sum_{j = 1}^{p} \mathbbm{1} ( {W}_{j} \leq -t) } {1 \vee  \sum_{j = 1}^{p}\mathbbm{1} ( {W}_{j} \geq t)  } \leq q \bigg\},
 	\end{equation}
where  $|{\mathcal{W}}|: = \{|W_j|: 1 \leq j \leq p\}$ with $\mathcal{W} := \{ W_j: 1 \leq j \leq p\}$. Here, $T^*$ is defined as $\infty$ if the set in \eqref{knockoff-threshold} is empty and the corresponding $S^*=\emptyset$.   
 \end{enumerate}
It has been shown in \cite{candes2018panning} that the above model-X knockoff procedure achieves exact FDR control in finite samples without requiring any model assumptions on how the response $Y $ depends on the covariates $X$.

\subsection{Sensitivity in Model-X knockoff} \label{sensi-set-knockoff}
As will be shown in Algorithm \ref{alg:mirror peeling} in Section \ref{Sec:DPknockoff_theory},  the sensitivity (Cf. Definition \ref{def:sensitivity}) of the knockoff statistics $W_j$'s plays a crucial role in determining the noise level required to preserve privacy. In view of \eqref{knockoff-W}, $W_j$'s depend on the augmented design matrix $[\bX, \wt{\bX}]$ and the response vector $\by$. Due to the randomness in generating the knockoff variable matrix $\wt{\bX}$, a key challenge in analyzing the sensitivity  is the non-unique realizations of knockoff variables. To address this, we assume that the knockoff variables  $ \wt{\bX}_{i, \cdot}$ depend on the original data vector $\bX_{i,\cdot}$ via the following functional form:
\begin{equation} \label{W_i-random-function}
    \wt{\bX}_{i, \cdot} = f( {\bX}_{i, \cdot}, \boldsymbol{R}_i), 
\end{equation} 
for $i\in [n]$, where  $f$ is a deterministic function,  and  $\boldsymbol{R}_i, \, i\in [n]$ are i.i.d. random variables (vectors) capturing the exogenous randomness in generating the knockoff variable matrix and are independent of all other variables. To fix the exogenous randomness, we set a seed for generating the random variables $\{\boldsymbol{R}_i\}_{i=1}^n$ and then apply \eqref{W_i-random-function} to construct the knockoff variable matrix $\wt{\bX}$. For any observed design matrix $\bX$ and a neighboring matrix $\bX'$, the same seed is applied to ensure that the same realization of $\{\boldsymbol{R}_i \}_{i=1}^n$ is used in \eqref{W_i-random-function} to construct $\wt{\bX}$ and $\wt{\bX}'$. 
Therefore, for identical rows in the neighboring design matrices $\bX$ and $\bX'$, the corresponding rows in the generated knockoff variable matrices $\widetilde{\bX}$ and $\widetilde{\bX}'$ will also be identical. 
Overall, when considering the sensitivity of knockoff statistics, we treat $[\bX, \wt{\bX}, \by]$ as the input dataset and the above procedure ensures that the neighboring datasets $[\bX, \wt\bX, \by]$ and $[\bX', \wt{\bX}', \by']$ differ only in a single row. To illustrate assumption \eqref{W_i-random-function}, we examine two commonly used knockoff variable generators. 
\begin{enumerate}
    \item[i)] {\it Gaussian knockoffs}. The Gaussian knockoff has been presented in \eqref{exam-Gaussian-knockoff}. The i.i.d. random variables $\bR_i \stackrel{d}{\sim} N({\bf 0}, I_p)$ and the function $f(\cdot, \cdot)$ can be written as 
    \begin{equation*}
        f(\bX_{i, \cdot}, \bR_i) =  \bX_{i, \cdot} (I_p - \bOmega \diag (\boldsymbol{r})) + \bR_i^T (2 \diag(\boldsymbol{r}) - \diag(\boldsymbol{r}) \bOmega  \diag(\boldsymbol{r})  )^{1/2}. 
    \end{equation*}

    \item[ii)] {\it Sequential conditional independent pairs}. Algorithm 1 in \cite{candes2018panning} provides a general approach to generating knockoff variables for arbitrary known covariate distribution. Specifically, it samples $\widetilde{X}_1$ from the conditional distribution $\mathcal{L} (X_1 | X_{-1})$ of $X_1$ given $X_{-1}$, and then sequentially samples $\widetilde{X}_j$ from the conditional distribution $\mathcal{L} (X_j | X_{-j}, \wt{X}_{1:(j-1)})$ for $j=2,\cdots p$. Here, $X_{-j}$ is the subvector of $X$ with $j$th entry removed for $j\in [p]$, and $\wt{X}_{1:(j-1)}$ is the subvector formed with entries in set $\{1,\cdots, j-1\}$ for $j=2,\cdots, p$.   We can perform the following procedure to implement the algorithm. For each row $\bX_{i, \cdot}$ in $\bX$, 
    we first generate a  random vector $\bR_i = (\bR_{i, 1}, \ldots, \bR_{i, p})^T$ with i.i.d. uniform components $\bR_{i, j} \stackrel{d}{\sim } U(0, 1)$. Then for $j \in [p]$,  the $j$th knockoff variable $\wt{\bX}_{i, j}$ can be constructed as $F_j^{-1}(\bR_{i,j})$, where  $F_j$ is the cumulative distribution function of the conditional distribution $\mathcal{L}(X_j| X_{-j} = \bX_{i, -j}, \wt{X}_{1:(j-1)} = \wt{\bX}_{i, 1:(j-1)})$, and $F_j^{-1} $ is the generalized inverse of $F_j$. Here to simplify the notation, we suppress $F_j$'s dependence on the realizations $\bX_{i, -j}$ and $\wt{\bX}_{i, 1:(j-1)}$. It is seen that  the dependence of $\wt{\bX}_{i, \cdot} $ on $(\bX_{i, \cdot}, \bR_i)$ can be summarized in the form of \eqref{W_i-random-function}. 
\end{enumerate}

With the above discussed mechanisms, we impose sensitivity assumption (Cf. Definition \ref{def:sensitivity}) on knockoff statistics $W_j:=w_j([\bX, \wt{\bX}],\by)$, that is, we assume $\max_{1\leq j\leq p}\Delta_2(W_j)\leq \Delta_n$ with some $\Delta_n\rightarrow 0$ as $n\rightarrow \infty$. Building on \eqref{W_i-random-function}  and setting the same random seed for generating knockoff variables, we analyze the sensitivity of various constructions of $W_j$'s in Section \ref{Sec6.Sensitivity}. 
	
\section{Differentially Private Knockoff Inference} \label{Sec:DPknockoff_theory}

 Recall the FDR definition in \eqref{def:DFR}. Let $\mathcal{H}_0 = \mathcal{S}_0^c$ be the set of null features. We assume the relevant set $\mathcal{S}_0$ is sparse in the sense that $s = |\mathcal{S}_0| = o(n \wedge p)$. Denote $p_0 = |\mathcal{H}_0|$. Our goal is to perform the model-X knockoff inference for variable selection with guaranteed FDR control while protecting differential privacy.
In scenarios where the covariate $X$ is of high dimension, disclosing the knockoff statistics for all $p$ features can result in a substantially increased variance, which poses a challenge to simultaneously protecting privacy and achieving high power. Exploiting the sparsity assumption allows us to selectively disclose only the $W_j$'s which are significantly nonzero, and such $W_j$'s likely correspond to important features in $\mathcal S_0$. This alleviates the increased variance concern without sacrificing much selection power. 
To achieve this, we propose a new mirror-peeling Algorithm \ref{alg:mirror peeling} for knockoff inference under DP. The privacy guarantee of Algorithm \ref{alg:mirror peeling} is summarized in Theorem \ref{lem:peeling}. 

\begin{algorithm}[!t] 
\caption{Knockoff Inference with Differential Privacy via Mirror Peeling}
\begin{algorithmic}[1]
\Require  Dataset $(\bX, \by)$, knockoff statistic construction functions $\{w_j\}_{j \in [p]}$ (Cf. \eqref{knockoff-W}) each with sensitivity at most $\Delta_n$, privacy parameter $\mu$, peeling size $m$, targeted FDR level $q$, seed number $a$.
\State Generate knockoff variable matrix $\wt{\bX}$ for $\bX$ using seed $a$ to fix the exogenous randomness (Cf. Section \ref{sensi-set-knockoff}).
\State Construct knockoff statistics $W_j := w_j([\bX, \wt{\bX}], \by)$, for $j \in [p]$. 
\State Let $\mathcal{S} = \{1, 2, \ldots, p\}$ be the index set of knockoff statistics.
\For {$j=1,2,\dots,m$}
\State Let 
$i_j = \argmax_{k \in \mathcal{S}} \{ |W_k| + Z_{k,j} \},$
where $Z_{k,j}$ is an independent sample from $\Lap(2\Delta_n/\varepsilon_{\mu, m})$, where $\varepsilon_{\mu,m} = \log \big(\Phi(\frac{\mu} {2 \sqrt{2m}}) / \Phi(-\frac{\mu} {2 \sqrt{2m}}) \big)$.
\State Let $\widetilde W_{i_j} = W_{i_j} + \widetilde Z_{j}$, where $\widetilde Z_{j}$ is an independent sample from $N(0, 2m\Delta_n^2/\mu^2)$. 
\State Update $\mathcal{S} = \mathcal{S}\setminus \{i_j\}$.
\EndFor
\State Let $D_m :=\{i_1, \ldots, i_m \}$ be the index set of filtered features, $\widetilde{\mathcal{W}}_{m} := \{ \widetilde{W}_{j}: j \in D_m \}$, and $|\widetilde{\mathcal{W}}_{m}| := \{ | \widetilde{W}_{j} | : j \in D_m \}$.
\State Compute the knockoff threshold 
\begin{equation} \label{def-widetilde-T}
    \widetilde{T}:= \inf \bigg\{t \in |\widetilde{\mathcal{W}}_m|: \frac{1 + \sum_{j  \in D_m} \mathbbm{1} (\widetilde{W}_{j} \leq -t) } {1 \vee  \sum_{j \in D_m} \mathbbm{1} (\widetilde{W}_{j} \geq t)  } \leq q \bigg\}
\end{equation}
and $\widetilde{T} = \infty$ if the above set is empty.
\Ensure Selected relevant set $ \widetilde{S} := \{ j \in D_m: \widetilde{W}_j \geq \widetilde{T}\} $.
\end{algorithmic}
\label{alg:mirror peeling}
\end{algorithm}

\begin{theorem}
    \label{lem:peeling}
    Algorithm \ref{alg:mirror peeling} is $\mu$-GDP.
\end{theorem}

Algorithm 1 incorporates several established privacy-preserving techniques into the knockoff framework. 
To achieve DP, random noise can be injected either into the observed data $(\bX,\by)$ (e.g., \cite{tao2025differentially}) or to the knockoff statistics $\{W_k\}$. We adopt the latter approach here. Although adding noise to the centralized data matrix may also yield a valid central-DP mechanism if calibrated properly, it is typically an input-perturbation strategy and shares the inefficiency of local-privacy-style procedures, since the raw high-dimensional observations are privatized before the statistical task is performed. For null features $j\in \mathcal H_0$, the model-X knockoff statistics $W_j$ are symmetrically distributed around 0, a key property for ensuring the exact finite-sample FDR control. This naturally motivates us to add Gaussian noise directly to $W_j$'s to preserve this property while simultaneously achieving the privacy guarantee. The peeling algorithm is a widely used iterative selection mechanism in the privacy literature \citep{dwork2021differentially,xia2023adaptive, xia2025differentially}. We adopt it here to overcome the difficulty of high dimensionality. The peeling step enables us to select the true signals with high probability while adding the noise required for privacy protection.  These intuitions will be made formal as we present our theoretical results  later.       

\subsection{Exact Finite-Sample FDR Control}
We establish the exact finite-sample FDR control of the DP model-X knockoff procedure presented in Algorithm \ref{alg:mirror peeling}. 

\begin{theorem} \label{thm-finite-FDR}
 For any $m \in [p]$, $\Delta_n > 0 $, $\mu >0 $, and for any sample size $n$, the output $\widetilde S$ from Algorithm  \ref{alg:mirror peeling} satisfies
  \(   \FDR (\widetilde{S}) \leq q. \)
\end{theorem}


Theorem \ref{thm-finite-FDR} shows that using the filtered and masked knockoff statistics $ \{\widetilde{W}_{j}: j \in D_m\} $ in knockoff procedure can still achieve \textit{exact finite-sample} FDR control, regardless of the parameter choices and the dependency of the response $Y$ on covariates $X$. This nice property stems from the mirror peeling algorithm, which filters the knockoff statistics based on their magnitudes, and the symmetry inherent in the added noises. However, the original proof of FDR control for non-DP knockoffs does not carry over directly  here because of two reasons: 1) the injection of privacy noise, and 2) the incorporation of the peeling procedure which depends on the knockoff statistics and affects the post-peeling DP-knockoff selection procedure. 

To overcome these challenges specific to the DP-knockoff inference, our proof relies on a conditioning argument and uses the following two key properties: (i) the added noises $ \bZ := \{Z_{k,j}:  k \in [p],  j\in [m]\} $ and $ \{\widetilde{Z}_{j}\}_{j = 1}^m$ are symmetric and independent of all other variables; and (ii) the filtered set $D_m$ produced by the peeling procedure  depends only on the absolute values  $\{|W_j|\}_{j = 1}^{p}$ and the independent Laplace errors $\bZ$.  Therefore, conditioning on the absolute values $\{|W_j|\}_{j=1}^p$ and $\bZ$, the filtered set $D_m$ becomes fixed and we can show that the perturbed knockoff statistics $\{\widetilde{W}_j\}_{j \in D_m \cap \mathcal{H}_0}$ satisfy the desired coin flip property, which in turn yields finite-sample FDR control. The proof of Theorem \ref{thm-finite-FDR} is presented in Section \ref{SecA.1} of the Supplement.

\subsection{Power Analysis}\label{sec:power}
We now develop a general framework for the power analysis of the DP-knockoff procedure described in Algorithm \ref{alg:mirror peeling}. We will establish sufficient conditions ensuring that this procedure asymptotically attains the same power as the original non-DP knockoff inference procedure, and consequently achieves asymptotic power one under some signal strength assumptions. 
The underlying intuition is that the original knockoff procedure only selects features with large enough knockoff statistics $W_j$'s.  Provided the added privacy noise remains relatively small, the filtered index set $D_m$ from the mirror peeling procedure will contain the majority of these features with high probability, thereby rendering the relative power loss asymptotically negligible.
Consequently, while allowing arbitrary noise variance in $Z_{i,j}$ guarantees exact finite-sample FDR control (cf.\ Theorem~\ref{thm-finite-FDR}), achieving asymptotically zero power loss requires an upper bound on the noise variance, which imposes constraints on the peeling size $m$ and the sensitivity level $\Delta_n$.

Let us first introduce some notation to streamline the presentation. Suppose that the original knockoff procedure using all the unmasked knockoff statistics $\mathcal{W} = \{ W_j: 1 \leq j \leq p\}$ computes the threshold ${T}^* $ and selects relevant subset $S^* = \{j: W_j \geq T^* \}$, where $T^*$ is defined in \eqref{knockoff-threshold}.
Recall that $\widetilde{S}$ represents the selection outcome of the DP-knockoff procedure in Algorithm \ref{alg:mirror peeling}. The {\it relative power loss} of $\widetilde{S} $ compared to the outcome $S^*$  is defined by $\e [\phi]$ with
\begin{equation} \label{formula-relative-power-loss}
     \phi :=  \frac{ | S^* \cap \mathcal{S}_0 | }{s} - \frac{ | \widetilde{S} \cap  \mathcal{S}_0 |  }{s}, 
\end{equation}
where $\mathcal{S}_0$ is the relevant set and $s = |\mathcal{S}_0|$. The total power loss is defined as 
$
1 - \mathrm{Power}(\widetilde{S}) := \frac{\e  [|\mathcal{S}_0 \cap \widetilde{S}^c| ]}{s}.
$
 
Define $b_n: = \frac{ 12 \Delta_n}{\mu} \sqrt{\pi m} \log p  $. We establish power guarantee of Algorithm \ref{alg:mirror peeling} when there exist enough relevant features with signal strength dominating both the injected privacy noise and the concentration fluctuations of the knockoff statistics, as formally summarized below. 

\begin{theorem} \label{thm-power-alter}
    Assume that  $   \mathbb{P} ( \max_{j \in [p]} |W_j - w_j| \geq \ell_n  ) \leq \kappa_n $ for some $\ell_n > 0$ and $\kappa_n \to 0$. Suppose further that $ w_j = 0 $ for $j \in \mathcal{H}_0$ and denote by $\mathcal{S}_1 := \{j \in \mathcal{S}_0:  w_j >  b_n + 2 \ell_n\}$ the set of strong signals. 
    If $ |\mathcal{S}_1| /(1 +s) \geq 1/(1 + q) $ and the peeling size $m > s$, then  the total power loss satisfies 
    \begin{equation}
        1 - \mathrm{Power}(\widetilde{S}) \leq  (1 - |\mathcal{S}_1|/s)  + \kappa_n  + O(p^{-1}),
    \end{equation}
    and the same upper bound also applies to relative power loss $\e[\phi]$.
\end{theorem}

We now discuss the implications of Theorem \ref{thm-power-alter}. 
It has been shown in \cite{fan2023ark} and \cite{fan2025asymptotic} that, for popularly studied constructions such as marginal correlation and regression coefficient difference, the knockoff statistics $W_j$ are concentrated around their population quantity $w_j$ at the rate $\ell_n = O(\sqrt{n^{-1} \log p })$ with high probability. 
Theorem \ref{thm-power-alter} ensures that if the signal strength satisfies $w_j >  b_n + 2\ell_n $ for $j \in \mathcal{S}_1$, then the proposed DP-knockoff procedure achieves asymptotic power $|\mathcal S_1|/s$. Here,  the term $2 \ell_n$ accounts for stochastic fluctuations of knockoff statistics and ensures that such variation does not void the separation between relevant  and null knockoff statistics. Therefore, $w_j > 2 \ell_n$ for $j \in \mathcal{S}_1$ guarantees that the non-DP knockoff procedure can detect all strong signals, achieving asymptotic power $|\mathcal S_1|/s$. 
The term $b_n$ stems from the privacy constraint and represents a high-probability uniform upper bound of the added privacy noises; $w_j > b_n$ for $j \in \mathcal{S}_1$ guarantees robustness of the selection outcome to the injected privacy noise.  It is worth noting that Theorem \ref{thm-power-alter}  allows the number of relevant features $s $ to be either fixed or diverging with the sample size. 

A natural question concerns the optimality of the signal strength condition in Theorem \ref{thm-power-alter}. 
To the best of our knowledge, the minimax signal strength required in large-scale multiple testing to achieve asymptotic power one under FDR control remains largely open even in the classical, non-private setting, except in a few special cases. The DP consideration imposes additional algorithmic and information-theoretic constraints, further complicating the problem. We establish in Theorem \ref{thm:minimax-signal} below a minimax lower bound on the signal strength necessary for achieving power arbitrarily close to one in high-dimensional linear models with privacy guarantee.

\begin{theorem}\label{thm:minimax-signal}
Fix $q\in(0,2/5)$ and $\mu>0$. Consider the Gaussian linear model
\(
\by = \bX\bbeta + \boldsymbol{\varepsilon},\ \boldsymbol{\varepsilon}\sim \mathcal N(0,\sigma^2 I_n),
\)
where the rows of $\bX\in\mathbb R^{n\times p}$ are i.i.d.\ isotropic sub-Gaussian with
$\E[\bX_i\bX_i^T]=I_p$. Assume $1\le s \le c p$ for a small constant $c > 0$. Let
\(
\Theta(a) :=\ \big\{\bbeta\in\mathbb R^p:\ \|\bbeta\|_0=s,\ \bbeta_j\in\{0,\pm a\}\big\}
\) with $a\in \mathbb R^+$, and let $\mathcal M$ be any selection procedure yielding estimated feature set $\widehat S$ which is $\mu$-GDP and controls
FDR at level $q$.
Moreover, assume that for arbitrarily small $\eta\in(0,1)$, $\mathcal M$ achieves power at least $1-\eta$ 
uniformly over $\Theta(a)$, that is, 
 \(
 \liminf_{n\to\infty}\ \inf_{\scriptstyle\bbeta\in\Theta(a)}\ 
 \E_{\scriptstyle\bbeta}\!\left(\frac{|\widehat S\cap \mathcal S_0|}{s} \right) >  1 - \eta.
 \)
Then it is necessary that
\begin{equation}\label{eq:minimax-signal-ineq-gdp}
a \gtrsim \sigma \max\bigg\{ \sqrt{\frac{\log (p/s)}{n}} , \frac{\sqrt{\log (p/s)}}{ \mu n  } \bigg\}.
\end{equation}
\end{theorem}


The proof of Theorem \ref{thm:minimax-signal} is presented in Section \ref{Sec:pf-minimax}. The main challenge is that DP constrains the distribution of the output $\widehat S$.
To prove the lower bound, we construct a large packing of $s$-sparse support set
$\{S_u\}_{u\in\mathcal V}$ with $\log|\mathcal V|\gtrsim s\log(p/s)$ and plant signals $\bbeta^{(u)}\in\Theta(a)$ with $\mathrm{supp}(\bbeta^{(u)})=S_u$, $u\in \mathcal V$. Assume the true parameter is $\bbeta^{(U)}$ with $U$ uniformly sampled from $\mathcal V$.
If a $\mu$-GDP algorithm outputs $\widehat S$ with controlled FDR and has high power uniformly over $\Theta(a)$, then $\widehat S$ must overlap with
the true support $S_U$ substantially while containing few nulls. This reduces the support recovery problem to a simple decoding problem of recovering the planted index $U$ with small error. Then
Fano's inequality can be applied which enforces a lower bound of the mutual information $I(U;\widehat S)\gtrsim \log|\mathcal V|$.  On the other hand, we upper bound $I(U;\widehat S)$ by some average pairwise KL divergence, which depends on the signal amplitude $a$ and can be controlled using data processing for the Gaussian model together with a privacy-driven KL bound implied by $\mu$-GDP.
Comparing the resulting upper and lower bounds gives the claimed lower bound of $a$.

The lower bound in \eqref{eq:minimax-signal-ineq-gdp} consists of the maximum of two terms: the standard non-private minimax rate and the cost from the privacy consideration. When the privacy budget $\mu$ is constant, the non-private term dominates and the privacy cost is negligible. Notably, the sparsity parameter $s$ appears only within the logarithmic factor. A possible explanation is that we consider a linear model with independent covariates,  in which setting the total signal strength accumulates linearly with $s$. In contrast, our DP-knockoff is not restricted to independent covariates or linear models. This result shows that, when $s$ is of constant order, the signal strength requirement in Theorem \ref{thm-power-alter} with marginal-correlation-based knockoff statistics is minimax-optimal up to constant factors; see Section \ref{Sec:sensi-mar-corr} for more detailed reasoning.    

Under a weaker signal strength condition  $w_j > 2 \ell_n$ for $j \in \mathcal{S}_0$ without requiring it to dominate $b_n$, we can establish a parallel power analysis for Algorithm~\ref{alg:mirror peeling} and derive an explicit bound on the power loss as a function of $b_n$. Due to space limitations, this result is presented in Theorem \ref{thm-power-general} in the Supplementary Material, with a detailed discussion on the corresponding convergence rate of the power loss  in Section \ref{Sec:addition-power-algo1}.


\section{Sensitivity Analysis of Knockoff Statistics}  \label{Sec6.Sensitivity}
Section \ref{Sec:DPknockoff_theory} shows that the proposed Algorithm \ref{alg:mirror peeling} achieves finite-sample FDR control as long as symmetrically distributed noise is added to the knockoff statistics, regardless of the noise level. However, Theorem \ref{thm-finite-FDR} suggests that larger sensitivity $\Delta_n$ (hence larger $b_n$) requires stronger signal to achieve asymptotic power one. Additionally, the relative power analysis indicates that $b_n$ related to the noise level needs to be sufficiently small to guarantee asymptotically vanishing power loss, which in turn requires effective control of the sensitivity level $\Delta_n$  (see \eqref{eq-b_n-constraints} for constraints on $\Delta_n$ in an example).  Combining these, it is important to understand the sensitivity level $\Delta_n$ for various choices of knockoff statistics. It is also important to note that the sensitivity requirement in the privacy context needs to hold \textit{almost surely} (Cf. Definition \ref{def:sensitivity}), making many existing results in the robust statistics literature inapplicable because they are mostly probabilistic statements.

Adopting the framework established in Section \ref{sensi-set-knockoff} to construct the knockoff variable matrix using a preset random seed, we obtain two neighboring datasets $D:=[\bX, \widetilde{\bX}, \by]$ and $D':=[\bX', \wt{\bX'}, \by']$ differing only in a single row. 
We assume the covariate vector satisfies $\| X \|_{\infty}:= \max_{j \in [p]} |X_j| \leq C_x$  and the response variable satisfies $|Y| \leq C_y$ almost surely, where the upper bounds $C_x$ and $C_y$ can grow slowly with $n$. Since the knockoff variable vector $\wt X$ and $X$ share the same marginal distribution, this automatically ensures the same upper bound for the knockoff variables.

In this section, we analyze the sensitivity of some popularly studied knockoff statistics in Sections \ref{Sec:sensi-mar-corr} -- \ref{Sec:sensi-SGD}, including those based on marginal correlation, Hilbert-Schmidt independence criterion (HSIC), ridge regression coefficient difference, and coefficient difference obtained via stochastic gradient descent.

\subsection{Marginal Correlation Knockoff Statistics} \label{Sec:sensi-mar-corr}

Given dataset $D = (\bX, \wt{\bX}, \by)$,  the marginal correlation knockoff statistic is defined as $W_j(D) =  n^{-1} ( |\bX_j^T \by | -  |\widetilde{\bX}_j^T \by | )$, for $j \in [p]$. 
The following lemma presents the sensitivity level of the knockoff statistics $W_1, \ldots, W_p$ with respect to the dataset $D$. The proof of Lemma \ref{lemma-W-sensi} is provided in Section \ref{Sec.B4}. 
\begin{lemma} \label{lemma-W-sensi}
      For neighboring datasets $D$ and $D'$, it holds that $\max_{1 \leq j \leq p}| W_j(D) - W_j(D') | \leq \Delta_n := 4 n^{-1} C_x C_y$. 
\end{lemma} 

We discuss the signal strength required by Theorem \ref{thm-power-alter} to achieve asymptotic power one with marginal correlation knockoff statistics.
The variance of $W_j$ is in the order of $O(n^{-1})$ for $j \in [p]$, leading to $\ell_n \asymp \sqrt{n^{-1} \log p}$, where the term $\sqrt{\log p}$ is to account for the effect of high dimensionality. Assuming 
$m \asymp s$, we obtain from Lemma \ref{lemma-W-sensi} that $b_n \asymp (\mu n )^{-1}\sqrt{s} \log p$. Therefore, the signal strength requirement translates into $w_j \gtrsim \sqrt{n^{-1} \log p} + (\mu n )^{-1}\sqrt{s} \log p$ in the current setting. 
This shows that when $s$ and $\mu$ are finite and $\log p = O(n)$, the marginal-correlation knockoff statistic attains the minimax-optimal signal strength in the setting of Theorem~\ref{thm:minimax-signal}.

\subsection{Knockoff Statistics Based on HSIC}\label{Sec:sensi-hsic}
To detect non-linear dependence between covariates and the response, a popularly used method in the machine learning literature is the Hilbert-Schmidt independence criterion (HSIC) proposed in \cite{gretton2007kernel}. 
Under some regularity condition, it is well known that the population HSIC between $X_j$ and $Y$ is zero if and only if $X_j \independent Y$, making HSIC a powerful tool for capturing dependence in a wide spectrum of applications. Specifically, given two kernel functions $k(\cdot,\cdot)$ and $l(\cdot,\cdot)$, the empirical HSIC between $X_j$ and $Y$ is defined as
\begin{align*}
     {\mbox{HSIC}_j} =& \frac{1}{n^2}\sum_{i_1, i_2 = 1}^{n}k(Y_{i_1}, Y_{i_2})l(\bX_{i_1, j}, \bX_{i_2, j}) + \frac{1}{n^4}\sum_{i_1, i_2 , i_3, i_4= 1}^{n}k(Y_{i_1}, Y_{i_2})l(\bX_{i_3, j}, \bX_{i_4, j}) \\
    &- \frac{2}{n^3}\sum_{i_1, i_2 , i_3= 1}^{n}k(Y_{i_1}, Y_{i_2})l(\bX_{i_1, j}, \bX_{i_3, j}).
\end{align*}
Similarly, we can define the empirical $\wt{\mbox{HSIC}}_j$ between $\wt{X}_j$ and $Y$.
An important observation is that distance correlation \citep{Szekely2007} can be represented as a special case of HSIC. In particular, when one chooses the distance-induced kernels, the corresponding HSIC reduces to the squared distance covariance, and hence distance correlation can be viewed within the same kernel-based framework.
Given dataset $ D = (\bX, \wt{\bX}, \by) $, the knockoff statistics based on HSIC is given by 
\begin{equation}
    W_j(D) = | \mbox{HSIC}_j (D) | - |\wt{\mbox{HSIC}}_j (D)| \quad \mbox{for}~ j \in [p]. 
\end{equation}
The following lemma characterizes the sensitivity of HSIC knockoff statistics. The proof can be found in Lemma 6 of \cite{kim2023differentially}.
\begin{lemma}\label{lem: hsic}
    Assume the kernel functions are bounded $k(\cdot,\cdot)\leq K$ and $l(\cdot,\cdot)\leq L$. $K$ and $L$ are positive constants. For neighboring datasets $D$ and $D'$, the sensitivity of HSIC knockoff statistics is bounded as $\max_{1 \leq j\leq p} | W_j(D)- W_j(D')|\leq \Delta_n:=\frac{8}{n}\sqrt{KL}.$
\end{lemma}

We connect to Theorem \ref{thm-power-alter} under the signal strength requirement. According to \cite{gretton2007kernel}, the concentration order for empirical HSIC varies depending on whether the two variables are independent. Specifically, it is  of order $O(n^{-1/2})$ when  $X_j$ and  $Y$ are dependent, and of order $O(n^{-1})$ when they are independent. This indicates that $\ell_n$ may be of order $n^{-1/2}$, up to some polynomial order of $\log p$ for accounting the dimensionality. Thus,  the signal strength requirement is $\asymp n^{-1}\sqrt{s} \log p + n^{-1/2}$ (up to a polynomial order of $\log p$) if we assume that $\mu$ is constant and $m\asymp s$. 


\subsection{Ridge Regression Coefficients} \label{Sec:sensi-reg-coef-dif}
Given data $D = (\bX, \wt{\bX}, \by)$, denote the augmented design matrix as $ \breve{\bX} = [\bX, \wt{\bX}] $ and define  the ridge regression coefficient as 
\begin{equation}\label{def:ridge-coeff}
    \widehat{\bbeta}(D) = \big(n^{-1} \breve\bX ^T \breve\bX + \lambda I_{2p} \big)^{-1}  n^{-1} \breve\bX^T \by.
\end{equation}
The knockoff statistics based on ridge regression coefficient difference is given by 
\begin{equation}\label{eq:ridge}
    W_j(D) = |\widehat{\bbeta}_j(D)| -  |\widehat{\bbeta}_{j+p} (D)|,  \quad \mbox{for} ~ j \in [p]. 
\end{equation}
Denote $\mathbf{W}(D) = (W_1(D), \ldots, W_p(D))$.
The following result establishes the sensitivity level for  $\mathbf{W}(D)$. The proof of Lemma \ref{le-sensi-ridge} is postponed to Section \ref{pf-le-ridge}. 
\begin{lemma} \label{le-sensi-ridge}
For neighboring datasets $D$ and $D'$, we have
\begin{equation}
\begin{split}
      \| \mathbf{W} (D) -   \mathbf{W}(D') \|_2  & \leq \Delta_n , 
\end{split}
\end{equation}
with 
\begin{align*}
\Delta_n & = 2\sqrt{2} C_x^2 C_y  n^{-1} p  \cdot \min\{\lambda^{-3/2},  \sigma_{\min}^{-1}(n^{-1} (\breve\bX') ^T \breve\bX' ) \sigma_{\min}^{-1/2}(n^{-1} \breve\bX^T \breve\bX ) \}   \\
     & \quad + 4 C_x C_y n^{-1} p^{1/2} \cdot  \min\{\lambda^{-1}, \sigma_{\min}^{-1} (n^{-1} (\breve\bX')^T \breve\bX' ) \}
\end{align*}
where $\sigma_{\min}(\cdot)$ denotes the smallest eigenvalue of a matrix. 
\end{lemma}

\begin{remark}
The dimensionality $p$ appears in the upper bound. 
To handle the situation where $n \lesssim p$,  we propose a framework in Section \ref{Sec:highdim-knockoff} on high-dimensional knockoff inference with differential privacy by incorporating an initial screening procedure prior to the DP-knockoff inference.  
\end{remark}

 \begin{remark}
 It is seen that a large $\lambda$ reduces the value of $\Delta_n$ which has a positive effect on power preservation. However, since large $\lambda$ can create bias in $\widehat{\bbeta}_j(D)$ as a variable importance measure, it can also harm the selection power. The optimal $\lambda$ for the highest power is challenging and we leave it for future study.   
 \end{remark}


For this construction, assuming $C_{x}$ and $C_{y}$ are constants,  we obtain that the sensitivity level of knockoff statistic $\Delta_n$ is of order $ O(n^{-1} p \lambda^{-3/2}) $. Moreover, the order of $\ell_n $ is $n^{-1/2}$ (up to some polynomial order of $\log p$ for accounting dimensionality) if the smallest eigenvalue of the covariance matrix $ n^{-1} \breve\bX^T \breve\bX $ is lower bounded by a positive constant with high probability. Therefore, connecting to Theorem \ref{thm-power-alter}, the signal strength condition is $n^{-1/2} + n^{-1} p \lambda^{-3/2}\sqrt{s} \log p$, up to some polynomial order of $\log p$, if $\mu$ is a constant and $m\asymp s$.

We note that although sparsity has been popularly considered in the literature to address the challenge of high dimensionality, we do not take this approach here to construct knockoff statistics. As explained at the beginning of this section, the robustness analyses of high dimensional sparse regression such as Lasso are mostly probabilistic. So, these results cannot be directly applied here to characterize the sensitivity of knockoff statistics based on sparse regression.      
 
\subsection{Stochastic Gradient Descent} \label{Sec:sensi-SGD}

We now consider knockoff statistics based on regression coefficients estimated using stochastic gradient descent (SGD) for minimizing a convex objective function. 
    Recent studies \citep{hardt2016train, kissel2022high} showed that the output of SGD has bounded sensitivity when only one dataset entry is altered. Specifically, we consider regressing the response $\by$ on the augmented design matrix $[\bX, \widetilde{\bX}]$ and estimate the regression coefficient $\bbeta \in \mathbb{R}^{2p}$ by the M-estimation: $\wh\bbeta = \underset{\scriptstyle{\bbeta} \in \mathbb{R}^{2p}}{\argmin} \sum_{i=1}^{n}\psi(\bbeta, D_i).$
	Here $\psi$ is a convex objective function, and $D_i$ denotes the $i$-th data point in $D = (\bX, \wt{\bX}, \by)$. SGD estimates the parameters by iteratively updating $\bbeta^{(t+1)} = \bbeta^{(t)} - \eta_{t}\psi'(\bbeta^{(t)}, D_{t+1})$, where $t =   0, \dots, n-1$. Here, $\eta_{t}$ is the step size and $\psi'(\bbeta^{(t)}, D_{t+1})$ is the derivative of the objective function $\psi(\bbeta^{(t)}, D_{t+1})$ with respect to the parameter $\bbeta^{(t)}$. The final estimated coefficient is $\wh\bbeta(D) : = \bbeta^{(n)}$.  For $j \in [p]$, the knockoff statistic is defined by $W_j:= |\wh\bbeta_j(D)| - |\wh\bbeta_{j+p}(D)| $, where $\wh\bbeta_j(D)$ is the $j$th component in the coefficient vector $\wh\bbeta(D)$.  Denote $\mathbf{W}(D) = (W_1(D), \ldots, W_p(D))$.
    We first define the strongly convex functions.
    \begin{definition}
    A function \( g: \mathbb{R}^n \rightarrow \mathbb{R} \) is said to be $\kappa$-strongly convex if there exists a constant \( \kappa > 0 \) such that for all \( z_1, z_2 \in \mathbb{R}^n \) and for all \( \alpha \in [0, 1] \), it holds that
    \begin{equation*}
    g(\alpha z_1 + (1-\alpha) z_2) \leq \alpha g(z_1) + (1-\alpha) g(z_2) - \frac{\kappa}{2} \alpha (1-\alpha) \|z_1 - z_2\|^2.
    \end{equation*}
    \end{definition}
    
	\begin{lemma}\label{lem: sgd}
		Assume  that given any data point $D_i$, $\psi(\bbeta, D_i)$ is $\kappa$-strongly convex and $L_1$-Lipschitz, and $\psi'(\bbeta, D_i)$ is $L_2$-Lipschitz. If $\eta_{t} = t^{-c}/L_2$ for $1 \leq t \leq n-1$ with some $0<c<1$ and $\kappa/L_2\geq \frac{c(1-c)}{1-2^{-(1-c)}}\frac{\log n}{n^{1-c}}$, then $\|\mathbf{W}(D) - \mathbf{W}(D') \|_2 \leq \Delta_n :=\frac{4L_1}{L_2}n^{-c}.$   
	\end{lemma}
    Lemma \ref{lem: sgd} is an immediate corollary of Proposition 5.1 in \cite{kissel2022high}.
    We next illustrate Lemma \ref{lem: sgd} through a commonly used example of ridge regression objective function with penalization parameter $\lambda >0$. For simplicity, assume $  \max_{j \in [p]} |X_j| \leq M$, $ |Y| \leq M $ and $\bbeta \in \mathbb{B}(R_{\beta})$, for some quantities $M > 0$ and $R_{\beta} > 0$, where $\mathbb{B}(R_{\beta})$ denotes the Euclidean ball with radius $R_{\beta}$. Example 3 in \cite{kissel2022high} showed that the ridge regression objective function is $\kappa$-strongly convex with $\kappa = \lambda$ and $L_1$-Lipschitz with $ L_1 = M^2(2p+1) (1+R_\beta)+\lambda R_\beta $, and its first derivative is $L_2$-Lipschitz with $L_2 = M^2(2p+1) + \lambda$. 
    Note that $L_1/L_2 \leq 1 + R_{\beta}$. 
    Therefore, if $\eta_t = t^{-c} / (M^2(1 + 2p) + \lambda) $ and $\lambda \geq  \frac{2 c(1 - c)}{1 - 2^{-(1 - c)}}  M^2(1 + 2p)   \frac{\log n}{n^{1 - c}} $, then it holds for large $n$ that 
    $$
    \|\mathbf{W}(D) - \mathbf{W}(D') \|_2 \leq \Delta_n:= 4 (1 + R_{\beta}) n^{-c}. 
    $$
    
    {In this case, when the concentration rate of $\bbeta_j(D)$ is $O(n^{-c/2})$ for $j \in [2p]$, then $\ell_n\asymp n^{-c/2}$ (up to a logarithmic term of $\log p$ for accounting the high dimensionality), and the signal strength assumption in Theorem \ref{thm-power-alter} becomes $n^{-c/2}+ n^{-c}\sqrt{s} \log p$, up to some logarithmic term of $\log p$, if $\mu$ is constant and $m\asymp s$. }

\section{High-dimensional Differentially Private Knockoff Inference} \label{Sec:highdim-knockoff}

The examples in Section  \ref{Sec6.Sensitivity} demonstrate that the sensitivity level can be dimension-free when knockoff statistics are constructed based on marginal dependence measures, such as marginal correlation and HSIC. However, the sensitivity becomes dimension-dependent if joint dependence measures such as regression coefficients are used to construct knockoff statistics, restricting that the dimension $p$ cannot be too large compared with sample size $n$. Yet, knockoff statistics based on joint dependence measures are generally more powerful than the ones based on marginal dependence measures. 
Motivated by this, we introduce a two-step framework that begins with a differentially private screening procedure, followed by the private knockoff procedure applied to the filtered features. We will show that our two-step procedure achieves unconditional FDR control while preserving privacy. Since the initial screening step is used to reduce the dimensionality, we recommend using knockoff statistics with dimension-free sensitivity. For example, we may use marginal correlation or the HSIC statistics introduced in Section 4. The dimension-free sensitivity allows us to preserve DP by adding very small noise to the data regardless of the dimensionality. For the second step, we recommend constructing knockoff statistics using joint dependence measures for better power performance.

Algorithm \ref{algo:dp-screen} introduces a general DP feature screening procedure. 
\begin{algorithm}[h]
\caption{Feature Screening with Privacy Guarantee}
\begin{algorithmic}[1]
\Require Dataset $(\bX, \by) \in \mathbb{R}^{n \times (p+1)}$, screening size $K_n$, privacy parameter $\mu$.
\medskip
\State Construct screening statistics $U_j := u(\bX_j, \by)$, for $j \in [p]$, each with sensitivity at most $\Delta_{u}$. 
\State Let $\mathcal{S} = \{1, 2, \ldots, p\}$ be the index set of screening statistics.
\For {$j=1,2,\dots,K_n$}
\State Let 
$$i_j = \argmax_{k \in \mathcal{S}} \{ U_k + Z_{k,j} \},$$
where $Z_{k,j}$ is an independent sample from $\Lap(2\Delta_u /\varepsilon_{\mu, K_n})$, where $\varepsilon_{\mu,K_n} = \log \big(\Phi(\frac{\mu} {2 \sqrt{K_n}}) / \Phi(-\frac{\mu} {2 \sqrt{K_n}}) \big)$.
\State Update $\mathcal{S} = \mathcal{S}  \setminus \{i_j\}$.
\EndFor
\State Let $\mathcal{C} :=\{i_1, \ldots, i_{K_n} \}$ be the index set of filtered features
\Ensure  $\mathcal{C}$.
\end{algorithmic}
\label{algo:dp-screen}
\end{algorithm}

\begin{proposition}\label{prop: screening-dp}
    Algorithm \ref{algo:dp-screen} is $\mu$-GDP.
\end{proposition}

Given dataset $ (\bX, \by)$, we randomly split it into two disjoint subsets $ (\bX_{i, \cdot},  \by_{i})_{i \in I_1}$ and $ (\bX_{i, \cdot}, \by_{i})_{i \in I_2} $ with subsample size $n_1 := |I_1|$ and $n_2:= |I_2|=n-n_1$, respectively. For simplicity, we use the shorthand notations $(\bX_{I_l,\cdot}, \by_{I_l})$, $l=1, 2$,  to denote these two subsets.  In the first step, given a prespecified screening size $K_n$, we propose an initial screening procedure  with privacy guarantee (as described in Algorithm \ref{algo:dp-screen}) and apply it to the sub-dataset $(\bX_{I_1,\cdot}, \by_{I_1})$ to obtain a filtered feature subset $\mathcal{C}$  with $|\mathcal{C}| = K_n$. Proposition \ref{prop: screening-dp} shows that the filtered feature subset satisfies $\mu$-GDP.



In the second step, we construct the knockoff variable matrix $\wt{\bX}_{I_2, \cdot}$ for $\bX_{I_2, \cdot}$ using a prespecified seed to fix the exogenous randomness (Cf. Section \ref{sensi-set-knockoff}). We then compute the knockoff statistics $W_j := w_j([\bX_{I_2, \mathcal{C}}, \wt{\bX}_{I_2, \mathcal{C}}], \by_{I_2}) $, $j \in \mathcal{C}$,  based on the filtered features, where $\bX_{I_2, \mathcal{C}}$ represents the submatrix of $\bX_{I_2, \cdot}$ formed by extracting out all the columns in $ \mathcal{C}$, and  $\wt{\bX}_{I_2, \mathcal{C}}$ is defined similarly.  
In the last step, we add Gaussian noise to the filtered knockoff statistics $\{W_j\}_{j \in \mathcal{C}}$ derived in the second step and select the set of relevant features via knockoff inference. The detailed procedure is provided in Algorithm \ref{algo:screen-knockoff-single}.

\begin{algorithm}[h]
\caption{Post-screening DP-Knockoff Inference with Single Split}
\begin{algorithmic}[1]
\Require Dataset $(\bX, \by) \in \mathbb{R}^{n \times (p+1)}$, subsample sizes $n_1$ and $n_2 = n - n_1$, screening size $K_n$,   privacy parameter $\mu$, target FDR level $q$, seed numbers $a$ and $b$.
\medskip
\State Randomly split (using seed number $a$ to fix the randomness) dataset $(\bX, \by)$ into two parts: $(\bX_{I_1, \cdot}, \by_{I_1})$ and $(\bX_{I_2, \cdot}, \by_{I_2}) $ with $|I_1| = n_1$ and $|I_2 | = n_2$.
\State Conduct the DP screening Algorithm \ref{algo:dp-screen} using data $(\bX_{I_1, \cdot}, \by_{I_1})$ with privacy parameter set as $\mu/\sqrt 2 $
and obtain a filtered subset of features $\{ X_j \}_{j \in \mathcal{C}}$ with size $|\mathcal{C}| = K_n$.
\State Construct a knockoff variable matrix $\widetilde{\bX}_{I_2, \cdot} \in \mathbb{R}^{n_2 \times p}$ for the data matrix $\bX_{I_2, \cdot}$, using  seed number $b$ to fix the exogenous randomness.
\State Compute knockoff statistics $\{W_j\}_{j \in \mathcal{C}}$ for filtered features based on data $(\bX_{I_2, \mathcal{C}}, \widetilde{\bX}_{I_2, \mathcal{C}}, \by_{I_2}) $; Suppose the   $l_2$ sensitivity level of the filtered knockoff statistics $\{W_j \}_{j \in \mathcal{C}}$ is at most $\Delta_w$. 
\State For $j \in \mathcal{C}$, let $\widetilde{W}_j = W_{j} + Z_{j}$, where $ Z_j$ is independently sampled from $N(0, 2 \Delta_w^2  / \mu^2 )$.   
\State  Compute the threshold 
\begin{equation}
\widetilde{T}  :=  \inf \bigg\{t > 0: \frac{1 + \sum_{j \in  \mathcal{C}}  \mathbbm{1} (\widetilde{W}_j  \leq -t) } {1 \vee  \sum_{j \in \mathcal{C} } \mathbbm{1} (\widetilde{W}_j  \geq t)  } \leq q \bigg\}
\end{equation}
and $\widetilde{T} = \infty$ if the above set is empty.
\Ensure The selected set $\widehat{\mathcal{S}}_1 := \{j \in \mathcal{C}:  \widetilde{W}_j \geq \widetilde{T} \}$. 
\end{algorithmic}
\label{algo:screen-knockoff-single}
\end{algorithm}

Theorems \ref{thm-GDP-single} and \ref{thm-FDR-single} establish the differential privacy guarantee and finite-sample FDR control of Algorithm \ref{algo:screen-knockoff-single}, respectively. We also remark that the unconditional FDR control in Theorem \ref{thm-FDR-single} is achieved by constructing knockoff matrix $\widetilde{\bX}_{I_2, \cdot} $ for all $p$ features jointly; if  knockoff variables are only constructed for features in subset  $\mathcal C$, then there is no guarantee of unconditional FDR control due to randomness of $\mathcal{C}$.

\begin{theorem} \label{thm-GDP-single}
    Algorithm \ref{algo:screen-knockoff-single} is $\mu$-GDP. 
\end{theorem}
\begin{theorem} \label{thm-FDR-single}
   For Algorithm \ref{algo:screen-knockoff-single}, we have  $\FDR(\widehat{\mathcal{S}}_1) \leq q$.
\end{theorem}

Note that Algorithm \ref{algo:screen-knockoff-single} uses a peeling procedure for initial screening at the first step. Therefore, we only add Gaussian noises to mask the knockoff statistics without mirror peeling in the  step of post-screening knockoff inference. Owing to the dimension reduction achieved by the screening procedure, the required noise level for privatizing the knockoff statistics is substantially reduced. Specifically, the variance of the added noise $Z_j$ in step 5 of Algorithm \ref{algo:screen-knockoff-single} is $2 \Delta_w^2  / \mu^2$, where $\Delta_w$ is the $l_2$ sensitivity which may depend on the reduced dimension $K_n$ rather than the original data dimension $p$.
To make full use of data, we further extend Algorithm \ref{algo:screen-knockoff-single} to Algorithm \ref{algo:screen-knockoff-multi} with multiple splits in Section \ref{SecA:Multi-split} of the Supplementary Material.  

Finally, we remark that the screening size $K_n$ balances the trade-off between the power of the initial screening procedure and the power of the post-screening knockoff procedure. On the one hand, a larger screening size in the first step enhances the power of the initial DP screening procedure, allowing more features in the filtered set. On the other hand, in the second step, if the knockoff statistics are constructed based on a joint dependence measure with dimensionality-dependent sensitivity, a larger $K_n$ leads to a higher sensitivity level $\Delta_w$, which in turn increases the variance of the added normal noise to the knockoff statistics and can potentially reduce the power of the post-screening DP-knockoff inference. In Section \ref{Sec:power-Algo-single-split} of the Supplementary Material,  Theorem \ref{thm-power-single-split} establishes power guarantees of Algorithm \ref{algo:screen-knockoff-single}  under appropriate  signal strength conditions.  

\section{Simulation}\label{sec:simu}

    In this section and Section \ref{Sec:add-simu} of the Supplementary Material, we conduct simulations to verify the performance of the proposed algorithms. Consider the linear model  $Y = X^T \beta +\varepsilon$, with $\varepsilon\asymp N(0,1)$. The covariate vector $X$ is drawn from a $p$-dimensional Gaussian distribution with mean vector zero and covariance matrix $\Sigma$, which has an autoregressive structure with the $(i,j)$-th element being $\Sigma_{i,j} = 0.5\times 0.3^{|i-j|}$. Because the algorithms require bounded support of the data, we truncate $X$ to be within $[-C_x, C_x]$, and set $C_x = 1.5$. We truncate $Y$ to be within $[-C_y, C_y]$, and set $C_y = C_1 \sqrt{\log(n)}$ 
    with $C_1=1.5$. 
    Here the choice of $\sqrt{\log(n)}$ is motivated by standard maximal inequalities for sub-Gaussian random variables.
	
	We set the first ten elements of $\beta$ to be non-zero, and the remaining to be 0.  Thus, only the first 10 predictors are related to the response variables. To implement Algorithm \ref{algo:screen-knockoff-single}, we start by splitting the samples into two parts $D_1$ and $D_2$ with equal sample sizes. First, we apply Algorithm \ref{algo:dp-screen} to $D_1$, using the marginal correlation as the screening statistic and allocating a privacy budget of $\mu/\sqrt{2}$. The number of predictors selected during this screening step, denoted as $K_n$, is set to 20. Subsequently, we generate the knockoff matrix using the dataset $D_2$. To ensure the unconditional FDR control, we generate the knockoff copy using all the predictors, then extract the knockoff variables corresponding to the screened predictors. 
    The knockoff statistics are constructed as \eqref{eq:ridge} using ridge regression. 
    To protect privacy on $D_2$, we add Gaussian noises drawn from $   N(0, 2 \Delta_n^2/\mu^2)  $ to the knockoff statistics, where $\Delta_n$ is the sensitivity of the knockoff statistics constructed using ridge regression coefficients and $\Delta_n \leq 2\sqrt{2} C_x^2 C_y  n^{-1} p  \lambda^{-3/2}  + 4 C_x C_y n^{-1} p^{1/2}  \lambda^{-1}$ based on Lemma \ref{le-sensi-ridge}.  
		
	To select the regularization parameter $\lambda$ in ridge regression, it is easy to derive that under the orthogonal setting, the choice of $\lambda$ that achieves the smallest mean square error of the estimated coefficients is proportional to $s/\|\beta\|^2$, where $s=10$ is the number of relevant variables in the model. In the simulation, we directly set $\lambda = s/\|\beta\|^2$ for simplicity.

    The target FDR level is set to $q=0.2$ for all experiments. We evaluate the performance of three methods: the differentially private knockoff method (DP), its non-private counterpart (NP), and the method proposed in \cite{cai2023private}, which is referred to as SLR (Simple Linear Regression). Note that the method in \cite{cai2023private} is only suitable for high-dimensional linear model and does not have finite-sample FDR control. For each method, we report the average FDR and statistical power based on 100 repetitions. In the non-private knockoff procedure, we similarly follow a three-step approach: (1) split the data, (2) perform feature screening to select relevant variables, and (3) apply the knockoff procedure with ridge regression coefficients difference knockoff statistics to control the FDR. For the non-private ridge regression, the regularization parameter $\lambda$ is selected by cross-validation. The SLR method follows the same settings as in \cite{cai2023private}. The code to reproduce the simulation results is publicly available at \url{https://github.com/zhanruicai/DP-knockoff}.

\begin{figure}[!h]
		\centering
		\begin{subfigure}[b]{0.48\textwidth}
			\centering
			\includegraphics[width=\textwidth]{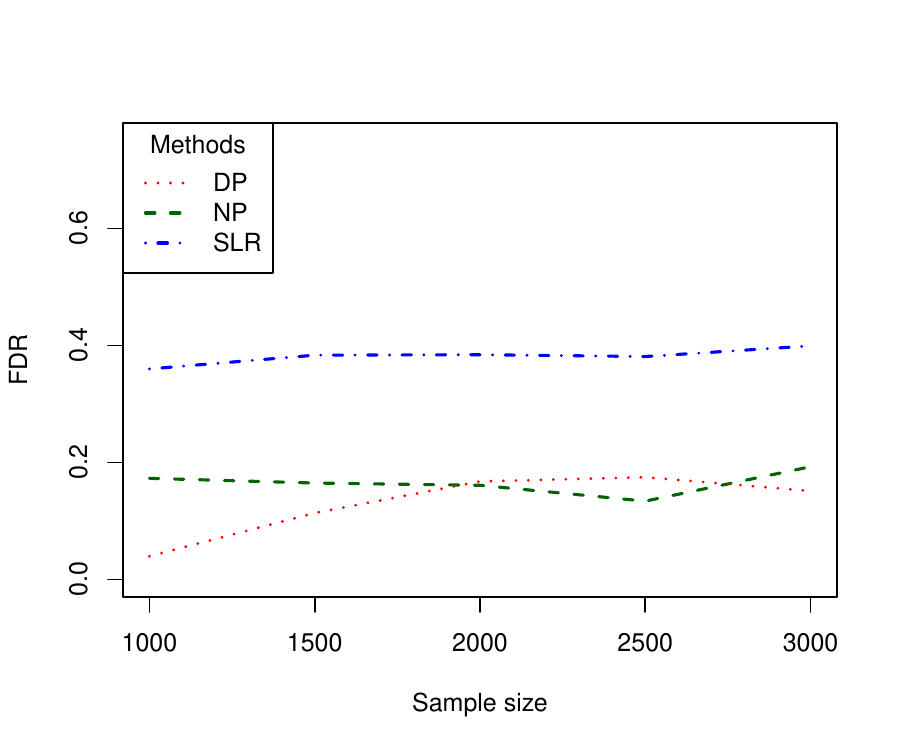}
		\end{subfigure}
		\hfill
		\begin{subfigure}[b]{0.48\textwidth}
			\centering
			\includegraphics[width=\textwidth]{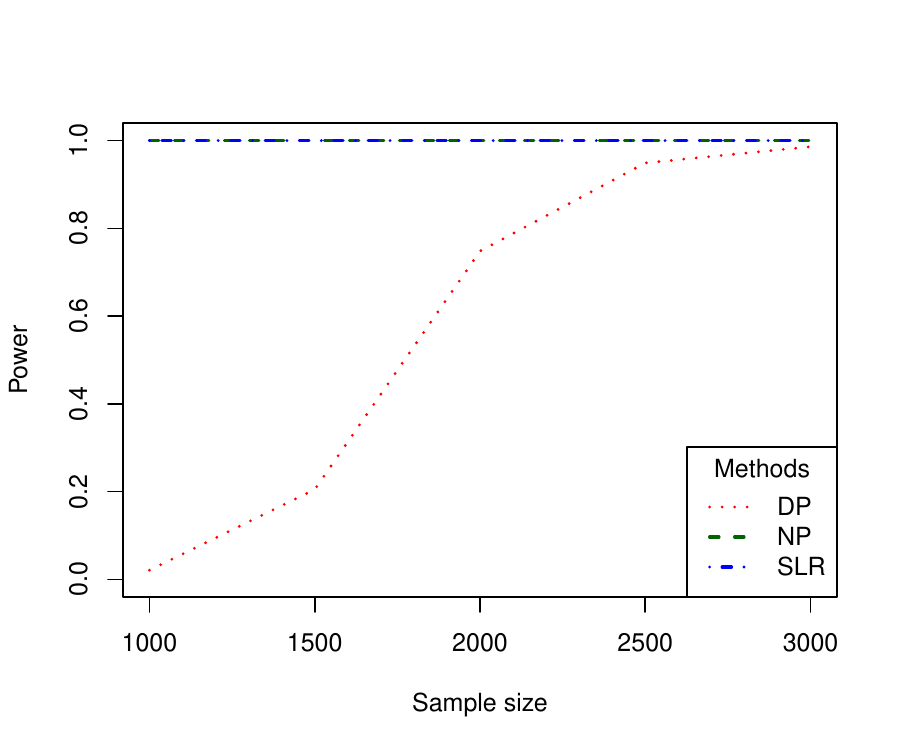}
		\end{subfigure}
		\caption{The FDR and Power of the DP-knockoff (DP), the non-private knockoff (NP), and the SLR method proposed by \cite{cai2023private}. 
        $p$ is fixed at $1000$. $\beta = 1$, and the privacy budget $\mu = 1$.} 
		\label{fig:dp.np.sig6}
	\end{figure}

\begin{figure}[!h]
		\centering
		\begin{subfigure}[b]{0.48\textwidth}
			\centering
			\includegraphics[width=\textwidth]{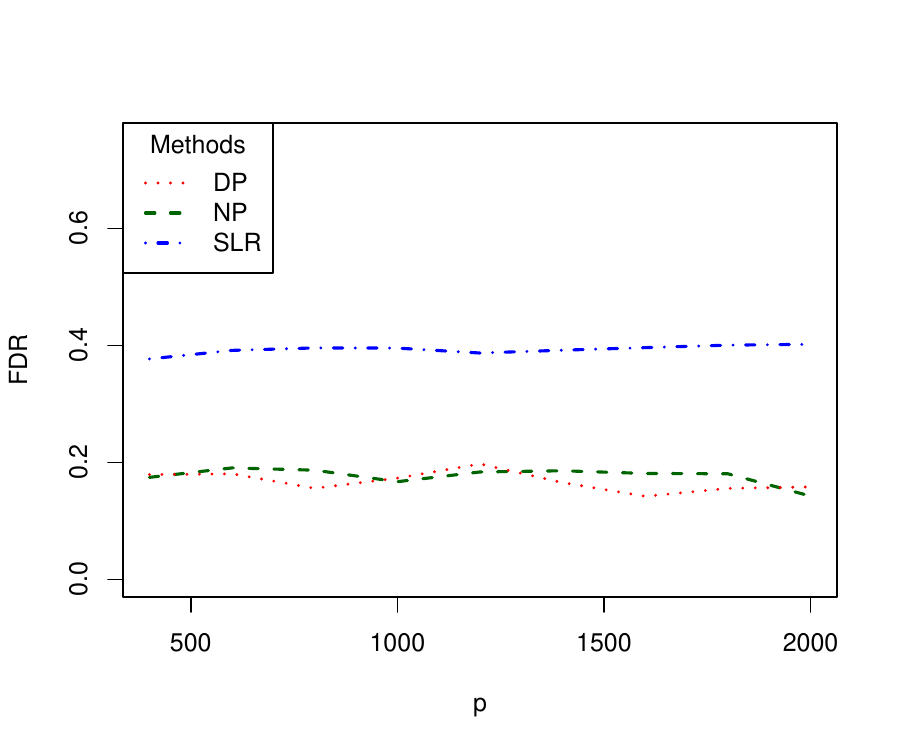}
		\end{subfigure}
		\hfill
		\begin{subfigure}[b]{0.48\textwidth}
			\centering
			\includegraphics[width=\textwidth]{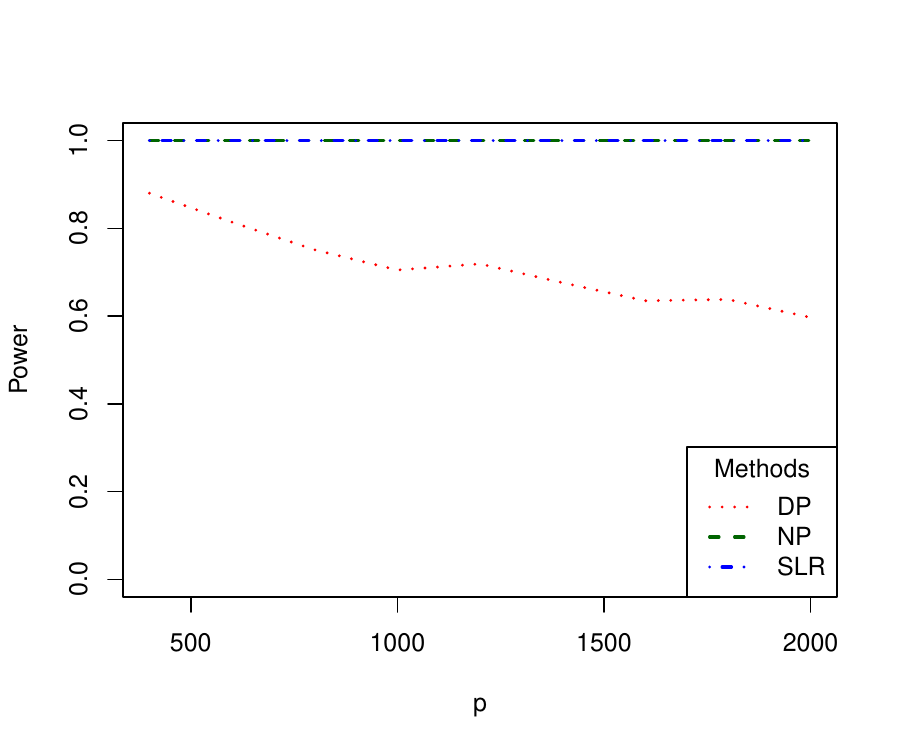}
		\end{subfigure}
		\caption{The FDR and Power of the DP-knockoff (DP), the non-private knockoff (NP), and the SLR method proposed by \cite{cai2023private}. $p$ increases from 400 to 2000, and $n$ is fixed at $2000$.  $\beta = 1$ and the privacy budget $\mu = 1$. }
		\label{fig:P_increase}
	\end{figure}

\begin{figure}[!h]
		\centering
		\begin{subfigure}[b]{0.48\textwidth}
			\centering
			\includegraphics[width=\textwidth]{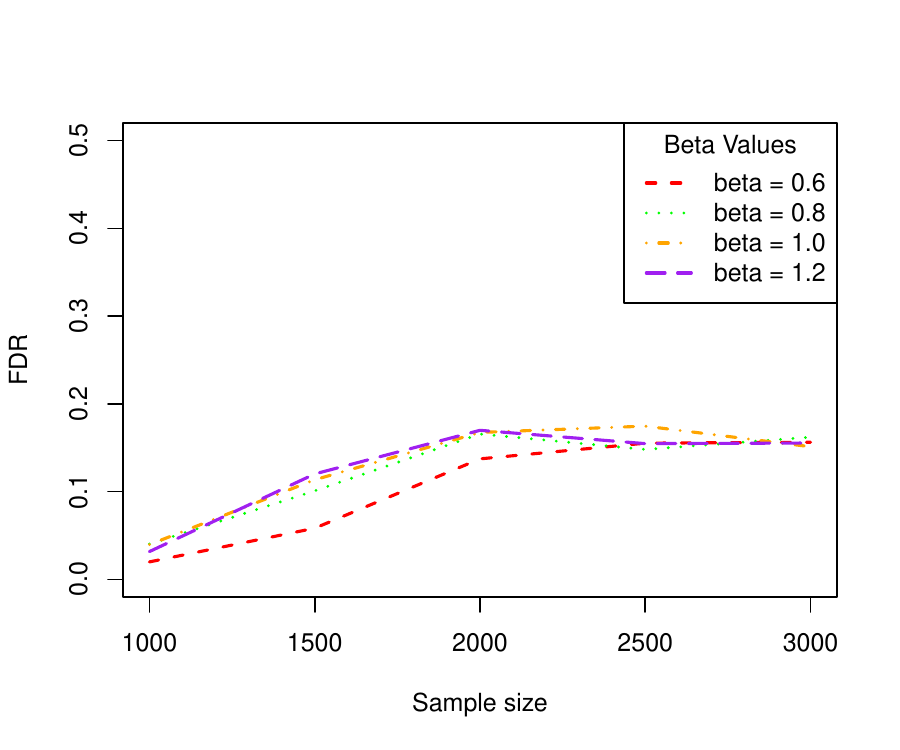}
		\end{subfigure}
		\hfill
		\begin{subfigure}[b]{0.48\textwidth}
			\centering
			\includegraphics[width=\textwidth]{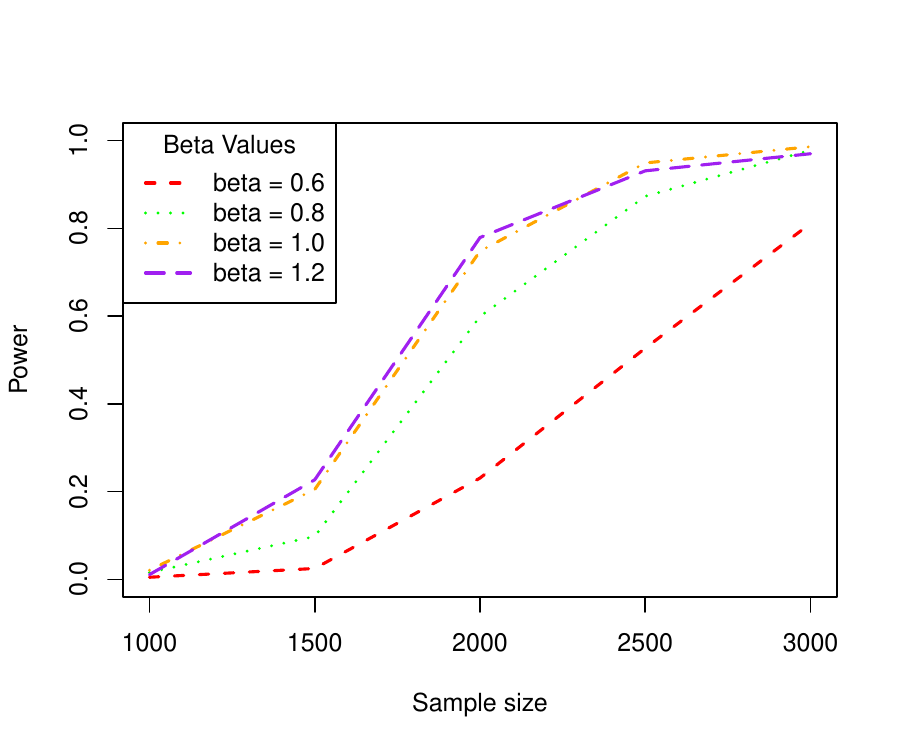}
		\end{subfigure}
		\caption{The FDR and Power of the DP-knockoff among different $\beta$ values. The dimension $p = 1000$, and the privacy budget $\mu = 1$.  }
		\label{fig:beta_increase}
	\end{figure}

    The simulation results demonstrate the performance of the DP-knockoff procedure under various settings.  Figure \ref{fig:dp.np.sig6} shows that when the dimension is fixed at $p=1000$, both the DP and non-private knockoff procedures maintain FDR control near the nominal 20\% level as the sample size $n$ increases from 1000 to 3000. In contrast, the SLR method has an FDR of roughly 45\%, primarily because it lacks finite-sample FDR guarantees. For reference, the simulation settings of SLR \citep{cai2023private} consider much larger sample sizes, on the order of 10,000. In terms of power, the non-private knockoff procedure achieves consistently higher power than the DP-knockoff, especially at smaller sample sizes.    Figure \ref{fig:P_increase} demonstrates the effect of increasing the dimensionality $p$ while fixing the sample size at $n=2000$. Both the DP and non-private knockoff procedures remain robust to dimension growth, indicating that feature screening is effective in reducing model complexity. Similar to Figure \ref{fig:dp.np.sig6}, however, the SLR method fails to control FDR, with values around 0.35.   Figure \ref{fig:beta_increase} investigates the effect of varying signal strengths on the DP-knockoff method. The results indicate that stronger signals (larger $\beta$) lead to improved power while maintaining FDR control.

    Additional simulation results are presented in Section G of the Supplementary Material. We first examine the effect of the privacy budget $\mu$. Figures \ref{fig:budget_N} and \ref{fig:budget_P} show that the proposed DP-knockoff procedure controls the FDR under different privacy budgets, while the power improves as $\mu$ increases, since less noise is required for privacy protection. We also compare different screening statistics in the first step of Algorithm 3. Figures \ref{fig:MD-linear} and \ref{fig:MD-non-linear} show that the distance-covariance-based screening statistic can improve power under both linear and nonlinear models, while still maintaining FDR control. In addition, Figure \ref{fig:SGD-log} studies the performance of the proposed procedure with SGD-based knockoff statistics under logistic regression models, where the DP-knockoff method continues to control the FDR and its power increases with the sample size. Finally, Figures \ref{fig:mm} and \ref{fig:Kn} investigate the sensitivity of the procedure to the tuning parameters, including the peeling size $m$ and the screening size $K_n$. The results show that FDR control is stable across the considered choices, while larger tuning parameters may lead to some power loss due to the increased amount of privacy noise.

	\section{Discussions}\label{sec:discussion}  

    We have proposed a private knockoff framework that enables model-free variable selection while ensuring privacy protection and controlling the FDR. Notably, this private algorithm achieves exact finite-sample FDR control irrespective of the scale of added noise, while retaining high statistical power relative to the non-private knockoff counterpart. To address challenges in high-dimensional settings, we propose a two-step procedure: an initial dimensionality reduction step followed by variable selection with FDR control, both incorporating privacy-preserving mechanisms.
    
    Several promising directions remain for future research. One intriguing property of the proposed framework is that the addition of noise enhances the stability of the knockoff process in controlling FDR because the DP-knockoff procedure is insensitive to modifications of individual data points. It would be interesting to further investigate the algorithm stability \citep{luo2024algorithmic, sun2025general}.  Another interesting observation is that the DP-knockoff framework accommodates certain inaccuracies in the estimation of knockoff statistics. Since the added noise is drawn from symmetric distributions, the coin-flipping property, and consequently FDR control, remains intact as long as the noise scale sufficiently dominates estimation errors in the knockoff statistics. This tolerance introduces greater flexibility in steps such as generating knockoff copies and estimating variable importance. Building on this, it would be worthwhile to extend methodologies in \cite{fan2023ark, fan2025asymptotic} to further explore the robustness of the DP-knockoff.  Another possible direction is to exploit local sensitivity for sparse-regression-based knockoff statistics. For example, a Propose-Test-Release type strategy could first privately certify that the realized dataset satisfies favorable sparsity or stability conditions, and then calibrate the noise to a smaller local sensitivity bound. This may improve power for stable Lasso-type statistics, but would require an additional private testing step and would likely lead to approximate-DP-type guarantees rather than the Gaussian-DP framework considered here. We leave this direction for future work.

\section{Acknowledgments}
We thank associate editor and the anonymous referees for their valuable comments and suggestions. Zhanrui Cai was supported in part by the Hong Kong Research Grants Council (Grant No.27301925) and the National Natural Science Foundation of China (Grant No.12501386).

\bibliographystyle{abbrvnat}
\bibliography{references}

\newpage

\setcounter{page}{1}
\setcounter{section}{0}
\setcounter{equation}{0}

\renewcommand{\thesection}{\Alph{section}}
\renewcommand{\thesubsection}{\thesection.\arabic{subsection}}
\renewcommand{\theequation}{\thesection.\arabic{equation}}

\begin{appendices}  
\begin{center}{\bf \Large \bigskip Supplement to ``Knockoff Inference under Privacy Constraints"}
		
\bigskip
		
Zhanrui Cai, Yingying Fan, and Lan Gao 
\end{center}

\noindent 

This Supplement contains extended power analysis for Algorithm \ref{alg:mirror peeling}, a power analysis for Algorithm \ref{algo:screen-knockoff-single}, an extension of Algorithm \ref{algo:screen-knockoff-single}, robustness analyses of FDR control under approximate covariate distributions, and all technical proofs. 
Section \ref{Sec:addition-power-algo1} presents additional power analysis for Algorithm \ref{alg:mirror peeling}.
Section \ref{Sec:power-Algo-single-split} studies the power of Algorithm \ref{algo:screen-knockoff-single}.
Section \ref{SecA:Multi-split} introduces a multi-split extension of Algorithm \ref{algo:screen-knockoff-single}.
Section \ref{sec:asym-FDR} establishes robustness of the proposed differentially private knockoff framework, showing that it continues to control the FDR when the covariate distribution is estimated or misspecified.
Section \ref{Sec:pf-thms} contains the proofs of Theorems \ref{lem:peeling}--\ref{thm:asymptotic-FDR}.
Section \ref{Sec:pf-lemmas} provides proofs of supporting lemmas and additional technical details.
Section \ref{Sec:add-simu} reports additional simulation results.
All the notation is the same as defined in the main manuscript.

\section{Extended Power Analysis for Algorithm \ref{alg:mirror peeling}} \label{Sec:addition-power-algo1}
In this section, we present a parallel power analysis for Algorithm \ref{alg:mirror peeling} under weaker signal strength conditions than those required in Theorem \ref{thm-power-alter}. 

Recall that $b_n := \frac{12 \Delta_n}{\mu}\sqrt{\pi m}  \log p$ represents a high-probability uniform upper bound for added noises in Algorithm \ref{alg:mirror peeling}. Denote the average tail distribution for the null knockoff statistics as 
\begin{equation} \label{eq-G-def}
G(t) := p_0^{-1} \sum_{j \in \mathcal{H}_0} \mathbb{P} (W_j \geq t) = p_0^{-1} \sum_{j \in \mathcal{H}_0} \mathbb{P} (W_j \leq - t),
\end{equation}
where $p_0 = |\mathcal{H}_0| $.

We introduce two conditions on the error level $b_n$ and  signal strength to facilitate the power analysis. 

\begin{condition}  \label{cond-power-1}
     It holds that 
    \begin{equation*}
       \sup_{t \in (0, G^{-1} (\frac{1}{p}) )} \frac{G(t - b_n) - G(t +  b_n)}{ G(t)} \to 0, \quad \text{as } n\rightarrow \infty.
    \end{equation*}
\end{condition}

\begin{condition} \label{cond-power-2}
    Assume $  \sum_{j = 1}^p \mathbb{P} ( | W_j - w_j | \geq \ell_n) \leq \kappa_n $ for some sequences $\ell_n > 0$ and $ \kappa_n \to 0 $. Here $ \{w_j\}_{j \in [p]} $ are fixed quantities such that $ w_j > 2 \ell_n $  for $j \in \mathcal{S}_0$ and $ w_j = 0  $ for $j \in \mathcal{H}_0$. 
\end{condition}

Condition \ref{cond-power-1} requires the noise upper bound $b_n  = \frac{12 \Delta_n}{\mu}\sqrt{\pi m}  \log p$ to decay sufficiently fast, which in turn imposes restrictions on the sensitivity level $\Delta_n$ of the knockoff statistics and the peeling size $m$. Condition \ref{cond-power-2} assumes that the knockoff statistics $W_j$'s concentrate around their population counterparts $w_j$'s at rate $\ell_n$ with high probability. For the relevant set $j \in \mathcal{S}_0$, the signal strength $w_j$  is assumed to dominate the concentration fluctuation $\ell_n$ to ensure a clear separation between the relevant and null knockoff statistics. This signal strength condition is weaker than the requirement $w_j > b_n + 2 \ell_n$ imposed in Theorem \ref{thm-power-alter}. 
In addition, Condition \ref{cond-power-2} requires the signals $w_j$ remain zero for null set $j \in \mathcal{H}_0$,  which is consistent with the inherent symmetry of null knockoff statistics. 
    
  The following result establishes a general power loss upper bound for the DP-knockoff  procedure. 

\begin{theorem} \label{thm-power-general}
     Assume $s \geq q^{-1}$,  the function $G(t)$ is continuous, and Conditions \ref{cond-power-1}--\ref{cond-power-2} hold. For any $\varepsilon \in (0, \frac{1 - q} {2 (1 + q)} \land \frac{1}{4})$, assume the peeling size satisfies $m >  s \big[1 +  \frac{6(1 + \varepsilon)q } {1 - q}\big]     $. Then we have that 
     \begin{equation} \label{general-power-lower bound}
        \mathrm{Power}(\widetilde{S}) := \e \Big[\frac{\widetilde{S} \cap \mathcal{S}_0}{s} \Big] \geq 1 - 2 \kappa_n - \e [\phi],
    \end{equation}
    where $\e [\phi]$ is the relative power loss  satisfying
    \begin{equation} \label{general-power-loss}
    \begin{split}
        \e [\phi]  & \leq s^{-1} \E \Big[ \sup_{t\in (0, G^{-1}(\frac{qs}{2p}))}\sum_{j \in \mathcal{S}_0 } \mathbbm{1} (W_j \in [t,t + b_n ]) \Big]  +   \mathbb{P} ( \breve{T} \leq b_n ) \\
        & \quad+ 3 \mathbb{P} (\mathcal B_{1,\varepsilon})  + \mathbb{P} (\mathcal{B}_2) +  O(\kappa_n + p^{-1} ),
    \end{split}
    \end{equation}
 with $\breve{T}$ the knockoff threshold of defined in \eqref{thre-D_m-orginal}, and the events $ \mathcal{B}_{1, \varepsilon}$ and $ \mathcal{B}_{2}$ defined as 
    \begin{align*} 
        \mathcal{B}_{1, \varepsilon}& := \bigg\{\sup_{t \in (0,\, G^{-1} ( \frac{ q s }{2 p})] } \bigg( \bigg| \frac{\sum_{j \in \mathcal{H}_0} \mathbbm{1} (W_{j} \geq t) } {\sum_{j \in \mathcal{H}_0} \mathbb{P} (W_j \geq t)} - 1 \bigg| \vee \bigg| \frac{\sum_{j \in \mathcal{H}_0} \mathbbm{1} (W_{j} \leq -t) } {\sum_{j \in \mathcal{H}_0} \mathbb{P} (W_j \leq -t)} - 1 \bigg| \bigg)> \varepsilon  \bigg\}   
    \end{align*}   
  and 
      \begin{equation*}
         \mathcal{B}_{2} := \bigcup_{j \in [p]}\bigg\{ \min_{  k \neq j } \big\{\big| W_k - W_j\big|  \land \big| W_k + W_j\big| \big\} \leq b_n \mbox{~and~}  |W_j| \geq G^{-1} \Big( \frac{3  m }{4 p }  \Big)   \bigg\} . 
    \end{equation*}



\end{theorem}


Theorem \ref{thm-power-general} provides both the absolute power lower bound in \eqref{general-power-lower bound} and the relative power loss upper bound compared to original non-DP knockoff inference in \eqref{general-power-loss}.  In the proof,  we first obtain \eqref{general-power-lower bound} by providing a lower bound for the power of the original knockoff procedure under Condition \ref{cond-power-2} on signal strength.
The proof of the relative power loss in \eqref{general-power-loss} is more technically involved and consists of three main steps. All the arguments below are in the sense of high probability. First, we locate the threshold $T^*$ of the original knockoff procedure by showing that $T^*$ is lower bounded by $G^{-1}  (\frac{2 q s}{p_0 (1 - q)})$ and upper bounded by $G^{-1} (\frac{q s}{2 p})$ under the event $\mathcal{B}_{1, \varepsilon}^c$. We then prove that once the peeling size $m$ exceeds a certain threshold, any knockoff statistics in $\{ j: |W_j| \geq T^*\}$ can pass through the mirror peeling procedure, ensuring  $\{j: |W_j| \geq T^*\}\subset D_m$.
These knockoff statistics  have largest magnitudes and determine the selection outcome $S^*$ of the original knockoff procedure. Finally, we show that the added Gaussian noises $\{\widetilde{Z}_{j}\}_{j \in [m]}$ have a negligible impact on the selection outcome $\wt S$, as long as the noise level is not too large. 
The proof is provided in Section \ref{Sec:pf-thm-power}.

Theorem \ref{thm-power-general} derives an explicit bound on the relative power loss and shows how it depends on $b_n$. The first four terms in the upper bound in  \eqref{general-power-loss} vanish  asymptotically under mild conditions.  
Specifically, we can establish the following convergence rate for $\mathbb{P} (\mathcal B_{1, \varepsilon} )$ under weak dependence between null knockoff statistics and for $\mathbb{P} (\mathcal B_{2}  )$ under some probability bound of appropriately scaled knockoff statistics. The proof of Lemmas \ref{lemma-B_1-rate} and \ref{lemma-B_2-rate} is provided in Sections \ref{Sec:pf-lemma-B_1-rate}--\ref{Sec:pf-lemma-B_2-rate}.  

\begin{lemma} \label{lemma-B_1-rate}
   Assume $s \to \infty$ as $n \to \infty$ and  $ \Var( \sum_{j \in \mathcal{H}_0} \mathbbm{1} (W_j \geq t) ) \leq C L_n p_0 G(t)  $ for $t \in [0, G^{-1}(\frac{qs}{2p})]$ and $L_n = o(s)$. 
Then we have for any $\varepsilon > 0$ and sufficiently large $n$ that 
   $$\mathbb{P}  (\mathcal B_{1, \varepsilon} ) \lesssim \varepsilon^{-2} \sqrt{L_n/s}.  $$
\end{lemma}

\begin{lemma} \label{lemma-B_2-rate}
    Assume that for any $k, j \in [p]$ and $ k \neq j $, and $\mathbb{P}(d_n W_k \in [t  , t + 2 d_n b_n] \,|\,W_j) \leq C d_n b_n $ for a constant $C$ and a scaling sequence $d_n > 0$. Then we have $\mathbb{P} ( \mathcal{B}_2) \lesssim m p d_n b_n $.  
\end{lemma}

The following lemma establishes sufficient conditions under which the first two terms in the upper bound in \eqref{general-power-loss} asymptotically vanish. The proof is presented in Section \ref{Sec:pf-lemma-power-1st-error}. 

\begin{lemma} \label{lemma-power-1st-error}
Denote $M_n := G^{-1} (\frac{q s}{2p})$ and $v_n:=  \sup_{t \in (0, M_n + b_n)} \Var(\sum_{j \in \mathcal{S}_0} \mathbbm{1}(W_j > t) )$. Assume the following condition: \\
(i) there exist a sequence $d_n > 0$ and a constant $C > 0$ such that for all $j \in [p]$, it holds that  
$\mathbb{P}(d_n W_j \in [t, t + d_nb_n]) \leq C d_n b_n $ for any $t \in [-d_nM_n, d_n M_n]$. \\
Then we have 
\begin{equation} \label{eq-lemma3-result1}
\mathbb{P} (\breve{T} \leq b_n) \leq \sum_{j = 1}^p \mathbb{P} (|W_j| \leq b_n) \lesssim p d_n b_n.
\end{equation}
If, in addition, the following conditions are satisfied: \\
(ii) $ s^{-2} v_n d_n M_n   \to 0$ and  (iii) $ d_n b_n \to 0 $.  \\
Then we have 
    \begin{equation} \label{power-le-result}
        \begin{split}
            &  s^{-1} \e \Big[ \sup_{t \in (0, G^{-1}(\frac{q s}{2p}))}  \sum_{j \in \mathcal{S}_0} \mathbbm{1} (W_j \in [t, t+ b_n]) \Big]  \lesssim [   s^{-2} v_n d_n (M_n + b_n)]^{1/4} + d_n b_n .
        \end{split}
    \end{equation}
\end{lemma}
 
We provide some discussions on the three conditions imposed in Lemma \ref{lemma-power-1st-error}. 
A sufficient condition for Condition (i) is that $d_n W_j$ has density function upper bounded by a constant $C$. We refer to Section \ref{Sec:discuss-cond(i)} for detailed discussion on condition (i) for some popular constructions of $W_j$'s considered in the main paper.  
The scaling order $d_n$ often satisfies $ d_n \asymp  ( \Var(W_j) )^{-1/2} $.
For example, $d_n$ is of order $O(\sqrt n)$ for the marginal correlation knockoff statistics $W_j = n^{-1} (|\bX_{\cdot, j}^T \by| - |\wt\bX_{\cdot, j}^T \by|) $. 
Condition (ii) imposes an upper bound constraint on the product of the variance $ \Var\big(\sum_{j \in \mathcal{S}_0} \mathbbm{1} (W_j > t) \big) $ and the scaled threshold upper bound $d_n G^{-1}(\frac{q s}{ 2p})$. This condition indicates that if $W_j$'s for null variables are tightly concentrated around 0 so that $d_n G^{-1}(\frac{q s}{ 2p})$ is small, the dependence among $\mathbbm{1} (W_j > t)$ for relevant variables $j\in \mathcal S_0$ can be large; it can be easily satisfied if there is a clear separation between the distributions of the null features and relevant features. 
Condition (iii) essentially requires
 that the noise level $b_n$ imposed for preserving privacy be small enough.  

\begin{example} \label{power-example}
We assume that the rescaled average tail probability function \linebreak $ \wt{G}(t) := p_0^{-1} \sum_{j \in \mathcal{H}_0 } \mathbb{P} ( d_{n} W_j > t) $ has a Gaussian tail  $\wt{G}(t)  \asymp \exp (- t^2/2 )$, where $\wt{G}(t)$ is related to $G(t )$ in that $ G(t) = \wt{G}(d_{n} t)$ and $d_{n}$ is the scaling order for $W_j$ with $j \in \mathcal{H}_0$.  
Then  we can obtain that $  G^{-1} (\frac{ q s }{2p }) \asymp d_{n}^{-1} \sqrt{\log (p/s)} $. 
In addition, we assume that $ \Var(\sum_{j \in \mathcal{S}_0} \mathbbm{1}(W_j > t) ) = O(s^{2 - \gamma})$ for some $\gamma \in (0, 1]$. 
In this case, Condition (ii) is equivalent to $s^{-\gamma}   \sqrt{\log(p/s)}  = o(1  )$. 
Additionally, recalling the definition that $b_n  = \frac{12 \Delta_n}{\mu}\sqrt{\pi m}  \log p$ with $\Delta_n$ the sensitivity level of knockoff statistics and $ m $ the peeling size in Algorithm \ref{alg:mirror peeling}, we obtain that Condition (iii) can be satisfied if 
\begin{equation} \label{eq-b_n-constraints}
 \frac{ \Delta_n d_n  }{\mu} \sqrt {m } \log p= o(1), 
\end{equation}
which imposes a constraint   on $\Delta_n$ and $m$. Note that Theorem~\ref{thm-power-general} shows that choosing the peeling size as $m = Cs$ for a sufficiently large constant $C$ guarantees asymptotically zero power loss. In this case, the requirement \eqref{eq-b_n-constraints} reduces to $\frac{ \Delta_n d_n  }{\mu} \sqrt {s }\log p = o(1)$. 
We provide detailed discussions on the sensitivity $\Delta_n$ for various constructions of knockoff statistics in Section \ref{Sec6.Sensitivity}. 
\end{example}

\subsection{Verification of Condition (i)} \label{Sec:discuss-cond(i)}
In this subsection, we discuss condition (i) in Lemma \ref{lemma-power-1st-error} for several constructions of knockoff statistics considered in the main paper. In general, condition (i) is satisfied if, after appropriate scaling, the statistics used to construct the knockoff statistics either admit a bounded density or converge to a limiting distribution with a bounded density with negligible approximation error. For example, it follows from the central limit theorem that the standardized marginal correlation and regression coefficient are asymptotically normal with a Berry-Esseen bound of order $O(n^{-1/2})$. Since the limiting normal distribution has bounded density, condition (i) can be verified for these knockoff statistics. The proof of Lemmas \ref{lemma-cond(i)-marg} and \ref{lemma-cond(i)-ridge} is deferred to Sections \ref{Sec:pf-density-marg}--\ref{Sec:pf-density-ridge}. 

\begin{lemma} \label{lemma-cond(i)-marg}
              For any $j \in [p]$,  assume $  \mathbb{E}[Y^2 \Var(\widetilde{X}_j | X)] > c_1 $, $  \Var (Y^2 \Var(\widetilde{X}_j | X))  < c_2 $, and $ \mathbb{E}[ |Y X_j|^3  ] \leq c_3$. For the marginal correlation knockoff statistics $\{W_j\}_{j \in [p]}$, we have that for any $j \in [p]$, 
              $$
              \mathbb{P} ( t \leq \sqrt{n} W_j \leq t + \sqrt{n} b_n )  \leq C \sqrt n b_n. 
              $$
\end{lemma}
\begin{lemma} \label{lemma-cond(i)-ridge}
    Assume data $(\bX, \by)$ is generated from a linear model $Y = X^T \bbeta + {\varepsilon}$ with $\varepsilon\stackrel{d}{\sim} N(0, \sigma^2)$. 
    Denote by $\breve{\bX} = [\bX, \widetilde{\bX}]$ the augmented data matrix with $\widetilde{\bX}$ being the knockoff matrix. Suppose that $ \mathbb{P} ( c_1 \leq \lambda_{\min} ( n^{-1} \breve{\bX}^T \breve{\bX}) \leq  \lambda_{\max} ( n^{-1} \breve{\bX}^T \breve{\bX}) \leq c_2 ) \geq 1 - O(n^{-1/2})$. For the ridge coefficient difference knockoff statistics $\{W_j\}_{j \in [p]}$, we have that for any $j \in [p]$, 
    $$
    \mathbb{P} ( t \leq \sqrt{n} W_j \leq t + \sqrt{n} b_n )  \leq C \sqrt n b_n. 
    $$
\end{lemma}

Condition (i) can be verified for the SGD-based statistics considered in Section \ref{Sec:sensi-SGD} using arguments analogous to those employed in the proofs of Lemmas \ref{lemma-cond(i)-marg} and \ref{lemma-cond(i)-ridge}. Specifically, with a suitably chosen step size, the SGD algorithm converges to the corresponding minimizer of the loss function (e.g., the ridge loss), and the resulting optimization error becomes negligible after sufficiently many iterations. Moreover, since such M-estimators typically satisfy asymptotic normality with a bounded density, condition (i) can be established in a similar manner.

HSIC-based knockoff statistics take more complicated forms. It has been established in the literature (\cite{gretton2007kernel,huang2022statistically}) that the HSIC estimator considered in Section \ref{Sec:sensi-hsic} can be formulated as a V-statistic and its limiting distribution is either  normal  or a weighted chi-square distribution, depending on whether the kernel is non-degenerate or degenerate. Thus, we expect condition (i)  to hold for HSIC-based knockoff statistics. 
      
      However, compared to marginal correlation and ridge regression coefficients,  SGD and HSIC statistics have significantly more complex forms. For SGD-based statistics, such verification would require a detailed analysis of the stability and asymptotic normality of the optimization output, and new results on the rate of convergence to the limit distribution. For HSIC-based statistics,  establishing an explicit convergence rate to the corresponding limiting distribution  would also require substantially more involved techniques and new theoretical results. We therefore view this as an interesting direction for future work.

\section{Power Analysis of Algorithm \ref{algo:screen-knockoff-single}} \label{Sec:power-Algo-single-split}
In this section, we establish the power analysis for the high-dimensional post-screening knockoff inference with single split presented in Algorithm \ref{algo:screen-knockoff-single}. 
Define the total power loss of the output $\widehat{S}$ of Algorithm \ref{algo:screen-knockoff-single} as 
$$
1 - \mathrm{Power}(\widehat{S}) := \mathbb{E}\Big[\frac{ |\mathcal{S}_0 \cap \widehat{S}^c| }{s} \Big]
$$
and the relative power loss compared with the original knockoff inference procedure without DP guarantee as $\mathbb{E}[\phi(\widehat{S})]$ with 
$$
 \phi(\widehat{S}) :=   \frac{ | \mathcal{S}_0 \cap S^* | }{s}  -  \frac {|\mathcal{S}_0 \cap \widehat{S}|  } {s}  ,
$$
where $S^*$ is the selected set by the original non-DP knockoff inference.

Assume the following conditions.
\begin{condition} \label{cond-single-split-power-1}
The screening statistics $\{U_j\}_{j \in [p]}$ obtained using the sub-dataset $\{(\bX_{I_1, \cdot}, \by_{I_1})\}$ satisfy that $\mathbb{P} (\max_{j \in [p]} | U_j - u_j | \geq \ell_n ) \leq \kappa_n $ for some $\ell_n >0 $ and $\kappa_n \to 0$, with $\{u_j\}_{j \in [p]}$ being fixed quantities and $u_j = 0$ for $j \in \mathcal{H}_0$.  
\end{condition}
\begin{condition}  \label{cond-single-split-power-2}
Given any filtered feature index set $\mathcal{C}$ independent of the sub-dataset $(\bX_{I_2, \cdot}, \by_{I_2})$, there exist fixed quantities $\{w_j\}_{j \in \mathcal{C}}$ with  $w_j = 0 $ for $j \in \mathcal{H}_0 \cap \mathcal{C}$, such that  the knockoff statistics $\{W_j\}_{j \in \mathcal{C}}$ constructed using the filtered augmented sub-dataset $(\bX_{I_2, \mathcal{C}}, \widetilde{\bX}_{I_2, \mathcal{C}}, \by_{I_2})$ satisfy that $\mathbb{P} (\max_{j \in \mathcal{C}} |W_j - w_j| \geq \ell_n ) \leq \kappa_n $. 
\end{condition}

Denote $b_{n,1}=   \frac{12 \Delta_u}{\mu}\sqrt{\pi K_n }  \log p $ and $b_{n,2}  =  \frac{  8 \Delta_w \sqrt{\log p}}{\mu} $. We establish the following result on the power analysis for Algorithm \ref{algo:screen-knockoff-single}. The proof is provided in Section \ref{Sec:pf-thm-power-single}.
\begin{theorem} \label{thm-power-single-split}
Assume Conditions \ref{cond-single-split-power-1} and \ref{cond-single-split-power-2} are satisfied. 
Denote by $\mathcal{S}_1 := \{j \in \mathcal{S}_0: u_j > b_{n,1} + 2 \ell_n  ~ \mbox{and} ~ w_j > b_{n,2}+ 2 \ell_n \} $. If $|\mathcal{S}_1|/(1 + s) \geq  1/(1 +q)$ and the screening size $K_n > s$, then for Algorithm \ref{algo:screen-knockoff-single},  we have the total power loss satisfies
$$
 1 - \mathrm{Power}(\widehat{S}_1)\leq (1 - |\mathcal{S}_1|/s) +  O( p^{-1}) + 2 \kappa_n,
$$
and the same upper bound applies to the relative power loss $\e[\phi(\widehat{S}_1)]$. 
 
\end{theorem}

\section{High-dimensional Differentially Private Knockoff Inference via Multiple Splits} \label{SecA:Multi-split}
To boost power and make full use of data, we further extend the privacy-preserved post-screening knockoff inference procedure with single split in Algorithm \ref{algo:screen-knockoff-single} to multiple splits.
Specifically, we perform the random splitting and Algorithm \ref{algo:screen-knockoff-single} for a total of $B$ repetitions and then apply the e-BH procedure (\cite{ren2024derandomised, wang2022false}) to aggregate the selected outcome from each repetition to achieve finite-sample FDR control. The details are presented in Algorithm \ref{algo:screen-knockoff-multi}. 


\begin{algorithm}[!t]
\caption{Post-screening DP-knockoff Inference with Multiple Splits}
\begin{algorithmic}[1]
\Require Dataset $(\bX, \by) \in \mathbb{R}^{n \times (p+1)}$, subsample sizes $n_1$ and $n_2 = n - n_1$, repetition number $B$, screening size $K_n$,   privacy parameter $\mu$, target FDR level $q$, parameter $\alpha_{kn} \in (0, 1)$, seed numbers $a$ and $b$.
\State Use seed number $a$ to fix the randomness and create $B$ different splits of $[n]$: $\{ (I_{b, 1}, I_{b, 2}) \}_{b = 1}^B$, where $|I_{b,1} | = n_1$, $|I_{b,2}| = n_2$, $I_{b, 1} \cap I_{b, 2} = \emptyset$,  and $I_{b, 1} \cup I_{b, 2}  = [n]$.  
\For {$b=1,2,\dots,B$}
\State Use the split sub-datasets $(\bX_{i, \cdot}, Y_i)_{i \in I_{b,1}}$ and $(\bX_{i, \cdot}, Y_i)_{i \in I_{b,2}}$ and perform Step 1--5 in Algorithm \ref{algo:screen-knockoff-single} with privacy parameter $\mu/\sqrt B$. Obtain corresponding filtered feature subset $\mathcal{C}_b$, masked knockoff statistics $\{\widetilde{W}_j^{(b)} \}_{j \in \mathcal{C}_b}$. 
\State Compute the threshold 
\begin{equation}
\widetilde{T}^{(b)}  :=  \inf \bigg\{t > 0: \frac{1 + \sum_{j \in  \mathcal{C}_b}  \mathbbm{1} (\widetilde{W}_j^{(b)}  \leq -t) } {1 \vee  \sum_{j \in \mathcal{C}_b } \mathbbm{1} (\widetilde{W}_j^{(b)}  \geq t)  } \leq \alpha_{kn} \bigg\}
\end{equation}
and $\widetilde{T}^{(b)} = \infty$ if the above set is empty.
Then calculate the e-values 
\begin{equation}
    e_j^{(b)} = 0 ~ \mbox{for}~ j \notin \mathcal{C}_b ~~\mbox{and}~~  e_j^{(b)} = p \cdot \frac{\mathbbm{1}(\widetilde{W}_{j} ^{(b) } \geq \widetilde{T}^{(b)} ) }{1 + \sum_{j \in \mathcal{C}_b}  \mathbbm{1}(\widetilde{W}_{j}^{(b)} \leq - \widetilde{T}^{(b)} ) } ~ \mbox{for}~ j \in \mathcal{C}_b .
\end{equation}
\EndFor
\State Compute the averaged e-values $ e_j^{\avg} = \frac{1}{B} \sum_{b = 1}^B e_j^{(b)}$ for each $j \in [p]$
\State Compute $\widehat{k} = \max\{k: e_{(k)}^{\avg} \geq p/(qk ) \}$ or $\widehat{k} = 0$ if this set is empty, where $ e_{(1)}^{\avg} \geq \cdots \geq e_{(p)}^{(\avg)}$ are ordered statistics of the $e_{j}^{\avg}$'s.
\Ensure The selected set $\widehat{\mathcal{S}}_2 := \{j \in [p]: e_j^{\avg} \geq p/(q \widehat{k}) \}$. 
\end{algorithmic}
\label{algo:screen-knockoff-multi}
\end{algorithm}

Theorems \ref{thm-GDP-multi} and \ref{thm-FDR-multi} show that Algorithm \ref{algo:screen-knockoff-multi} preserves  differential privacy and achieves finite-sample FDR control. 
\begin{theorem}  \label{thm-GDP-multi}
    Algorithm \ref{algo:screen-knockoff-multi} is $\mu$-GDP. 
\end{theorem}

\begin{theorem}  \label{thm-FDR-multi}
   For Algorithm \ref{algo:screen-knockoff-multi}, $\FDR(\widehat{\mathcal{S}}_2) \leq q$.
\end{theorem}

\section{DP-knockoff Procedure with Unknown Covariate Distribution} \label{sec:asym-FDR}
In this section, we discuss the implementation of DP-knockoff inference  when there is no exact knowledge of the covariate distribution. To simplify the presentation, we focus on the special case where $X \stackrel{d}{\sim} N(\mu, \bSigma)$ with mean $\mu = \mathbf{0}_p$ and covariance matrix $\bSigma$. Our analysis of the estimated covariate distribution can be extended to general distributions; see \cite{fan2023ark, fan2025asymptotic}  for relevant developments. 
Let $\bOmega := \bSigma^{-1}$ denote the precision matrix. When $\bOmega$ is known, ideal knockoff variables can be constructed by applying the formula in \eqref{exam-Gaussian-knockoff} for generating Gaussian knockoff variables.
When $\bOmega$ is unknown, we consider two scenarios in the sequel: estimating the precision matrix $\bOmega$ using out-of-sample unlabeled data and in-sample data, respectively.

\subsection{Out-of-Sample Estimation}
In many applications, additional unlabeled data may be available for estimating the covariate distribution (\cite{candes2018panning, ren2024derandomised}), which involves no  privacy cost.
In this case, one may compute a high-quality estimator $\widehat{\bOmega} $ of $\boldsymbol{\Omega}$ using classical covariance or precision-matrix estimation methods, and then plug it into the Gaussian knockoff construction without affecting the privacy budget of the main dataset.
Specifically, the approximate Gaussian knockoff variable matrix $\wh{\bX} \in \mathbb{R}^{n \times p}$ can be generated by  plugging in the estimator $\widehat{\bOmega}$ into \eqref{exam-Gaussian-knockoff}: 
\begin{equation} \label{Gaussian-knockoff-matrix-est}
    \wh{\bX} =   \bX (I_p - \wh\bOmega\diag (\boldsymbol{r})) +  \bZ(2 \diag(\boldsymbol{r}) - \diag(\boldsymbol{r})  \wh\bOmega \diag(\boldsymbol{r})  )^{1/2},
\end{equation}
where $\diag(\boldsymbol{r}) = \diag(r_1, \ldots, r_p)$ is a deterministic diagonal matrix ensuring that $2 \diag(\boldsymbol{r}) - \diag(\boldsymbol{r}) \wh\bOmega  \diag(\boldsymbol{r})  $ is positive definite, and $\bZ \in \mathbb{R}^{n \times p}$ consists of i.i.d.  standard Gaussian entries and is independent of $\bX$.
Using the approximate knockoff matrix $\wh{\bX}$, we can construct approximate knockoff statistics  $\{\wh{W}_j\}_{j \in [p]}$, and conduct Step 3--10 of Algorithm \ref{alg:mirror peeling} with $\{\wh{W}_j\}_{j \in [p]}$ to perform DP approximate knockoff inference. 

Theorem \ref{thm-finite-FDR} has established that Algorithm \ref{alg:mirror peeling} achieves finite-sample FDR control when applying ideal knockoff statistics $\{W_j\}_{j \in [p]}$ constructed under the true covariate distribution. Since the selected set is determined directly by the knockoff statistics, it is natural to expect that Algorithm \ref{alg:mirror peeling} will continue to achieve asymptotic FDR control provided that these approximate statistics are sufficiently accurate to some ideal counterparts. This intuition motivates us to assume a coupling between the approximate and ideal knockoff statistics such that their realizations are close.
\begin{condition}[Coupling accuracy] \label{cond:gene-W-cou-accu}
    There exist realizations $\{W_j\}_{j \in [p]}$ of ideal knockoff statistics such that for some sequence $ \rho_n \to 0$,
    \begin{equation*}
        \mathbb{P} \left( \max_{1 \leq j \leq p} | \wh{W}_j -  {W}_j | \geq \rho_n \right) \to 0.
    \end{equation*}
\end{condition}

Condition \ref{cond:gene-W-cou-accu} was also imposed in \cite{fan2023ark} for the general robustness analysis of model-X knockoff procedures when knockoff variables are generated from an approximate covariate distribution.
Construction of the coupled ideal knockoff statistics $\{W_j\}_{j \in [p]}$ and verification of Condition \ref{cond:gene-W-cou-accu} can be carried out through constructing the coupled ideal knockoff variables and analyzing their coupling accuracy with the ideal knockoff variables. 
Specifically, $\wh{\bX}$ generated by \eqref{Gaussian-knockoff-matrix-est} can be coupled with an ideal knockoff variable matrix  
\begin{equation} \label{Gaussian-knockoff-matrix}
    \wt{\bX} :=   \bX (I_p -  \bOmega\diag (\boldsymbol{r})) +  \bZ(2 \diag(\boldsymbol{r}) - \diag(\boldsymbol{r})  \bOmega \diag(\boldsymbol{r})  )^{1/2},
\end{equation}
where importantly, $\bZ$ and $\diag(\boldsymbol{r})$ are identical to those used in \eqref{Gaussian-knockoff-matrix-est} for generating $\wh{\bX}$. In this case, Proposition 3 in \cite{fan2023ark} shows that if 
\begin{equation} \label{eq: Omega-error}
\mathbb{P} (\|\wh{\bOmega} - \bOmega \|_2 \leq u_n ) \to 1
\end{equation} 
for a sequence $u_n \to 0$, then under regular sparsity conditions, 
\begin{equation}  \label{eq: knockoff-X-error}
\mathbb{P} \Big(\max_{j \in [p]} n^{-1/2} \|\widehat{\bX}_j - \wt{\bX}_j\|_{2} \leq Cu_n \Big)  \to 1,
\end{equation}
where $C$ is a constant. 
Therefore, the approximate knockoff statistics $\wh{W}_j = w_j ([\bX, \widehat{\bX}], \by)$ can be naturally coupled with the ideal knockoff statistics ${W}_j = w_j ([\bX, \wt{\bX}], \by)$, where the coupling accuracy of $\wh{W}_j$ is governed by that of the approximate knockoff variable matrix $\wh{\bX}$. 
Consequently, Condition \ref{cond:gene-W-cou-accu} can be justified for a variety of knockoff statistics when \eqref{eq: Omega-error} holds. 
For instance, for marginal correlation difference knockoff statistics, Lemma 5 of \cite{fan2023ark} establishes that Condition \ref{cond:gene-W-cou-accu} is satisfied with $\rho_n = u_n$. For knockoff statistics through ridge regression coefficient differences, we show that, in a linear model setting, Condition \ref{cond:gene-W-cou-accu} is satisfied with $\rho_n = (\lambda   +  \sqrt{ \frac{\log n}{ n }})u_n $, where $\lambda$ is the regularization parameter in ridge regression; see, Lemma \ref{lemma: knockoff-couling-ridge} in Section \ref{Sec: Lemma-knock-accu-ridge}. 
We refer to Sections 3 and 4 of \cite{fan2023ark} for additional discussion on verification of  Condition \ref{cond:gene-W-cou-accu}.
In addition, we emphasize that the theoretical analysis only requires existence of the coupled ideal knockoff statistics and variables, and their actual construction is not invoked in practical implementation.

Algorithm \ref{alg:mirror peeling} based on the approximate knockoff statistics can then be coupled with the corresponding algorithm based on the coupled ideal knockoff statistics, where the same Gaussian noises $\{Z_{k, j}\}_{k \in [p], j \in [m]}$ and $\{\wt{Z}_{j}\}_{j \in [m]}$ are injected in both procedures. Therefore, Algorithm \ref{alg:mirror peeling} with approximate knockoff statistics achieves asymptotic FDR control provided that the approximate knockoff statistics are sufficiently accurate.
Next we introduce additional technical assumptions and a condition on the coupling accuracy $\rho_n$ of the knockoff statistics. 

Define $b_n := \frac{12 \Delta_n}{\mu}\sqrt{\pi m}  \log p$ and let the average tail distribution for the null ideal knockoff statistics be $G(t) := p_0^{-1} \sum_{j \in \mathcal{H}_0} \mathbb{P} (W_j \geq t) = p_0^{-1} \sum_{j \in \mathcal{H}_0} \mathbb{P} (W_j \leq - t)$, where $p_0 = |\mathcal{H}_0 |$. Denote by $\wh{T} $ the knockoff threshold defined similar to \eqref{def-widetilde-T} for Algorithm \ref{alg:mirror peeling} based on approximate knockoff variable matrix $\wh\bX$, and let $ \wh{S} $ be the corresponding selected set. 

\begin{condition} \label{cond:locationT}
    There exists a sequence $ M_n > 0$ such that 
    $
        \mathbb{P} ( \wh{T} < M_n ) \to 1.
    $
\end{condition}
 
\begin{condition} \label{cond:approx-indicator}
It holds that 
   $$
    \sup_{t \in (0, M_n + \rho_n + b_n)} \left\{ \left| \frac{\sum_{j \in \mathcal{H}_0} \mathbbm{1}( {W}_j \geq t) } {\sum_{j \in \mathcal{H}_0} \mathbb{P}( {W}_j \geq t)} - 1 \right| \vee \left| \frac{\sum_{j \in \mathcal{H}_0} \mathbbm{1}( {W}_j \leq -t) } {\sum_{j \in \mathcal{H}_0} \mathbb{P}( {W}_j \leq -t)} - 1 \right| \right\} = \mathfrak{o}_\mathbf{P}(1). 
   $$
\end{condition}
\begin{condition} \label{cond:rho-n}
    It holds that 
    \begin{equation*}
       \sup_{t \in (0, M_n \vee G^{-1} (\frac{1}{p_0}) )} \frac{G(t - \rho_n - b_n) - G(t + \rho_n + b_n)}{ G(t)} \to 0.
    \end{equation*}
\end{condition}

Condition \ref{cond:locationT} imposes an upper bound on the knockoff threshold, ensuring that the selection rule remains within a regime where its asymptotic behavior is analytically tractable. Condition \ref{cond:approx-indicator} assumes a weak dependence structure on the null knockoff statistics $\{W_j\}_{j \in [p]}$,
which guarantees that the empirical distribution of null statistics concentrates around its expectation. Moreover, Condition \ref{cond:rho-n} requires that the coupling error $\rho_n$ and the privacy-induced noise level $b_n$ decay sufficiently fast.
These conditions, or their analogs, have appeared and been verified under various constructions of knockoff statistics in prior works  on robustness analyses of FDR control for knockoff inference (\cite{fan2023ark}, \cite{fan2025asymptotic}).

The theorem below establishes asymptotic FDR control for DP-knockoff inference using an approximate covariate distribution.
\begin{theorem} \label{thm:asymptotic-FDR}
    Assume Conditions \ref{cond:gene-W-cou-accu}--\ref{cond:rho-n} hold and the peeling size satisfies $m \geq  \frac{(2q +1 +c)s}{1 - q}$  for a constant $c > 0$. 
    When approximate knockoff variable matrix $\widehat{\bX}$ generated by \eqref{Gaussian-knockoff-matrix-est} is applied in Algorithm \ref{alg:mirror peeling}, we have 
    \begin{equation}
        \limsup_{n \to \infty} \FDR(\widehat{S}) \leq q,  \quad  \mbox{and} \quad \mathbb{P} \left(\FDP (\wh{S}) \leq q + \varepsilon \right) \to 1 ~ \mbox{for any} ~ \varepsilon > 0.
    \end{equation}    
\end{theorem}
Theorem \ref{thm:asymptotic-FDR} provides a general framework for analyzing the robustness of Algorithm \ref{alg:mirror peeling} to approximate covariate distributions in achieving FDR control. Although the required conditions resemble those in existing work on the robustness of the knockoff framework (\cite{fan2023ark}, \cite{fan2025asymptotic}), the proof of Theorem \ref{thm:asymptotic-FDR} entails new technical developments to handle the additional peeling procedure in Algorithm \ref{alg:mirror peeling}.
The complete proof is provided in Section \ref{Sec:pf-thm8}. 

\subsection{In-Sample Estimation}
In this section, we consider the setting where the covariance matrix is estimated from the same dataset used for knockoff inference, thereby it requires a DP method for estimating the covariance (or precision) matrix. 

In recent years, there have been notable advances in the theory and methodology of DP covariance and precision matrix estimation under various DP notions. While covariance or precision matrix estimation under Gaussian differential privacy (GDP) has not yet been explicitly studied (to the best of our knowledge),  Theorem 5.1 of \cite{liu2022identification} shows that any $(\sqrt{\frac{2}{\pi}} \mu, 0)$-DP algorithm is also $\mu$-GDP.  Consequently, it suffices for our purposes to focus on $ (\varepsilon, 0)$-DP (a.k.a., $\varepsilon$-DP) algorithms for covariance or precision matrix estimation. For example,  \cite{alabi2023privately} proposed an $\varepsilon$-DP  algorithm  for estimating the Gaussian covariance matrix, which achieves the Frobenius error rate $ \| \bSig^{-1/2} \widehat{\bSig} \bSig^{-1/2} - I_p \|_{F} \leq \widetilde{O}( \frac{p}{\sqrt{n \varepsilon}} + \frac{p^{3/2} \sqrt{\log \kappa} }{n \varepsilon } + \frac{p}{n \varepsilon} ) $ (see Theorem 4.4 therein), where $\kappa$ is the condition number of population covariance matrix $\bSig$. 
On the other hand, recent work by \cite{amin2019differentially},  \cite{dong2022differentially} and \cite{d2025purely} developed $\varepsilon$-DP covariance estimators and established the associated Frobenius error bound under settings of general covariates with bounded $ \ell_2$ norm.  
There are also recent developments under other privacy notions, such as $(\varepsilon, \delta)$-DP \citep{dwork2014algorithmic, wang2021differentially, wang2018differentially, cai2024optimal} and concentrated differential privacy \citep{amin2019differentially, dong2022differentially, biswas2020coinpress}.

The aforementioned results typically require the dimensionality $p$ to grow sublinearly with the sample size $n$. An exception is the work of \cite{wang2021differentially}, which proposed an $(\varepsilon,\delta)$-DP algorithm for high-dimensional ($p \gg n$) sparse covariance matrix estimation and achieved an estimation error that grows polynomially in the sparsity level and logarithmically on the dimension $p$. However, we are not aware of any existing work that studies the differentially private covariance estimation in high-dimensional settings under GDP or $\varepsilon$-DP notions, which provides stronger privacy guarantee and remains an important  direction for future research.

In light of established results in the literature,  we assume that there exists an estimator $ \widehat{\bOmega}$ of the precision matrix $\bOmega$ which is $\mu$-GDP and satisfies
\begin{equation}   \label{eq: private-omega-error}  
    \| \widehat{\bOmega} - \bOmega\|_{2} \leq \delta_{n,1} (\mu) 
\end{equation}
with asymptotic probability one, where $\delta_{n,1}(\mu) \to 0$ is the estimation accuracy bound. 
The estimated precision matrix $\widehat{\bOmega}$ can then be applied in \eqref{Gaussian-knockoff-matrix-est} to generate approximate knockoff variable matrix $\wh{\bX}$ and subsequently construct approximate knockoff statistics $ \{\wh{W}_j\}_{j \in [p]} $. Finally, the approximate knockoff inference is performed through steps 3-10 using $\{\wh{W}_j\}$ and with privacy parameter $\mu$. It immediately follows from the combination property of GDP that such approximate knockoff procedure is $(\sqrt{2}\mu)$-GDP.

Next we proceed with the asymptotic FDR analysis for the DP approximate knockoff inference. We first introduce a condition on the coupling accuracy of the approximate knockoff statistics parallel to Condition \ref{cond:gene-W-cou-accu}.

\begin{condition} \label{thm9-cond1}
       There exist coupled realizations $\{W_j\}_{j \in [p]}$ of ideal knockoff statistics such that   
    \begin{equation*}
        \mathbb{P} \left(\max_{j \in [p]} | \widehat{W}_j - W_j | \leq \delta_{n,2} (\mu) \right)  \to 1,
    \end{equation*}
    where $\delta_{n,2}(\mu) \to 0$ is a sequence depending on the error $\delta_{n,1}(\mu)$ of the DP estimate $\widehat{\bOmega}$ and the specific construction of knockoff statistics.  
\end{condition}

Following the same argument as in the verification of Condition \ref{cond:gene-W-cou-accu}, Condition \ref{thm9-cond1} can be similarly verified for commonly used knockoff statistics when \eqref{eq: private-omega-error} holds. For instance, for marginal-correlation-difference knockoff statistics, Condition \ref{thm9-cond1} is satisfied with $\delta_{n,2}(\mu)=\delta_{n,1}(\mu)$. For knockoff statistics based on differences in ridge regression coefficients, Condition \ref{thm9-cond1} is satisfied with $ \big(\lambda   +  \sqrt{ \frac{\log n}{ n }}\big)\delta_{n,1}(\mu) $, where $\lambda$ is the regularization parameter in ridge regression. Theorem \ref{thm-fdr-est-moment}  establishes the asymptotic FDR control under in-sample estimation for covariance matrix. The proof of Theorem \ref{thm-fdr-est-moment} is omitted, as it follows from the same arguments used to prove Theorem \ref{thm:asymptotic-FDR}.

\begin{theorem} \label{thm-fdr-est-moment}
    Assume Condition \ref{thm9-cond1} holds and Conditions \ref{cond:locationT} -- \ref{cond:rho-n} are satisfied with $\rho_n$ replaced with $\delta_{n,2}(\mu)$. Suppose $m \geq  \frac{(2q +1 +c)s}{1 - q}$  for a constant $c > 0$.  When approximate knockoff variable matrix $\widehat{\bX}$ is applied in Algorithm \ref{alg:mirror peeling} and the selected set $\widehat{S}$ is output, we have 
    \begin{equation}
        \limsup_{n \to \infty} \FDR(\widehat{S}) \leq q,  \quad  \mbox{and} \quad \mathbb{P} \left(\FDP (\wh{S}) \leq q + \varepsilon \right) \to 1 ~ \mbox{for any} ~ \varepsilon > 0.
    \end{equation}    
\end{theorem}


\section{Proof of Theorems and Propositions} \label{Sec:pf-thms}
\subsection{Proof of Theorem \ref{lem:peeling}}

By the proof of Proposition \ref{prop: screening-dp}, reporting the index $i_j$ in the $j$-th step of the peeling procedure is $\mu/{\sqrt{2m}}$-GDP. In addition,  reporting $\widetilde{W}_{i_j}$ is also $\mu/{\sqrt{2m}}$-GDP. By the composition theorem of GDP (Corollary 2 of \cite{dong2019gaussian}), reporting $\{i_j, \widetilde{W}_{i_j}\}$ at step $j$ of the peeling procedure is $\mu/\sqrt{m}$-GDP. Thus Algorithm \ref{alg:mirror peeling} is $\mu$-GDP by applying the composition  theorem again for the $m$ steps.

\subsection{Proof of Theorem \ref{thm-finite-FDR}} \label{SecA.1}
  The main idea of the proof is applying the symmetry in knockoff statistics $\{W_j\}_{j \in \mathcal{H}_0}$ and the added Gaussian noises, sharing a similar spirit with the proof of finite-sample FDR control of the model-X knockoff inference. 
  Observe that the filtered index set $D_m = \{i_1, \ldots, i_m \}$ generated by Algorithm \ref{alg:mirror peeling} only depends on the absolute values $|\mathcal{W}|= \{|W_1|, \ldots, |W_p| \}$ and the independent  errors $ \bZ := \{Z_{k, j}: 1 \leq k \leq p, 1 \leq j \leq m \}$. The FDR of the selected set $\widetilde{S}$ based on $ \widetilde{W}_{m} := \{ \widetilde{W}_{j}: j \in D_m\} $ can be written as 
  \begin{equation} \label{pf-thm1-1}
  \begin{split}
  \FDR(\widetilde{S}) & =  \e \bigg[ \frac{ | \widetilde{S} \cap \mathcal{H}_0 | } { 1 \vee |\widetilde{S}| } \bigg] =   \e \bigg[ \frac{ \sum_{j \in \mathcal{H}_0 \cap D_m } \mathbbm{1} (\widetilde{W}_j \geq \widetilde{T}) } { 1 \vee  \sum_{j \in D_m} \mathbbm{1} (\widetilde{W}_j \geq \widetilde{T}) } \bigg] \\
  & \leq q  \cdot \e \bigg[ \frac{\sum_{j \in \mathcal{H}_0 \cap D_m} \mathbbm{1} (\widetilde{W}_j \geq \widetilde{T} ) } {1 + \sum_{j \in \mathcal{H}_0 \cap D_m} \mathbbm{1} (\widetilde{W}_j \leq - \widetilde{T} )}  \bigg] \\
  & =  q  \cdot \e \Bigg\{ \e \bigg[ \frac{\sum_{j \in \mathcal{H}_0 \cap D_m} \mathbbm{1} (\widetilde{W}_j \geq \widetilde{T} ) } {1 + \sum_{j \in \mathcal{H}_0 \cap D_m} \mathbbm{1} (\widetilde{W}_j \leq - \widetilde{T} )} \bigg| |\mathcal{W}|, \bZ  \bigg] \Bigg\}
  \end{split}
  \end{equation}
In what follows, all the technical arguments are made conditional on $|\mathcal{W}|$ and $\bZ$. Note that $D_m $ is determined given $|\mathcal{W}|$ and $\bZ$.  
Without loss of generality, we assume that $|\widetilde{W}_{i_1}| \geq |\widetilde{W}_{i_2}| \geq \cdots \geq |\widetilde{W}_{i_m}|$ (the order of magnitude does not affect the proof procedure).
Let $V^+ (k) = \sum_{j \in \{ i_1, \ldots, i_k\} \cap \mathcal{H}_0} \mathbbm{1}(\widetilde{W}_j \geq 0)  $, $V^- (k) = \sum_{j \in \{ i_1, \ldots, i_k\}\cap \mathcal{H}_0} \mathbbm{1}(\widetilde{W}_j <  0) $ and $\sigma$-field $\mathscr{F}_k $ be the filtration defined by knowing all the non-null $\widetilde{W}_j$'s as well as $V^{\pm} (k')$ for $k' \geq k$. 
Noting that if $i_k\in \mathcal H_0$, ${W}_{i_k} $ and $\widetilde{Z}_{i_k}$ are both symmetric given $|\mathcal{W}|$ and $\bZ$ and $\widetilde{Z}_{i_k}$ is independent of everything else,  we can see that $\widetilde{W}_{i_k} = W_{i_k} + \widetilde{Z}_{i_k}$ is symmetric given $|\mathcal{W}|$ and $\bZ$. Therefore, $\{\widetilde{W}_{i_k} \}_{i_k \in \mathcal{H}_0}$ still satisfy the coin-flip sign property (that is, the signs of $\{\widetilde{W}_{i_k}\}_{i_k\in \mathcal{H}_0}$ are i.i.d. Bernoulli variables with success probability $1/2$) given $|\mathcal{W}|$ and $\bZ$. We have $\mathbb{P} (\mathbbm{1}(\widetilde{W}_{i_k} \geq 0) | \mathscr{F}_k) = \frac{V^+ (k)} {V^+ (k) + V^- (k)}$ when $k\in \mathcal H_0$. Hence, following the same proof technique in \cite{barber2015controlling} yields that given $|\mathcal{W}|$ and $\bZ$, $ M(k) := \frac{V^+ (k)} {1 + V^{-}(k)}$ is a super-martingale running backward in time with respect to $\mathscr{F}_k$. 

In addition, given $|\mathcal{W}|$ and $\bZ$,  $\widetilde{T} = |\widetilde{W}_{i_{k^*}}|$ for some $1 \leq k^* \leq m$ and $\{k^* \geq k \} \in \mathscr{F}_k$ is a stopping time with respect to the backward filtration $\{\mathscr{F}_k\}$. Therefore, it holds that (denote $N:=|\mathcal{H}_0 \cap D_m|  \in \sigma(|\mathcal{W}|, \bZ)$)
\begin{equation} \label{pf-thm1-2}
\begin{split}
     & \e \bigg[ \frac{\sum_{j \in \mathcal{H}_0 \cap D_m} \mathbbm{1} (\widetilde{W}_j \geq \widetilde{T} ) } {1 + \sum_{j \in \mathcal{H}_0 \cap D_m} \mathbbm{1} (\widetilde{W}_j \leq - \widetilde{T} )} \bigg| |\mathcal{W}|, \bZ  \bigg] = \e [M( {k}^*) | |\mathcal{W}|, \bZ  ]\\
     & \leq \e [M(m) | |\mathcal{W}|, \bZ] =  \e \bigg[ \frac{\sum_{j \in \mathcal{H}_0 \cap D_m} \mathbbm{1} (\widetilde{W}_j \geq 0) } {1 + \sum_{j \in \mathcal{H}_0 \cap D_m} \mathbbm{1} (\widetilde{W}_j < 0 )} \bigg| |\mathcal{W}|, \bZ  \bigg] \\
     & = \e \bigg[ \frac{ X  } {1 + N - X }  \bigg| |\mathcal{W}|, \bZ  \bigg], 
\end{split}
\end{equation}
where $X \sim binomial(N, 1/2)$ given $|\mathcal{W}|$ and $\bZ$. Consequently, it follows by applying the same calculation for binomial distribution in \cite{barber2015controlling} that
\begin{equation} \label{pf-thm1-3}
    \begin{split}
        \e \bigg[ \frac{ X  } {1 + N - X }  \bigg| |\mathcal{W}|, \bZ  \bigg] & \leq 1, 
    \end{split}
\end{equation}
Combining \eqref{pf-thm1-1} -- \eqref{pf-thm1-3} yields $\FDR(\widetilde{S}) \leq q$. This completes the proof of Theorem \ref{thm-finite-FDR}.

\subsection{Proof of Theorem \ref{thm-power-alter}} 
 Note that 
    \begin{equation} \label{pf-thm4-new-eq1}
    \begin{split}
        \e [\phi] & \leq s^{-1} \e [ | S^* \cap \mathcal{S}_0  \cap \widetilde{S}^c| ]  \\
        & = s^{-1}  \e [ | S^* \cap \mathcal{S}_1  \cap \widetilde{S}^c| ]   + s^{-1}  \e [ | S^* \cap \mathcal{S}_0 \cap \mathcal{S}_1^c  \cap \widetilde{S}^c| ] \\
        & \leq s^{-1}  \e [ | S^* \cap \mathcal{S}_1  \cap \widetilde{S}^c| ]   + s^{-1} \e [ | \mathcal{S}_0 \cap \mathcal{S}_1^c |].
    \end{split}
    \end{equation}
     It remains to address the first error term  $s^{-1}  \e [ | S^* \cap \mathcal{S}_1  \cap \widetilde{S}^c| ] $ related to the set of strong signals. The proof consists of two steps. First, we show that the strong signal set $\mathcal{S}_1$ will pass through the mirror peeling algorithm in high probability when $m$ is chosen appropriately. In the second step, we prove that the variables in $\mathcal{S}_1$ will  be selected in the output $\widetilde{S}$.
     Define the event $\mathcal{C}_n := \{ \max_{j \in [p]} | W_j - w_j | < \ell_n  \}$. It follows from the assumption in Theorem \ref{thm-power-alter} that $\mathbb{P} (\mathcal{C}_n) \geq 1- \kappa_n  $.  Under event $\mathcal{C}_n$, observe that for any $j \in \mathcal{S}_1$, 
     \begin{align*}
          |W_j| > |w_j| - \ell_n \geq b_n + \ell_n ,
     \end{align*}
and for any $j \in \mathcal{H}_0$, 
\begin{align*}
    |W_j| < |w_j| + \ell_n \leq \ell_n.
\end{align*}
Thus, it holds that $ \sum_{j \in \mathcal{H}_0} \mathbbm{1} (  |W_j| \geq  \ell_n) = 0 $ and hence
$ \sum_{j \in [p]} \mathbbm{1} (  |W_j| \geq  \ell_n) \leq s $.
If the peeling size $m > s $, there there must exist some $j_0 \in D_m$ such that $|W_{j_0}| < \ell_n$. Under event $A_{1, n} \cap \mathcal{C}_n$ with $A_{1,n}$ defined below \eqref{pf-le-filter-2}, we have for any $j \in \mathcal{S}_1$ and $k \in [m]$ that 
$$
|W_j | + Z_{k, j} \geq |W_j | - b_n/2 > b_n/2 + \ell_n > b_n/2 + |W_{j_0}| > |W_{j_0}| + Z_{k, j_0},
$$
which implies $j \in D_m$ and hence $\mathcal{S}_1 \subset  D_m$.  Therefore, the strong signal set $\mathcal{S}_1$ will survive the mirror peeling algorithm. 

Next we proceed to prove $ \mathcal{S}_1 \subset \widetilde{S} $. Define $\widetilde{A}_n := \{\max_{j \in [m]} |\widetilde{Z}_{j}|  <  b_n /2 \}$. By a similar argument in \eqref{pf-le-filter-2}, we can obtain $\mathbb{P} (\widetilde{A}_n)  \geq 1 - O( p^{-1}) $. Under event $ A_{1, n} \cap \mathcal{C}_n \cap \widetilde{A}_n$, it holds that $W_j > w_j - \ell_n > b_n + \ell_n $ for $j \in \mathcal{S}_1$ and  $W_j > w_j - \ell_n \geq - \ell_n $ for $j \in \mathcal{H}_0$. Therefore,
\begin{align*}
 & \sum_{j \in D_m } \mathbbm{1}(\widetilde{W}_j \leq -\ell_n - b_n/2 ) 
  \leq \sum_{j \in D_m } \mathbbm{1}( {W}_j < -\ell_n   ) \\
& \leq | \mathcal{S}_0 \cap \mathcal{S}_1^c | +\sum _{j \in D_m \cap \mathcal{S}_1} \mathbbm{1}( {W}_j < -\ell_n   ) + \sum_{j \in D_m \cap \mathcal{H}_0} \mathbbm{1}( {W}_j < -\ell_n   )  
  = | \mathcal{S}_0 \cap \mathcal{S}_1^c | . 
\end{align*}
Similarly, under event $ A_{1, n} \cap \mathcal{C}_n \cap \widetilde{A}_n$, we can obtain that
\begin{align*}
     \sum_{j \in D_m } \mathbbm{1}(\widetilde{W}_j \geq \ell_n +   b_n/2 )  \geq   \sum_{j \in D_m } \mathbbm{1}( {W}_j > \ell_n  ) \geq |\mathcal{S}_1|.
\end{align*}
Recalling the assumption that $|\mathcal{S}_1|/(1 + s) \geq 1/(1 + q)$, we have  
\begin{equation*}
     \frac{ 1 + \sum_{j \in D_m} \mathbbm{1}(\widetilde{W}_j < - \ell_n - b_n / 2 ) } {  \sum_{j \in D_m} \mathbbm{1}( \widetilde{W}_j \geq  \ell_n + b_n /2 )   } \leq \frac{1 + | \mathcal{S}_0 \cap \mathcal{S}_1^c|} {|\mathcal{S}_1|} = \frac{ 1 + s -|\mathcal{S}_1| }{ |\mathcal{S}_1|} \leq q.
\end{equation*}
By definition of the threshold $\widetilde{T}$, it follows that $ \widetilde{T} \leq  \ell_n + b_n / 2 $ and thus $ W_j > b_n + \ell_n > \widetilde{T} $ for any $j \in \mathcal{S}_1$. This implies that $ \mathcal{S}_1 \subset \widetilde{S} $  under event $ A_{1, n} \cap \mathcal{C}_n \cap \widetilde{A}_n$. 

Combining the above arguments, we obtain that 
 \begin{equation}
    \begin{split}
         s^{-1}  \e [ | S^* \cap \mathcal{S}_1  \cap \widetilde{S}^c| ]  & \leq s^{-1}  \e [ | \mathcal{S}_1  \cap \widetilde{S}^c| ] \\
         & \leq    \mathbb{P} (A_{1, n}^c) + \mathbb{P} (\widetilde{A}_n^c) + \mathbb{P} (\mathcal{C}_n^c)\\
         & \leq   O( p^{-1} )+ \kappa_n, 
    \end{split}      
    \end{equation}
    which together with \eqref{pf-thm4-new-eq1} yields the desired result for $\mathbb{E}[\phi]$.
    
    For the total power loss $1 - \mathrm{Power}(\widetilde{S}): = s^{-1}\mathbb{E}\big[ |\mathcal{S}_0 \cap \widetilde{S}^c|   \big]$,  noting that 
    \begin{equation*}
        \begin{split}
             s^{-1} \mathbb{E}\big[ |\mathcal{S}_0 \cap \widetilde{S}^c|   \big] &  =    s^{-1} \mathbb{E}\big[ |\mathcal{S}_1  \cap \widetilde{S}^c|   \big] +    s^{-1} \mathbb{E}\big[ |\mathcal{S}_1^c  \cap \mathcal{S}_0 \cap \widetilde{S}^c|   \big] \\
             & \leq s^{-1} \mathbb{E}\big[ |\mathcal{S}_1  \cap \widetilde{S}^c|   \big] +    s^{-1} \mathbb{E}\big[ |\mathcal{S}_1^c  \cap \mathcal{S}_0  |  \big]
        \end{split}
    \end{equation*}
 we can obtain the same upper bound for $1 - \mathrm{Power}(\widetilde{S})$ as that for $\e [\phi]$.        
 This completes the proof of Theorem \ref{thm-power-alter}.

\subsection{Proof of Theorem \ref{thm:minimax-signal}} \label{Sec:pf-minimax}
We establish the minimax lower bound of signal strength by reducing high-power differentially private FDR control to a multi-hypothesis testing problem over a large packing of sparse alternatives, and then applying Fano’s inequality. Specifically,  differential privacy leads to an upper bound on the pairwise output KL divergence between hypotheses, and comparing this privacy-limited information with the combinatorial entropy of the packing yields the necessary signal-strength threshold. 

Let $\mathcal{V}$ index the largest packing set such that $ \{b^{(u)}: u \in \mathcal{V}\} \subset \{0, 1\}^p$ with $S_u:=\mathrm{supp}(b^{(u)})$ satisfying that $|S_u| = s$ for any element $u \in \mathcal{V}$ and $ |S_u \cap S_v| \leq s/4$ for $u \neq {v} \in \mathcal{V}$. It follows from the constant-weight code version of Varshamov-Gilbert bound that  $\log |\mathcal{V}| \gtrsim s \log (p/s)$. Specifically, we have 
$$
\log |\mathcal{V}|\geq \log  \frac{ {p \choose s} } { \sum_{i = 0}^{3s/4} {s \choose i} {p-s \choose i} } >   c_1 s \log (p/s),
$$
for some constant $c_1 > 0$. Given $u \in \mathcal{V}$, define $\bbeta^{(u)} = a b^{(u)} \in \Theta(a)$.

Given  $u \in \mathcal{V}$,  assume data $(\bX, \by)$ follows the linear model with parameter $ \bbeta^{(u)}$. Let $\widehat{S}$ be the output of an arbitrary selection procedure run on the data $(\bX, \by)$ with $\mu$-GDP guarantee and controlled FDR at level $q$. Define a decoder of the true support as 
\begin{equation} \label{eq:decoder-max-overlap}
\widehat{U} := \widehat{U}(\widehat S) = \argmax_{w \in \mathcal{V}} |\widehat{S} \cap S_w|.
\end{equation}
Let $\mathbb{P}_u$ and $\mathbb{E}_u$ denote probability and expectation under the data-generating distribution corresponding to the parameter $\bbeta^{(u)}$. Define the power of the selection $\widehat{S}$ as 
$$
\mathrm{Pow}_u(\widehat{S}) := \mathbb{E}_u\frac{ |\widehat{S} \cap S_u| }{ s }.
$$
Lemma \ref{le-decoder} below shows that high power under FDR control implies a high probability of successful decoding of the true support. 
\begin{lemma} \label{le-decoder}
 If the $\mathrm{Pow}_u(\widehat{S}) \geq 1 - \eta $ and $\FDR_u (\widehat{S}) \leq q$, then we have 
 \begin{equation} \label{eq-res-le-decoder}
     \mathbb{P}_u(\widehat{U} \neq u) \leq 4 \eta + \frac{5}{2}q .
 \end{equation}
\end{lemma} 

Assume $U$ is a uniform random variable  over the index set $\mathcal{V}$ and let data $D = (\bX, \by)$ be generated from the linear model with parameter $\bbeta^{(U)}$. Recall that $\widehat S$ is the output of a $\mu$-GDP algorithm based on data $D$, and $\widehat U$ is the decoder in \eqref{eq:decoder-max-overlap} based on $\widehat S$. Denote by $Q_u$ the probability law of $\widehat S$ given $U=u$. Define the mixture law $\bar Q:=\frac{1}{|\mathcal V|}\sum_{u\in\mathcal V}Q_u$.

Lemma \ref{le-decoder} converts the problem of support recovery to a decoding problem, where for the latter we can apply Fano's inequality to provide a lower bound on the probability of decoding error. 

Note that the mutual information of $(U, \widehat S)$ takes the form
\[
  I(U;\widehat S)
  \ :=\ \mathrm{KL}(\mathbb P_{U,\widehat S}\,\|\,\mathbb P_U\otimes \mathbb P_{\widehat S})
  \ =\ \frac{1}{|\mathcal V|}\sum_{u\in\mathcal V}\mathrm{KL}(Q_u\|\bar Q),
\]
where $\mathbb P_{U, \widehat S}$ is the joint probability law of $(U, \widehat S)$, and $\mathbb P_{U}$ and $\mathbb P_{\widehat S}$ are the marginal probability law of $U$ and $\widehat S$, respectively. 


The following lemma is the standard Fano inequality (e.g., Theorem 2.10.1 in \cite{CoverThomas2006}) applied to our decoding problem.

\begin{lemma} \label{le-fano}
Any decoder $\widehat U(\widehat S)$ satisfies
\[
  \inf_{\widehat U}\ \mathbb P\!\bigl(\widehat U(\widehat S)\neq U\bigr)
  \ \ge\
  1-\frac{I(U;\widehat S)+\log 2}{\log |\mathcal V|}.
\]
Moreover,
\[
  I(U;\widehat S)\ \le\ \frac{1}{|\mathcal V|^2}\sum_{u,u'\in\mathcal V} \mathrm{KL}(Q_u\|Q_{u'}).
\]
\end{lemma}
On the other hand, the following lemma provides an upper bound for the mutual information $I(U; \widehat{S})$. 
\begin{lemma} \label{le-bound-mutual}
We have
\begin{equation}
    I(U; \widehat{S}) \lesssim K(a):=\min \Big\{ \frac{2 n a^2 s} { \sigma^2 },  \mu^2 \Big( \frac{n^2 a^2 s}{ \sigma^2} + \frac{n a \sqrt{s}}{\sigma} \Big) \Big\}
\end{equation}
\end{lemma}

The proofs of Lemma \ref{le-decoder} and Lemma \ref{le-bound-mutual} are presented in Section \ref{Sec:pf-le-decoder} and Section \ref{Sec:pf-le-bound-mutual}, respectively. 

Combining Lemmas \ref{le-fano} and \ref{le-bound-mutual} and recalling  that $\log |\mathcal{V}| \gtrsim s \log (p/s)$, we obtain that 
\begin{equation} \label{eq-error-prob1}
     \mathbb{P}(\widehat{U} \neq U) \geq 1 - \frac{ C K(a) + \log 2 }{ s \log (p/s) } .
\end{equation}
Additionally, Lemma \ref{le-decoder} implies that 
\begin{equation} \label{eq-error-prob2}
\mathbb{P}(\widehat{U} \neq U) = \frac{1}{|\mathcal{V}|} \sum_{u \in \mathcal{V}} \mathbb{P}_u(\widehat{U} \neq u) \leq 4 \eta + \frac{5}{2}q,
\end{equation}
which together with \eqref{eq-error-prob1} requires
\begin{equation}
    4 \eta + \frac{5}{2}q \geq 1 - \frac{ C K(a) + \log 2 }{ s \log (p/s) }.  
\end{equation}
Therefore, for relatively small $\eta$ and $q$ such that $4 \eta + 5q/2 < 1$, we obtain that $K(a) \gtrsim s \log (p/s)$, which implies that 
\begin{equation}
    a \gtrsim \sigma  \cdot \max \Big\{ \sqrt{\frac{ \log (p/s)} {n}}, \frac{ \sqrt{\log(p/s)} }{ n \mu  }  \Big\} . 
\end{equation}
This completes the proof of Theorem \ref{thm:minimax-signal}.

\subsection{Proof of Proposition \ref{prop: screening-dp}}

Let $D=(\bX, \by)$ and $D'=(\bX', \by')$ be two neighboring datasets which differ in one datum. Recall that the
screening statistics $U_j=u(\bX_j, \by)$, $j\in[p]$, have sensitivity at most
$\Delta_u$, that is, $\max_{j\in[p]} |U_j(D)-U_j(D')|\leq \Delta_u .$

We first consider one step of the peeling procedure. For a fixed subset
$S\subset[p]$, define
\[
M(D)=\arg\max_{k\in S}\{U_k(D)+Z_k\},
\]
where $Z_k$'s are independent samples from $\operatorname{Lap}\left(\frac{2\Delta_u}{\varepsilon_{\mu,K_n}}\right).$
We show that $M$ is $\varepsilon_{\mu,K_n}$-DP.

Fix $i\in S$ and condition on
$Z_{-i}:=\{Z_k:k\in S,k\neq i\}$. Let
\[
T_i(D;Z_{-i})
=
\max_{k\in S,k\neq i}\{U_k(D)+Z_k\}-U_i(D).
\]
Then $M(D)=i$ if and only if $Z_i\geq T_i(D;Z_{-i}).$
By the sensitivity assumption, we have
\[
\begin{aligned}
T_i(D';Z_{-i})
&=
\max_{k\in S,k\neq i}\{U_k(D')+Z_k\}-U_i(D')  \\
&\leq
\max_{k\in S,k\neq i}\{U_k(D)+Z_k+\Delta_u\}
-\{U_i(D)-\Delta_u\} \\
&=
T_i(D;Z_{-i})+2\Delta_u .
\end{aligned}
\]
Since $Z_i\sim \operatorname{Lap}\left(\frac{2\Delta_u}{\varepsilon_{\mu,K_n}}\right),$ it holds for any $t\in\mathbb R$ that
\[
\mathbb P(Z_i\geq t+2\Delta_u)
\geq
e^{-\varepsilon_{\mu,K_n}}
\mathbb P(Z_i\geq t).
\]
Consequently,
\[
\begin{aligned}
\mathbb P(M(D')=i\mid Z_{-i})
&=
\mathbb P\{Z_i\geq T_i(D';Z_{-i})\mid Z_{-i}\}  \\
&\geq
\mathbb P\{Z_i\geq T_i(D;Z_{-i})+2\Delta_u\mid Z_{-i}\} \\
&\geq
e^{-\varepsilon_{\mu,K_n}}
\mathbb P\{Z_i\geq T_i(D;Z_{-i})\mid Z_{-i}\}  \\
&=
e^{-\varepsilon_{\mu,K_n}}
\mathbb P(M(D)=i\mid Z_{-i}).
\end{aligned}
\]
Taking expectation with respect to $Z_{-i}$, we obtain
\[
\mathbb P(M(D)=i)
\leq
e^{\varepsilon_{\mu,K_n}}\mathbb P(M(D')=i).
\]
Summing over $i\in E$ for any subset $E\subset S$, we have
\[
\mathbb P(M(D)\in E)
\leq
e^{\varepsilon_{\mu,K_n}}\mathbb P(M(D')\in E).
\]
Thus, one peeling step is $\varepsilon_{\mu,K_n}$-DP.

Next, we convert this pure-DP guarantee into GDP. By Lemma 13, every
$\varepsilon$-DP mechanism is
\[
2\Phi^{-1}\left(\frac{e^\varepsilon}{1+e^\varepsilon}\right)\text{-GDP}.
\]
By the definition
\[
\varepsilon_{\mu,K_n}
=
\log\left\{
\frac{\Phi(\mu/(2\sqrt{K_n}))}
{\Phi(-\mu/(2\sqrt{K_n}))}
\right\},
\]
we have
\[
\begin{aligned}
\frac{e^{\varepsilon_{\mu,K_n}}}
{1+e^{\varepsilon_{\mu,K_n}}}
&=
\frac{\Phi(\mu/(2\sqrt{K_n}))}
{\Phi(\mu/(2\sqrt{K_n}))+\Phi(-\mu/(2\sqrt{K_n}))} \\
&=
\Phi\left(\frac{\mu}{2\sqrt{K_n}}\right),
\end{aligned}
\]
where the last equality follows from $\Phi(x)+\Phi(-x)=1$. Therefore, each peeling
step is
\[
2\Phi^{-1}\left\{
\Phi\left(\frac{\mu}{2\sqrt{K_n}}\right)
\right\}
=
\frac{\mu}{\sqrt{K_n}}\text{-GDP}.
\]

Algorithm 2 performs $K_n$ adaptive peeling steps. At each step, conditional on the
previously selected indices, the remaining index set $S$ is fixed, and the same
argument above applies to the next selected index. Therefore, by the composition theorem of GDP, the ordered sequence $(i_1,\ldots,i_{K_n})$ satisfies $\sqrt{\sum_{j=1}^{K_n}
\left(\frac{\mu}{\sqrt{K_n}}\right)^2}=\mu$-GDP. Finally, the output
$C=\{i_1,\ldots,i_{K_n}\}$ is a deterministic function of the ordered sequence $(i_1,\ldots,i_{K_n})$. By the post-processing property of GDP, the output $C$ is also $\mu$-GDP. This completes
the proof of Proposition 5.

\begin{lemma}\label{lem: pure-to-mu}
Every $\varepsilon$-DP mechanism is $\mu$-GDP with
\[
\mu
=
2\Phi^{-1}\left(
\frac{e^\varepsilon}{1+e^\varepsilon}
\right).
\]
\end{lemma}
The proof of Lemma \ref{lem: pure-to-mu} is postponed to Section \ref{Sec: pf-lem-pure-to-mu}.

\subsection{Proof of Theorem \ref{thm-GDP-single}}
First, the screening procedure is $\mu/\sqrt{2}$-GDP. Moreover, noting that the $\ell_2$-sensitivity of $W_j$ is at most $\Delta_w$, Step 5 of Algorithm \ref{algo:screen-knockoff-single} is also $\mu/\sqrt{2}$-GDP by Theorem 1 of \cite{dong2019gaussian}.  Consequently, it follows by the composition theorem that Algorithm \ref{algo:screen-knockoff-single} is $\mu$-GDP. This completes the proof of Theorem \ref{thm-GDP-single}.

\subsection{Proof of Theorem \ref{thm-FDR-single}}
The main idea of proof is to leverage the symmetry of $\{  \widetilde{W}_j\}_{j \in \mathcal{C} \cap \mathcal{H}_0}$ given any subset $\mathcal{C} \subset [p]$ which is independent of the dataset $(\bX_{i, \cdot}, y_i)_{i \in I_2}$ for knockoff inference.
Note that 
\begin{equation}
    \begin{split}
        \FDR(\widehat{\mathcal{S}}_1) & = \e \bigg[ 
 \frac{ \sum_{j \in \mathcal{C} \cap \mathcal{H}_0} \mathbbm{1} (\widetilde{W}_j \geq  \widetilde{T} )  } {1 \vee  \sum_{ j \in \mathcal{C} } \mathbbm{1} (\widetilde{W}_j \geq \widetilde{T} ) } \bigg]  \\
 & = \e \bigg[  \frac{1 + \sum_{j \in \mathcal{C} \cap \mathcal{H}_0} \mathbbm{1} (\widetilde{W}_j \leq - \widetilde{T} )} { 1 \vee  \sum_{ j \in \mathcal{C} } \mathbbm{1} (\widetilde{W}_j \geq \widetilde{T} ) } \cdot  
 \frac{ \sum_{j \in \mathcal{C} \cap \mathcal{H}_0} \mathbbm{1} (\widetilde{W}_j \geq  \widetilde{T})  } {1 + \sum_{j \in \mathcal{C} \cap \mathcal{H}_0} \mathbbm{1} (\widetilde{W}_j \leq - \widetilde{T} ) } \bigg]  \\
 & \leq q \e \bigg[\frac{ \sum_{j \in \mathcal{C} \cap \mathcal{H}_0} \mathbbm{1} (\widetilde{W}_j \geq  \widetilde{T})  } {1 + \sum_{j \in \mathcal{C} \cap \mathcal{H}_0} \mathbbm{1} (\widetilde{W}_j \leq - \widetilde{T} ) } \bigg]    \\
 & =  q  \e \bigg\{ \e \bigg[\frac{ \sum_{j \in \mathcal{C} \cap \mathcal{H}_0} \mathbbm{1} (\widetilde{W}_j \geq  \widetilde{T})  } {1 + \sum_{j \in \mathcal{C} \cap \mathcal{H}_0} \mathbbm{1} (\widetilde{W}_j \leq - \widetilde{T} ) } \bigg| \mathcal{C} \bigg] \bigg\}.
 \end{split} 
\end{equation} 
Given any $\mathcal{C}$ which is independent of the training subsample $(\bX_{i, \cdot}, \widetilde{\bX}_{i, \cdot}, y_i)_{i \in I_2}$ for knockoff inference, the filtered knockoff statistics $\{W_j\}_{j \in \mathcal{C} \cap \mathcal{H}_0}$ still satisfy the coin-flip sign property (that is, the signs of $\{W_j\}_{j \in \mathcal{C} \cap \mathcal{H}_0}$ are i.i.d. Bernoulli variables with success probability $1/2$). In addition, recalling that the masked knockoff statistics $\widetilde{W}_j = W_j + Z_j$, where $Z_j $ is independently sampled from a mean-zero normal distribution. Thus, the masked knockoff statistics $\{\widetilde{W}_j\}_{j \in \mathcal{C} \cap \mathcal{H}_0}$ also satisfy the coin-flip property. 
Therefore,  by following the martingale technique in \cite{barber2015controlling}, we can obtain that conditional on $\mathcal{C}$,
\begin{equation} \label{pf-thm2-new-eq1}
    \e \bigg[ 
 \frac{ \sum_{j \in \mathcal{C} \cap \mathcal{H}_0} \mathbbm{1} (\widetilde{W}_j \geq   \widetilde{T})  } {1 + \sum_{j \in \mathcal{C} \cap \mathcal{H}_0} \mathbbm{1} (\widetilde{W}_j \leq - \widetilde{T}) } \bigg| \mathcal{C} \bigg] \leq 1 .
\end{equation}
Hence we have $\FDR(\widehat{\mathcal{S}}_1) \leq q$. 

\subsection{Proof of Theorem \ref{thm-power-general}} \label{Sec:pf-thm-power}
We first establish the lower bound on the power of the original knockoff procedure, and then derive the upper bound on the relative power loss of the differentially private knockoff procedure compared to the original inference. 
The proof of relative power loss consists of three steps. First, we will establish  upper and lower bounds for the knockoff threshold $T^*$ in the original knockoff procedure. Second, we will show that all the influential knockoff statistics, that is, the ones in  $ \{j: |W_j| \geq T^*  \}$, will pass through the mirror peeling algorithm when $m$ is chosen appropriately.  Therefore, the knockoff procedure based on the unmasked knockoff statistics $\{W_{i_j}: i_j \in D_m\}$ will yield a set $\breve{S}$ 
that contains $S^*$, where recall that $S^* = \{j: W_j\geq T^*\}$ is the set of variables selected by the original knockoff procedure. 
The last step is to prove that the selected set obtained by using the masked knockoff statistics $\{\widetilde{W}_{i_j}: i_j \in D_m \}$ aligns with the selected set by using the unmasked counterpart with high probability. Combining these three steps proves the desired result on relative power loss in \eqref{general-power-loss}. 

First, we need to  lower bound the power of the original knockoff procedure. Observe that under Condition \ref{cond-power-2} on signal strength, $w_j = 0$ for $j \in \mathcal{H}_0$ and $w_j > 2 \ell_n$ for $j \in \mathcal{S}_0$. Therefore, we obtain
\begin{equation}
    \mathbb{P} ( \cup_{j \in \mathcal{H}_0} \{ |W_j| \geq  \ell_n \} ) = \mathbb{P} ( \cup_{j \in \mathcal{H}_0} \{ |W_j - w_j| \geq \ell_n \} ) \leq \kappa_n, 
\end{equation}
and 
\begin{equation}
    \mathbb{P} ( \cup_{j \in \mathcal{S}_0} \{ W_j \leq  \ell_n \} ) \leq \mathbb{P} ( \cup_{j \in \mathcal{S}_0} \{ | W_j - w_j | \geq  \ell_n \} ) \leq \kappa_n. 
\end{equation}
This means that with probability at least $ 1 - 2 \kappa_n$, it holds that  $ |W_j| < \ell_n $ for all $j \in \mathcal{H}_0$ and $W_j >  \ell_n $ for all $ j \in \mathcal{S}_0$. Consequently, we have when $s \geq  q^{-1}$, 
\begin{equation}
    1 + \sum_{j = 1}^p \mathbbm{1} ( W_j \leq -  \ell_n ) = 1 \leq  q s = q \sum_{j = 1}^p \mathbbm{1} (W_j \geq  \ell_n ), 
\end{equation}
which implies the threshold $T^* \leq  \ell_n $ by definition of $T^*$. Therefore, the selection outcome satisfies $S^* = \{ j: W_j \geq T^* \} \supset \{j: W_j \geq  \ell_n \} = \mathcal{S}_0$ with probability $1 - 2 \kappa_n$ and thus we have for the power lower bound, 
\begin{equation} \label{eq-power-org-knockoffs}
    \e\Big[\frac{S^* \cap \mathcal{S}_0}{s} \Big] \geq 1 - 2 \kappa_n. 
\end{equation}
Recalling the definition of relative power loss $ \phi$ in \eqref{formula-relative-power-loss}, we have the power lower of the differentially private knockoff procedure satisfies 
\begin{equation}
    \e\Big[\frac{\widetilde{S} \cap \mathcal{S}_0}{s} \Big] =  \e\Big[\frac{ {S}^* \cap \mathcal{S}_0}{s} \Big] - \e [\phi] \geq 1 - 2 \kappa_n - \e [\phi]. 
\end{equation}
This completes the proof of \eqref{general-power-lower bound}.

Next we proceed to bound the relative power loss $\e[\phi]$.  Denote the threshold of the knockoff procedure utilizing the filtered but unmasked knockoff statistics $ \mathcal{W}_m :=\{W_{i_1}, \ldots, W_{i_m}\}  $ as 
\begin{equation} \label{thre-D_m-orginal}
    \breve{T}:=  \inf \bigg\{t \in | {\mathcal{W}}_m|: \frac{1 + \sum_{j = 1}^{m} \mathbbm{1} ( {W}_{i_j} \leq -t) } {1 \vee  \sum_{j = 1}^{m}\mathbbm{1} ( {W}_{i_j} \geq t)  } \leq q \bigg\}
\end{equation}
and the selected set $\breve{S} = \{j \in D_m: W_{j} \geq \breve{T} \}$,  where $|\mathcal{W}_m| :=\{|W_{i_1}|, \ldots, |W_{i_m}|\}$.

Recall the definition $S^* = \{j: W_j \geq T^*\}$ denotes the selected set of the original knockoff procedure and $\widetilde{S}$ represents the selection outcome of the differentially private knockoff procedure in Algorithm \ref{alg:mirror peeling}.  The following Lemmas \ref{le-1-pf-thm-power} -- \ref{le-4-pf-thm-power} outline the different steps in the proof and their proofs are postponed to Sections \ref{Sec.B.power.1} -- \ref{Sec.B.power.4}. 

    \begin{lemma} \label{le-1-pf-thm-power}      
   Assume Condition \ref{cond-power-2} holds. For some $\varepsilon \in (0, \frac{1 - q} {2 (1 + q)} \land \frac{1}{4})$, we have 
    $$ 
    \mathbb{P} \bigg( G^{-1} \Big( \frac{2 q s} {p_0 (1 - q)} \Big)  \leq  T^* \leq G^{-1} \Big( \frac{ q s }{ 2 p } \Big) \bigg) \geq 1 - \mathbb{P} (\mathcal{B}_{1, \varepsilon}) - O(\kappa_n). 
    $$    
    \end{lemma}

\begin{lemma}  \label{le-2-pf-thm-power}
   Let $\mathcal B_3 := \big\{ S^* \subset D_m \big\}$. Assume $m >  s \big(1 +  \frac{6(1 + \varepsilon)q } {1 - q}\big)     $ for some $\varepsilon \in (0, \frac{1 - q}{ 2 (1 + q)})$, we have 
    $$
    \mathbb{P} (\mathcal{B}_3  ) \geq 1 -  \mathbb{P} ( \mathcal{B}_{1, \varepsilon} )  - O(\kappa_n + p^{-1}). 
    $$ 
\end{lemma}

\begin{lemma}  \label{le-3-pf-thm-power}
Under event $\mathcal B_3 $, we have 
\begin{equation}
    \breve{T} \leq T^* \quad \mbox{and} \quad  S^* \subset \breve{S} .
\end{equation}
\end{lemma}

\begin{lemma} \label{le-4-pf-thm-power}
We have
\begin{equation}
\begin{split}
    & \mathbb{P} \big(\breve{S}  \cap \widetilde{S}^c \cap \mathcal{S}_0 \subset \{j \in D_m \cap \mathcal{S}_0: W_{j} \in [\breve{T}, \breve{T} + b_n ] \} \big)  \\
    & \geq 1 - \mathbb{P} (\mathcal{B}_2) - \mathbb{P}(\mathcal{B}_{1, \varepsilon} ) -  \mathbb{P}( \breve{T}   \leq b_n ) - O( p^{-1}) .
    \end{split}
\end{equation}
\end{lemma}

Conditional on $\mathcal B_3$ defined in Lemma \ref{le-2-pf-thm-power}, by Lemma \ref{le-3-pf-thm-power} we have
$
    |S^* \cap \widetilde{S}^c \cap \mathcal{S}_0|\leq |\breve{S} \cap \widetilde{S}^c \cap \mathcal{S}_0|$ and $\breve{T} \leq T^* $. Thus, by Lemma \ref{le-2-pf-thm-power}
\begin{align*}
    \E \Big[\frac{|S^* \cap \widetilde{S}^c \cap \mathcal{S}_0|}{s} \Big]&\leq \E \Big[\frac{|S^* \cap \widetilde{S}^c \cap \mathcal{S}_0|}{s} \times \mathbbm{1}(\mathcal B_3  ) \Big] +\mathbb P(\mathcal B_3^c)  \\
   & \leq \E \Big[\frac{|\breve{S} \cap \widetilde{S}^c \cap \mathcal{S}_0|}{s}   ) \times \mathbbm{1}(\mathcal B_3  ) \Big] +   \mathbb{P} (\mathcal{B}_{1, \varepsilon} )  + O( p^{-1}) + O(\kappa_n) .
\end{align*} 
Further by Lemma \ref{le-4-pf-thm-power} we have
\begin{align*}
   &  \e \Big[\frac{|\breve{S} \cap \widetilde{S}^c \cap \mathcal{S}_0|}{s} \times \mathbbm{1}(\mathcal B_3  ) \Big]  \\
   & \leq  \e \Big[\frac{\big|\{j \in D_m \cap \mathcal{S}_0: W_{j} \in [\breve{T}, \breve{T} + b_n ] \}\big|}{s} \times \mathbbm{1}(\mathcal B_3  ) \Big] \\
   & \quad + \mathbb{P} (\mathcal{B}_2) + \mathbb{P}(\mathcal{B}_{1, \varepsilon} )  + \mathbb{P} (  \breve{T} \leq b_n ) + O( p^{-1} )\\
   &\leq \E \Big[ \frac{\sum_{j \in \mathcal{S}_0 } \mathbbm{1} (W_j \in [\breve{T},\breve{T} + b_n ]) }{s} \times \mathbbm{1}(\mathcal B_3  ) \Big] \\
   & \quad + \mathbb{P} (\mathcal{B}_2) + \mathbb{P}(\mathcal{B}_{1, \varepsilon} ) + \mathbb{P} (  \breve{T} \leq b_n )  + O( p^{-1})\\
   &\leq s^{-1} \E \Big[\sup_{t\in (0,  G^{-1}(\frac{q s}{2p})]}\sum_{j \in \mathcal{S}_0 } \mathbbm{1} (W_j \in [t,t + b_n ]) \Big] \\
   & \quad + 2 \mathbb{P} ( \mathcal{B}_{1, \varepsilon} ) + \mathbb{P} (\mathcal{B}_2) + \mathbb{P} (  \breve{T} \leq b_n )   + O(\kappa_n)  + O( p^{-1} ),
\end{align*}
where in the last step  is obtained since $ \breve{T}  \leq T^*$ under $\mathcal{B}_{3}$ by Lemma \ref{le-2-pf-thm-power},  and $ T^* \leq G^{-1} (\frac{q s}{2 p}) $ with probability $ 1 - \mathbb{P} (\mathcal{B}_{1, \varepsilon} ) - O(\kappa_n)$ by Lemma \ref{le-1-pf-thm-power}. 

Combining the above two results and noting that the relative power loss $\e[\phi] \leq \mathbb{E}[\frac{|S^* \cap \widetilde{S}^c \cap \mathcal{S}_0|}{s}]$, we obtain the desired result in \eqref{general-power-loss}.
This completes the proof of Theorem \ref{thm-power-general}. 

\subsection{Proof of Theorem \ref{thm-power-single-split}} \label{Sec:pf-thm-power-single}
 
    The proof proceeds in two steps. First, we show that the strong-signal set $\mathcal{S}_1$ passes through the screening procedure. Second,  we prove that these features are selected by the downstream DP-knockoff inference procedure.  Similar to the proof in \eqref{pf-le-filter-2}, we can show that $\mathbb{P} (\max_{k\in [K_n], j \in [p]} |Z_{k, j}|\geq  b_{n,1}/2) \leq   p^{-1} $, where $Z_{k, j} \stackrel{i.i.d.}{\sim}  \Lap(   2 \Delta_u  / \varepsilon_{\mu, K_n})$ are added privacy noise in the screening procedure and $  \varepsilon_{\mu, K_n} =  \log ( \Phi(\frac{\mu}{2 \sqrt{2 K_n}}) / \Phi(- \frac{\mu}{2 \sqrt{2 K_n}}))$. In addition,  $\mathbb{P} (\max_{j\in  \mathcal{C}} |Z_{j}|\geq  b_{n,2}/2) \leq 2 p^{-1}$, where $ Z_j \stackrel{i.i.d.}{\sim} N(0, 2 \Delta_w^2 /\mu^2)$ are the added noise in the DP-knockoff procedure and $\mathcal{C}$ is any given subset of $[p]$ with $|\mathcal{C}| = K_n$. Therefore, all subsequent arguments are carried out conditional on the events $\{ \max_{k\in [K_n], j \in [p]} |Z_{k, j}|<  b_{n,1} /2 \}$ and  $ \{ \max_{j\in  \mathcal{C}} |Z_{j}| < b_{n,2} /2 \} $.

    Note that when $ K_n > s $, there must exist some $j^* \in \mathcal{H}_0$ such that it is retained in the screened index set $ \mathcal{C} $, that is, $ j^* \in \mathcal{C} $. In addition, under Condition \ref{cond-single-split-power-1}, it holds that with probability $1 - \kappa_n$, 
    \begin{equation} \label{eq1-pf-thm-power-single}
        U_j > u_j - \ell_n > b_{n,1} + \ell_n  ~\mbox{for any} ~ j \in \mathcal{S}_1
    \end{equation} 
    and 
    \begin{equation} \label{eq2-pf-thm-power-single}
        |U_j| < |u_j| + \ell_n \leq \ell_n ~\mbox{for any} ~ j \in \mathcal{H}_0.
    \end{equation}
   Therefore, for $j \in\mathcal{S}_1$ and any $k \in [K_n]$, 
   $$
   U_j + Z_{k,j} > b_{n,1} + \ell_n - b_{n,1}/2 \geq b_{n,1}/2 + \ell_n > U_{j^*}  + Z_{k, j^*}.
   $$
   This implies that $j \in \mathcal{C}$ for any $j \in {\mathcal{S}_1}$, by the screening rule in Algorithm \ref{algo:dp-screen}. 

   Next we prove that $\mathcal{S}_1$ will also be selected in the DP-knockoff procedure. Similar to \eqref{eq1-pf-thm-power-single} and \eqref{eq2-pf-thm-power-single}, we can show under Condition \ref{cond-single-split-power-2} that with probability $ 1 - 4 p^{-1} -\kappa_n$, for $j \in \mathcal{S}_1$,
   \begin{equation*}
       \widetilde{W}_j  > W_j - b_{n,2} /2 \geq w_j - \ell_n - b_{n,2}/2 > b_{n,2}/2 + \ell_n ,
   \end{equation*}
   and for $j \in \mathcal{H}_0$, 
   \begin{equation*}
       |\widetilde{W}_j|  < |W_j| + b_{n,2} /2 \leq |w_j| + \ell_n + b_{n,2}/2 \leq  b_{n,2}/2 + \ell_n .
   \end{equation*}
In addition, we have shown that $\mathcal{S}_1 \subset \mathcal{C}$ with probability $ 1- 4 p^{-1} -\kappa_n$. Therefore,  it follows that 
   \begin{equation}
       \sum_{j \in \mathcal{C}} \mathbbm{1}(\widetilde{W}_j \geq b_{n,2}/2 + \ell_n ) = \sum_{j \in \mathcal{C} \cap \mathcal{S}_1} \mathbbm{1}(\widetilde{W}_j \geq b_{n,2}/2 + \ell_n ) \geq |\mathcal{S}_1|
   \end{equation} 
   and 
   \begin{equation}
   \begin{split}
       1 + \sum_{j \in \mathcal{C}} \mathbbm{1}(\widetilde{W}_j \leq - b_{n,2}/2 - \ell_n ) & \leq 1 + \sum_{j \in \mathcal{C} \cap \mathcal{H}_0} \mathbbm{1}(\widetilde{W}_j \leq -b_{n,2}/2 - \ell_n ) + | \mathcal{S}_0 \cap \mathcal{S}_1^c | \\
       & = 1 + | \mathcal{S}_0 \cap \mathcal{S}_1^c |.
    \end{split}
   \end{equation} 
   Consequently, when $|\mathcal{S}_1|/(1 + s) \geq 1/(1 + q)$, we obtain 
   $$
  \frac{ 1 +\sum_{j \in \mathcal{C}} \mathbbm{1}(\widetilde{W}_j \leq - b_{n,2}/2 - \ell_n ) } { \sum_{j \in \mathcal{C}} \mathbbm{1}(\widetilde{W}_j \geq b_{n,2}/2 + \ell_n ) } \leq \frac{ 1 + (s - |\mathcal{S}_1|) } { |\mathcal{S}_1| } \leq q  
  $$
  and hence $  \widetilde{T} \leq b_{n,2} /2 + \ell_n$. This yields that $ \widetilde{W}_j > \widetilde{T} $ for any $j \in \mathcal{S}_1$ and thus $\mathcal{S}_1 \subset \widehat{S}$. Consequently, it holds that the total power loss of Algorithm \ref{algo:screen-knockoff-single} satisfies
  $$
  1 - \mathrm{Power}(\widehat{S}_1) = \mathbb{E}\frac{ | \mathcal{S}_0 \cap \widehat{S}_1^c | }{s} \leq \frac{ | \mathcal{S}_0 \cap  \mathcal{S}_1^c | }{s} + O( p^{-1}) + 2 \kappa_n.
  $$
  For the relative power loss $\e[\phi(\widehat{S}_1)]$, it holds that 
  $$
  \e[\phi(\widehat{S}_1)] \leq \e \frac{ |S^* \cap \mathcal{S}_0 \cap \widehat{S}_1^c| }{s} \leq 1 - \mathrm{Power} (\widehat{S}_1)
  $$
  and hence the same upper bound applies to $\e[\phi(\widehat{S}_1)]$. 
  This completes the proof of Theorem \ref{thm-power-single-split}.

\subsection{Proof of Theorem \ref{thm-GDP-multi}}

    Observe that each split of the algorithm is $\mu/\sqrt{B}$-GDP. By the composition theorem of GDP (Corollary 2 of \cite{dong2019gaussian}), Algorithm \ref{algo:screen-knockoff-multi} is $\mu$-GDP.

\subsection{Proof of Theorem \ref{thm-FDR-multi}}
   According to Theorem 2 in \cite{ren2024derandomised}, it suffices to prove that $\sum_{j \in \mathcal{H}_0} \e [e_j^{\avg}] \leq p$.  Observe that $e_j^{\avg} = \frac{1}{B} \sum_{b = 1}^B e_j^{(b)}$ which is the average of $\{ e_j^{(b)}\}_{b \in [B]}$, we only need to show $ \sum_{j \in \mathcal{H}_0} \e [e_j^{(b)}] \leq p$. Recalling the definition of $e_j^{(b)}$, we have
   \begin{equation}
   \begin{split}
       \sum_{j \in \mathcal{H}_0} \e [e_j^{(b)}] =  \sum_{j \in \mathcal{C}_b \cap \mathcal{H}_0} \e \bigg[\frac{ p \cdot \mathbbm{1}(\widetilde{W}_j^{(b)} \geq \widetilde{T}^{(b)})  } { 1 +  \sum_{j \in \mathcal{C}_b} \mathbbm{1}(\widetilde{W}_j^{(b)} \leq - \widetilde{T}^{(b)})   } \bigg] \\
       \leq  p \e \bigg[ \frac{\sum_{j \in \mathcal{C}_b \cap \mathcal{H}_0} \mathbbm{1}(\widetilde{W}_j^{(b)} \geq \widetilde{T}^{(b)})  } { 1 +  \sum_{j \in \mathcal{C}_b \cap \mathcal{H}_0} \mathbbm{1}(\widetilde{W}_j^{(b)} \leq - \widetilde{T}^{(b)})  } \bigg].
    \end{split}
   \end{equation}
   Applying the same arguments as in the proof of \eqref{pf-thm2-new-eq1}, we can obtain that 
   \begin{equation*}
   \begin{split}
       & \e \bigg[ \frac{\sum_{j \in \mathcal{C}_b \cap \mathcal{H}_0} \mathbbm{1}(\widetilde{W}_j^{(b)} \geq \widetilde{T}^{(b)})  } { 1 +  \sum_{j \in \mathcal{C}_b \cap \mathcal{H}_0} \mathbbm{1}(\widetilde{W}_j^{(b)} \leq - \widetilde{T}^{(b)})  } \bigg]  \\
       & = \e \bigg\{ \e \bigg[  \frac{\sum_{j \in \mathcal{C}_b \cap \mathcal{H}_0} \mathbbm{1}(\widetilde{W}_j^{(b)} \geq \widetilde{T}^{(b)})  } { 1 + \sum_{j \in \mathcal{C}_b \cap \mathcal{H}_0} \mathbbm{1}(\widetilde{W}_j^{(b)} \leq - \widetilde{T}^{(b)})  } \bigg| \mathcal{C}_b \bigg]  \bigg\} \leq 1. 
       \end{split}
   \end{equation*}
   Therefore, it is immediately seen that $\sum_{j \in \mathcal{H}_0} \e [e_j^{(b)}] \leq p$. This completes the proof of Theorem \ref{thm-FDR-multi}.

\subsection{Proof of Theorem \ref{thm:asymptotic-FDR}} \label{Sec:pf-thm8}
 The main idea of the proof is to connect the procedure with its counterpart based on the coupled ideal knockoff statistics. Since the procedure with ideal knockoff statistics attains exact FDR control, the approximate procedure achieves asymptotic FDR control provided that the coupling error $\rho_n$ is sufficiently small. Denote by $\wh{D}_m$ the filtered set in Step 9 and $ \wh{T} $ the knockoff threshold in Step 10 of Algorithm \ref{alg:mirror peeling} when approximate knockoff statistics are applied. Recall that $s = | \mathcal{S}_0 |$ denotes the number of relevant variables. 
 Define $\mathscr{A}_1 := \{ \max_{j\in [p]} |\wh{W}_j - W_j | \leq \rho_n \}$, $\mathscr{A}_2 := \{ \wh{T} < M_n\}$ and $\mathscr{A}_3 := \{ \max_{k\in[p]; j\in[m]} |Z_{k,j}| \leq b_n/2\} \cap \{ \max_{j\in[p]} |\wt{Z}_j| \leq b_n/2 \}$.
 For a small number $\varepsilon \in (0, 1/4)$,  define event 
$$\mathscr{A}_{4, \varepsilon} = \Bigg\{ \sup_{t \in (0, M_n + \rho_n)}  \bigg\{  \bigg| \frac{\sum_{j \in \mathcal{H}_0} \mathbbm{1}( {W}_j \geq t) } {\sum_{j \in \mathcal{H}_0} \mathbb{P}( {W}_j \geq t)} - 1  \bigg| \vee  \bigg| \frac{\sum_{j \in \mathcal{H}_0} \mathbbm{1}( {W}_j \leq - t) } {\sum_{j \in \mathcal{H}_0} \mathbb{P}( {W}_j \leq -t)} - 1  \bigg| \bigg\} < \varepsilon \Bigg\}.$$
Denote by $\mathscr{A} := \mathscr{A}_{1} \cap \mathscr{A}_{2} \cap \mathscr{A}_3 \cap \mathscr{A}_{4, \varepsilon}$. It follows from \eqref{pf-le-filter-2} and Conditions \ref{cond:gene-W-cou-accu} -- \ref{cond:approx-indicator} that $\mathbb{P} ({\mathscr{A} }) \to 1$.    We first present some lemmas that lead to the desired result. Proofs of Lemmas \ref{pf-thm8-lemma1} and \ref{pf-thm8-lemma2} are postponed to Sections \ref{Sec:pf-newle6} and \ref{Sec:pf-newle7}, respectively. 
 \begin{lemma} \label{pf-thm8-lemma1}
      Assume Condition \ref{cond:rho-n} is satisfied. For any constant $c_1 \in (0, 1)$, under event $\mathscr{A} $ with $\varepsilon < c_1/4$ and when $n$ is sufficiently large, it holds that $    \wh{D}_m   \subset \big\{j:  |\wh{W}_j | \geq G^{-1} ( \frac{(1 +c_1)m}{2p_0} ) \big\} $ and $  \wh{D}_m  \supset \big\{ j: |\wh{W}_j| \geq G^{-1} ( \frac{(1 - c_1)(m - s)}{2p_0} ) \big\} $.
 \end{lemma}

 \begin{lemma} \label{pf-thm8-lemma2}
     Assume Condition \ref{cond:rho-n} is satisfied. Suppose $m \geq  \frac{(2q +1 +c)s}{1 - q}$  for a constant $c > 0$ and $c_1 < \frac{2c(1-q)}{3(1+q)(1 + 2q) - 3(1-q) + 3c(1+q)}$. Under event $\mathscr{A}$ with a small $\varepsilon < c_1/4$ and when $n$ is sufficiently large, it holds that  $\wh{T} \geq G^{-1} ( \frac{(1 - c_1)(m - s)}{2p_0} )  $. 
 \end{lemma} 
 We choose a small constant $c_1 < \frac{2c(1-q)}{3(1+q)(1 + 2q) - 3(1-q) + 3c(1+q)}$ and $\mathscr{A} := \mathscr{A}_{1} \cap \mathscr{A}_{2} \cap \mathscr{A}_3 \cap \mathscr{A}_{4, \varepsilon}$ with a sufficiently small $\varepsilon < c_1/8$. Observe that 
 \begin{equation} \label{pf-thm8-eq1}
 \begin{split}
     \FDP(\wh{S}) & = \frac{\sum_{j \in \wh{D}_m \cap \mathcal{H}_0} \mathbbm{1}(\wh{W}_j + \widetilde{Z}_j \geq \wh{T}) } { 1 \vee \sum_{j \in \wh{D}_m} \mathbbm{1}(\wh{W}_j + \widetilde{Z}_j \geq \wh{T}) } \\
     & \leq \frac{\sum_{j \in \wh{D}_m \cap \mathcal{H}_0} \mathbbm{1}(\wh{W}_j + \widetilde{Z}_j \geq \wh{T}) } { 1 \vee \sum_{j \in \wh{D}_m} \mathbbm{1}(\wh{W}_j + \widetilde{Z}_j \geq \wh{T}) } \mathbbm{1} (\mathscr{A})  + \mathbbm{1} (\mathscr{A}^c)  \\
     & \leq q \frac{\sum_{j \in \wh{D}_m \cap \mathcal{H}_0} \mathbbm{1}(\wh{W}_j + \widetilde{Z}_j \geq \wh{T})} {1 + \sum_{j \in \wh{D}_m \cap \mathcal{H}_0} \mathbbm{1}(\wh{W}_j + \widetilde{Z}_j \leq - \wh{T})} \mathbbm{1} (\mathscr{A} ) + \mathfrak{o}_{\mathbf{P}}(1).
 \end{split}    
 \end{equation}
 It follows from Condition \ref{cond:rho-n} that when $n$ is sufficiently large, 
 \begin{equation*}
     \begin{split}
         G \left( G^{-1} \Big(\frac{(1 -c_1/2 )(m-s)}{2p_0} \Big) + b_n/2    \right) & = (1 + o(1)) G\left( G^{-1} \Big( \frac{(1- c_1)(m-s)}{2p_0} \Big)  \right) \\
        & = (1 + o(1)) \frac{(1 - c_1/2)(m-s)}{2p_0} \\
        & \geq  \frac{(1 - c_1 )(m-s)}{2p_0},
     \end{split}
 \end{equation*}
 which together with monotonicity of $G(\cdot)$ implies 
 \begin{equation*}
 \begin{split}
      G^{-1}\Big( \frac{(1 - c_1 )(m-s)}{2p_0}\Big) 
     & \geq G^{-1} \Big( \frac{(1 - c_1/2)(m-s)}{2p_0} \Big) + b_n /2 \\
     &>  G^{-1} \Big( \frac{(1 + 2 c_1)m}{2p_0} \Big) + b_n /2. 
    \end{split}
 \end{equation*}
 Under the event $\mathscr{A}$, it follows from Lemmas \ref{pf-thm8-lemma1} and \ref{pf-thm8-lemma2} that 
 \begin{equation} \label{pf-thm8-eq2}
 \begin{split}
     \sum_{j \in \wh{D}_m \cap \mathcal{H}_0} \mathbbm{1}(\wh{W}_j + \widetilde{Z}_j \geq \wh{T}) & = \sum_{j \in \mathcal{H}_0} \mathbbm{1} ( \wh{W}_j + \widetilde{Z}_j\geq \wh{T}, j \in \wh{D}_m ) \\
     & \leq \sum_{j \in \mathcal{H}_0} \mathbbm{1} \left(  \wh{W}_j   \geq \wh{T} - b_n/2  , |\wh{W}_j | \geq G^{-1} \Big( \frac{(1 + 2 c_1)m}{2p_0} \Big) \right) \\
     & = \sum_{j \in \mathcal{H}_0} \mathbbm{1}  (  \wh{W}_j    \geq \wh{T} ),
\end{split}
 \end{equation}
and 
 \begin{equation} \label{pf-thm8-eq3}
 \begin{split}
    &\sum_{j \in \wh{D}_m \cap \mathcal{H}_0} \mathbbm{1}(\wh{W}_j + \widetilde{Z}_j 
 \leq -\wh{T})  
 = \sum_{j \in \mathcal{H}_0} \mathbbm{1} ( \wh{W}_j + \widetilde{Z}_j  \leq -\wh{T}, j \in \wh{D}_m ) \\
     & \geq \sum_{j \in \mathcal{H}_0} \mathbbm{1} \Big( \wh{W}_j   \leq -\wh{T} - b_n/2,   |\wh{W}_j | \geq G^{-1} ( \frac{(1 - c_1/2 )(m-s)}{2p_0} )  \Big) \\
     & = \sum_{j \in \mathcal{H}_0} \mathbbm{1}  ( \wh{W}_j \leq -\wh{T}  ).
\end{split}
\end{equation}
Therefore, by substituting \eqref{pf-thm8-eq2} and \eqref{pf-thm8-eq3} into \eqref{pf-thm8-eq1}, we obtain  that 
\begin{equation} \label{pf-thm8-eq4}
    \begin{split}
        \FDP(\wh{S}) & \leq q \frac{\sum_{j \in   \mathcal{H}_0} \mathbbm{1}(\wh{W}_j   \geq \wh{T})} {1 + \sum_{j \in  \mathcal{H}_0} \mathbbm{1}(\wh{W}_j \leq - \wh{T})} \mathbbm{1} (\mathscr{A} ) +  \mathfrak{o}_{\mathbf{P}}(1) \\
        & \leq  q \frac{\sum_{j \in   \mathcal{H}_0} \mathbbm{1}( {W}_j \geq \wh{T} - \rho_n)} {1 + \sum_{j \in  \mathcal{H}_0} \mathbbm{1}( {W}_j \leq - \wh{T} - \rho_n)} \mathbbm{1} (\mathscr{A} ) +  \mathfrak{o}_{\mathbf{P}}(1) \\
        & \leq  q \sup_{t \in (0, M_n) }\frac{\sum_{j \in   \mathcal{H}_0} \mathbbm{1}( {W}_j \geq t - \rho_n)} {1 + \sum_{j \in  \mathcal{H}_0} \mathbbm{1}( {W}_j \leq - t - \rho_n)} \mathbbm{1} (\mathscr{A} ) +  \mathfrak{o}_{\mathbf{P}}(1) \\
        & \leq  \frac{q (1 + \varepsilon)}{1 - \varepsilon} \sup_{t \in (0,M_n) } \frac{\sum_{j \in \mathcal{H}_0} \mathbb{P}(  {W}_j \geq t - \rho_n ) } { \sum_{j \in  \mathcal{H}_0} \mathbb{P}( {W}_j \leq - t - \rho_n) } + \mathfrak{o}_{\mathbf{P}}(1) . 
    \end{split}
\end{equation}
Moreover, it follows from Condition \ref{cond:rho-n} that 
\begin{equation}  \label{pf-thm8-eq5}
 \begin{split}
     \sup_{t \in (0,M_n) } \frac{\sum_{j \in \mathcal{H}_0} \mathbb{P}(  {W}_j \geq t - \rho_n ) } { \sum_{j \in  \mathcal{H}_0} \mathbb{P}( {W}_j \leq - t - \rho_n) }  & = \sup_{t \in (0,M_n) } \frac{ G(t - \rho_n ) } {  G(  t +\rho_n)} \\
     & = \sup_{t \in (0,M_n) } \frac{ G(t) + G(t - \rho_n) - G(t)  } {  G( t) - [ G(t) -  G(t + \rho_n) ]} \\
     & = 1 + o(1). 
  \end{split}   
\end{equation}
Combining \eqref{pf-thm8-eq4} and \eqref{pf-thm8-eq5} yields that for any sufficiently small $\varepsilon > 0$, 
\begin{equation} \label{pf-thm8-eq6}
    \FDP(\wh{S}) \leq \frac{q (1 + \varepsilon)}{1 - \varepsilon} (1 + o(1)) + \mathfrak{o}_{\mathbf{P}}(1), 
\end{equation}
which leads to  $\mathbb{P} (\FDP(\wh{S}) \geq q + \varepsilon) \to 0$ for any $\varepsilon > 0$. In addition, $\limsup_{n\to \infty } \FDR(\wh{S})  = \limsup_{n \to \infty} \e [\FDP(\wh{S})]  \leq q$ follows immediately from \eqref{pf-thm8-eq6} and the fact that $\e[\mathfrak{o}_{\mathbf{P}}(1)] \leq \mathbb{P}(\mathscr{A}^c) \to 0$. This completes the proof of Theorem \ref{thm:asymptotic-FDR}.

\section{Proof of Lemmas} \label{Sec:pf-lemmas}

\subsection{Proof of Lemma \ref{lemma-W-sensi}} \label{Sec.B4}
    Without loss of generality, we assume the adjacent datasets $D = (\bX, \by)$ and $D' = (\bX', \by')$ differ only in the first row. That means, $\bX_{i, \cdot} = \bX'_{i, \cdot}$ for $2 \leq i \leq n$. It follows from the condition \eqref{W_i-random-function}  that the knockoff variables satisfy $\widetilde{\bX}_{i, \cdot} =  \widetilde{\bX}'_{i, \cdot}$, for $2 \leq i \leq n$. Hence the knockoff data matrix $\widetilde{\bX}$ and $\widetilde{\bX}'$ also differ only in the first row.
    Therefore, we obtain
    \begin{equation*}
        \widetilde{\bX} - \widetilde{\bX}' = \begin{pmatrix}
          \widetilde{\bX}_{1, \cdot} - \widetilde{\bX}_{1, \cdot}'\\
          {\bf 0}_{(n-1) \times p}.
         \end{pmatrix} 
    \end{equation*}
    Consequently, for $j \in [p]$, the sensitivity of $W_j$ satisfies
    \begin{equation*}
    \begin{split}
        |  W_j(D) - W_j(D') | & \leq n^{-1} \big|  \bX_j^T \by -  \bX_j'^T \by' \big| + n^{-1} \big| \widetilde\bX_j^T \by - \widetilde\bX_j'^T \by'  \big|  \\
        & \leq n^{-1} \big|  \bX_{1, j} y_1 -  \bX'_{1, j} y_1'  \big| + n^{-1} \big| \widetilde\bX_{1, j}^T y_1 - \widetilde\bX_{1, j}'^T y_1'  \big|\\
        & \leq 4 n^{-1} C_x C_y. 
    \end{split}
    \end{equation*}
    This completes the proof of Lemma \ref{lemma-W-sensi}.

\subsection{Proof of Lemma \ref{le-sensi-ridge}} \label{pf-le-ridge}
Let $\breve\bx_i$ and $\breve\bx_i'$ be the $i$th row (as vector) of the augmented data matrices $\breve\bX$ and $\breve\bX'$, respectively. 
To analyze the difference between $\widehat{\bbeta}(D)$ and $\widehat{\bbeta}(D')$, note that
\begin{equation} \label{dif-beta-ridge}
\begin{split}
    \|  \widehat{\bbeta}(D) -  \widehat{\bbeta}(D') \|_2 &\leq 
    \big\| \big[\big(n^{-1} \breve\bX ^T \breve\bX + \lambda I_{2p} \big)^{-1} - \big(n^{-1} (\breve\bX') ^T \breve\bX' + \lambda I_{2p} \big)^{-1}  \big] n^{-1} \breve\bX^T \by \big\|_2 \\
    & \quad +  \big\| \big(n^{-1} (\breve\bX') ^T \breve\bX' + \lambda I_{2p} \big)^{-1} n^{-1}(\breve\bX^T \by - (\breve\bX')^{T} \by' ) \big\|_2:= R_1 + R_2.
\end{split}
\end{equation}
As for $R_1$, it follows by applying the equality $A^{-1} - B^{-1} = B^{-1}(B - A)A^{-1}$ that 
\begin{equation}
    \begin{split}
        R_1 & =  \Big\|  \big(n^{-1} (\breve\bX') ^T \breve\bX' + \lambda I_{2p} \big)^{-1}\big[ n^{-1} (\breve\bX')^T \breve\bX' - n^{-1} \breve\bX^T \breve\bX \big]  \\
        & \quad \big(n^{-1} \breve\bX ^T \bX + \lambda I_{2p} \big)^{-1}  n^{-1} \breve\bX^T \by \Big\|_2 \\
        & \leq \big\| \big(n^{-1} (\breve\bX') ^T \breve\bX' + \lambda I_{2p} \big)^{-1} \big\|_2 \big\|  n^{-1} (\breve\bX')^T \breve\bX' - n^{-1} \breve\bX^T \breve\bX \big\|_2 \\
        & \quad \times \big\|  \big(n^{-1} \breve\bX ^T \breve\bX + \lambda I_{2p} \big)^{-1}   n^{-1} \breve\bX^T \by \big\|_2. 
    \end{split}
\end{equation}
It is easy to see that  $\big\| \big(n^{-1} (\breve\bX') ^T \breve\bX' + \lambda I_{2p} \big)^{-1} \big\|_2 \leq \min\{\lambda^{-1},  \sigma_{\min}^{-1}(n^{-1} (\breve\bX') ^T \breve\bX' ) \}$ and 
\begin{align*}
\big\|  n^{-1} (\breve\bX')^T \breve\bX' - n^{-1} \breve\bX^T \breve\bX \big\|_2 &  = \big\|  n^{-1} \breve\bx_1' (\bx_1')^T - n^{-1} \breve\bx_1 \breve\bx_1 ^T\big\|_2 \\
& \leq \max\{\| n^{-1} \breve\bx_1' (\bx_1')^T \|_2, \| n^{-1} \breve\bx_1 \breve\bx_1 ^T \|_2 \} \leq 2C_x^2 n^{-1} p.
\end{align*} 
In addition, assuming singular value decomposition $\breve\bX = U D V^T$, we can obtain 
\begin{equation}
    \begin{split}
        \big\|  \big(n^{-1} \breve\bX ^T \breve\bX + \lambda I_{2p} \big)^{-1}   n^{-1} \breve\bX^T \by \big\|_2 
        & = \big\|  n^{-1} V \big(n^{-1}  D^T D   + \lambda I_{2p} \big)^{-1}     D^T U^T \by \big\|_2 \\
        & \leq \big\|  n^{-1}   \big(n^{-1}  D^T D   + \lambda I_{2p} \big)^{-1}     D^T \big\|_{2} \| U^T \by \big\|_2 \\
        & = \big\| \big(n^{-1}  D^T D   + \lambda I_{2p} \big)^{-1}    n^{-1/2}     D^T \big\|_{2} n^{-1/2} \| U^T \by \big\|_2 \\
        & \leq \min\{\lambda^{-1/2}, \sigma_{\min}^{-1/2} (n^{-1} \breve\bX^T \breve\bX ) \} n^{-1/2} \| \by\|_2 \\
        & \leq C_y \min\{\lambda^{-1/2}, \sigma_{\min}^{-1/2} (n^{-1} \breve\bX^T \breve\bX ) \} .
    \end{split}
\end{equation}
Therefore, we have
\begin{equation} \label{R_1}
    \begin{split}
        R_1 & \leq 2C_x^2 C_y  n^{-1} p  \cdot \min\{\lambda^{-1},  \sigma_{\min}^{-1}(n^{-1} (\breve\bX') ^T \breve\bX' ) \}  \cdot \min\{\lambda^{-1/2},  \sigma_{\min}^{-1/2}(n^{-1} \breve\bX^T \breve\bX ) \} \\
        & = 2C_x^2 C_y  n^{-1} p \min\{\lambda^{-3/2},  \sigma_{\min}^{-1}(n^{-1} (\breve\bX') ^T \breve\bX' ) \sigma_{\min}^{-1/2}(n^{-1} \breve\bX^T \breve\bX ) \}. 
    \end{split}
\end{equation}

Next we proceed to address $R_2$. Note that
\begin{equation}
    \begin{split}
        R_2 & =  \big\| \big(n^{-1} (\breve\bX') ^T \breve\bX' + \lambda I_{2p} \big)^{-1} n^{-1} (y_1 \breve\bx_1 - y_1' \breve\bx_1') \big\|_2 \\
        & \leq \| \big(n^{-1} (\breve\bX') ^T \breve\bX' + \lambda I_{2p} \big)^{-1} \|_2 n^{-1} \| y_1 \breve\bx_1 - y_1' \breve\bx_1'\|_2 \\
        & \leq \min\{\lambda^{-1}, \sigma_{\min}^{-1}(n^{-1} (\breve\bX')^T\breve\bX' ) \}\cdot  2 n^{-1} C_x C_y (2p)^{1/2}  \\
        & = 2\sqrt{2} C_x C_y n^{-1} p^{1/2} \cdot  \min\{\lambda^{-1}, \sigma_{\min}^{-1}(n^{-1} (\breve\bX')^T \breve\bX' ) \},
    \end{split}
\end{equation}
which combining with \eqref{dif-beta-ridge} and \eqref{R_1} leads to 
\begin{align*}
  \| \bW(D) - \bW(D')\|_2  & \leq \sqrt{\sum_{j = 1}^p \big( |\widehat{\bbeta}_j(D) - \widehat{\bbeta}_j(D')| + |\widehat{\bbeta}_{j+p}(D) - \widehat{\bbeta}_{j+p} (D')| \big)^2 } \\
  & \leq \sqrt{2 \sum_{j = 1}^p \big(  |\widehat{\bbeta}_j(D) - \widehat{\bbeta}_j(D')|^2 + |\widehat{\bbeta}_{j+p}(D) - \widehat{\bbeta}_{j+p} (D')|^2 \big) } \\
 & \leq  \sqrt{2} \|  \widehat{\bbeta}(D) -  \widehat{\bbeta}(D') \|_2 \\
 &\leq 2\sqrt{2}C_x^2 C_y  n^{-1} p \min\{\lambda^{-3/2},  \sigma_{\min}^{-1}(n^{-1} (\breve\bX') ^T \breve\bX' ) \sigma_{\min}^{-1/2}(n^{-1} \breve\bX^T \breve\bX ) \} \\
 & \quad + 4 C_x C_y n^{-1} p^{1/2} \cdot  \min\{\lambda^{-1}, \sigma_{\min}^{-1}(n^{-1} (\breve\bX')^T \breve\bX' ) \}
\end{align*}
This concludes the proof of Lemma \ref{le-sensi-ridge}.

\subsection{proof of Lemma \ref{lemma-B_1-rate}} \label{Sec:pf-lemma-B_1-rate}
	Recall the definition $ G(t) = p_0^{-1} \sum_{j \in \mathcal{H}_0} \mathbb{P} ( {W}_j \geq t) $ and that $G(t)$ is a non-increasing, continuous function. Similar to the proof of Lemma 3 in \cite{fan2023ark}, the main idea is to partition the continuous interval $(0, G^{-1} (\frac{ q s} {2p})]$ into $k_n$ sub-intervals with end points $\{ t_i \}_{i = 0}^{k_n}$ such that $t_0\geq t_1\geq \cdots \geq t_{k_n}$ and $$|G(t_i)/ G(t_{i+1}) - 1 | \to 0$$ uniformly for $0 \leq i \leq k_n$ as $k_n\rightarrow \infty$. 
	Then the supreme over the continuous interval $(0, G^{-1} (\frac{ q s}{2 p})]$ can be reduced to the supreme over the set of discrete points $\{t_i\}_{i= 0}^{k_n}$, and therefore, we can apply the union bound to establish the desired result.  
	
	Specifically, we define  a sequence $  0 < z_0 < z_1 < \cdots < z_{k_n} = 1 $ and $$t_i := G^{-1} (z_i),$$ where $ z_0 = \frac { q s} {2 p}$, $z_i = \frac { q s} {2 p} + \frac { \sqrt{s L_n } i (\log (i+1) )^{2} } {p}$, and $k_n$ is defined such that $    k_n (\log (k_n + 1))^2    =  (\frac{p}{2} - \frac{qs}{2})/\sqrt{s L_n} $.  
    Under this partition, next we will  show that as $s \to \infty$,
	\begin{align} 
		& \sup_{0 \leq i \leq k_n} |G(t_{i+1})/G(t_{i}) - 1 |  \lesssim  \sqrt{L_n/s} (\log  (s/L_n) )^2 ,\label{pf-lemma2-G-grid-1} \\
        & \sup_{0 \leq i \leq k_n}|G(t_{i})/G(t_{i+1}) - 1 |  \lesssim \sqrt{L_n/s} (\log  (s/L_n) )^2 .\label{pf-lemma2-G-grid-2}
	\end{align} 
    Indeed, for $0 \leq i \leq k_n$, it holds that 
    \begin{equation*}
    \begin{split}
      | {G(t_{i+1}) } /{ G(t_i)} -1 | & =  | z_{i+1} /z_i - 1 | \\
      & = \frac{\sqrt{s L_n}  \{ (i+1) (\log (i+2))^2  - i (\log (i+1))^2 \} }{ qs /2 + \sqrt{s L_n}  i (\log (i+1))^2 } \\
      & = \frac{ \sqrt{s L_n}  \{ (\log (i+2))^2 + i [ (\log (i+2))^2 - (\log (i+1))^2 ]  \} }{qs /2 + \sqrt{s L_n}  i (\log (i+1))^2 } \\
      & <  \frac{ \sqrt{s L_n}  \{ (\log (i+2))^2 + 2 i   (\log (i+2)  -  \log (i+1) ) \log (i+2)   \} }{qs /2 + \sqrt{s L_n}  i (\log (i+1))^2 } .
    \end{split}
    \end{equation*}
In addition, we can obtain from the basic inequality $\log (1 + x) \leq x$ for $x \geq 0$ that $ \log (i+2) - \log (i+1) = \log (1 + (i+1)^{-1}) \leq (i+1)^{-1}  $. Therefore,  
\begin{equation*}
    \begin{split}
         | {G(t_{i+1}) } /{ G(t_i)} -1 | & \lesssim \frac{ \sqrt{s L_n}    (\log (i+1))^2   }{qs /2 + \sqrt{s L_n}  i (\log (i+1))^2 } \\
         & \lesssim   \min\{   \sqrt{\frac{L_n}{s}}  (\log i )^2  ,  i^{-1}  \}.
    \end{split}
\end{equation*}
Note that we have $ | {G(t_{i+1}) } /{ G(t_i)} -1 |    \lesssim \sqrt{\frac{L_n}{s}} (\log  (s/L_n) )^2$ when $i \leq \sqrt{s/L_n}$,  and $ | {G(t_{i+1}) } /{ G(t_i)} -1 |    \lesssim  i^{-1} \lesssim \sqrt{L_n/s}$ when $i > \sqrt{s/L_n}$. Consequently, \eqref{pf-lemma2-G-grid-1} follows. Similarly,  \eqref{pf-lemma2-G-grid-2} can be proved.

Now we are ready to prove the desired result in Lemma \ref{lemma-B_1-rate}.  For $t \in (0, G^{-1}( \frac{ q s} {2 p})]$, there exists some $0 \leq i \leq k_n - 1 $ such that $t \in [t_{i+1}, t_i]$.  It follows from the monotonicity of $ \mathbb{P} ( {W}_j \geq t) $ and $\mathbbm{1} ( W_j \geq t) $ that 
	\begin{equation*}  
		\begin{split}
			\bigg| \frac { \sum_{j \in \mathcal{H}_0 } \mathbbm{1} ( {W}_j \geq t) } { p_0 G(t) } - 1   \bigg| 
			& \leq  \max \bigg\{ \bigg| \frac{  \sum_{j \in \mathcal{H}_0 } \mathbbm{1} ( {W}_j \geq t_{i+1})} {p_0 G(t_i)} - 1  \bigg|, \\
			&\quad\bigg| \frac{ \sum_{j \in \mathcal{H}_0 } \mathbbm{1} ( {W}_j \geq t_{i})} {p_0 G(t_{i+1})} - 1  \bigg| \bigg\} .
		\end{split}
	\end{equation*}
	The two terms within the brackets on the right-hand side of the expression above can be bounded similarly and we will provide only the details on how to bound the first term for simplicity. 
	
	With the aid of the fact that $ | x y - 1 | \leq | x -1| |y - 1| + |x - 1| + |y -1| $ for all $x, y \in \mathbb{R}$, we can deduce that 
	\begin{equation*}   
		\begin{split}
			\bigg| \frac{  \sum_{j \in \mathcal{H}_0 } \mathbbm{1} ( {W}_j \geq t_{i+1})} {p_0 G(t_i)} - 1  \bigg|  & \leq \bigg|  \frac{\sum_{j \in \mathcal{H}_0 } \mathbbm{1} ( {W}_j \geq t_{i+1})} {p_0 G(t_{i+1})} - 1  \bigg| \cdot \sup_{0 \leq i \leq l_n} \bigg| \frac{G(t_{i})} {G(t_{i+1})} -1 \bigg| \\
			& \quad + \bigg|  \frac{\sum_{j \in \mathcal{H}_0 } \mathbbm{1} ( {W}_j \geq t_{i+1})} {p_0 G(t_{i+1})} - 1  \bigg| + \sup_{0 \leq i \leq l_n} \bigg| \frac{G(t_{i})} {G(t_{i+1})} -1 \bigg|\\
			& \lesssim \big[ 1 + \sqrt{L_n/s} (\log  (s/L_n) )^2 \big] \cdot \bigg|  \frac{\sum_{j \in \mathcal{H}_0 } \mathbbm{1} ( {W}_j \geq t_{i+1})} {p_0 G(t_{i+1})} - 1  \bigg| \\
            & \quad + \sqrt{L_n/s} (\log  (s/L_n) )^2  .
		\end{split}
	\end{equation*}
Since $L_n = o(s)$, it holds that $\sqrt{L_n/s} (\log  (s/L_n) )^2 \to 0 $ and hence 
$$
	\bigg| \frac{  \sum_{j \in \mathcal{H}_0 } \mathbbm{1} ( {W}_j \geq t_{i+1})} {p_0 G(t_i)} - 1  \bigg|  \lesssim \bigg|  \frac{\sum_{j \in \mathcal{H}_0 } \mathbbm{1} ( {W}_j \geq t_{i+1})} {p_0 G(t_{i+1})} - 1  \bigg|+ \sqrt{L_n/s} (\log  (s/L_n) )^2. 
$$

 Therefore, we can deduce from the union bound and Chebyshev's inequality that for any $\varepsilon > 0$ and sufficiently large $n$, 
	\begin{equation} \label{eq_d_n}
		\begin{split}
			& \mathbb{P} \bigg( \sup_{t \in (0,\, G^{-1} ( \frac{ q s }{2 p})] }  \bigg| \frac{\sum_{j \in \mathcal{H}_0} \mathbbm{1} (W_{j} \geq t) } {\sum_{j \in \mathcal{H}_0} \mathbb{P} (W_j \geq t)} - 1 \bigg| > \varepsilon )  \\
            & \leq \mathbb{P} \bigg( \sup_{ 0 \leq i \leq k_n}   \bigg| \frac{\sum_{j \in \mathcal{H}_0} \mathbbm{1} (W_{j} \geq t_{i}) } {\sum_{j \in \mathcal{H}_0} \mathbb{P} (W_j \geq t_i)} - 1 \bigg| +  \sqrt{L_n/s} (\log  (s/L_n) )^2 > \varepsilon \bigg)  \\
            & \leq \sum_{i = 0}^{k_n} \mathbb{P}\bigg(    \bigg| \frac { \sum_{j \in \mathcal{H}_0 } \{ \mathbbm{1} ( {W}_j \geq t_i) - \mathbb{P} (  {W}_j \geq t_i ) \} } { p_0 G(t_i) }   \bigg|   > \varepsilon - \sqrt{L_n/s} (\log  (s/L_n) )^2  \bigg)   \\
			& \lesssim   \sum_{i = 0}^{k_n} \frac {  \Var  \big( \sum_{j \in \mathcal{H}_0 }  \mathbbm{1} ( {W}_j \geq t_i) \big) } { \varepsilon^2 p_0^2 G^2 (t_i) } \lesssim \varepsilon^{-2}   L_n \sum_{i = 0}^{k_n} \frac{1}{p_0 G(t_i)}
		\end{split}
	\end{equation} 
  	Moreover, it holds that 
	\begin{equation} \label{pf-le2-3}
		\begin{split}
			\sum_{i = 0}^{k_n} \frac{ 1 } { p_0 G (t_i)} & =   p_0^{-1} \sum_{i = 0}^{k_n} \frac{ 1 } { z_i  }  = \frac{p} {p_0} \sum_{i = 1}^{k_n}   \frac {1} {  q s/2   + \sqrt{L_n s}  i (\log (i+1))^2  } +  \frac{2 p } {p_0 q s} \\
            & \leq \frac{p} {p_0} \sum_{i = 1}^{k_n}   \frac {1} {    \sqrt{L_n s}  i (\log (i+1))^2  } + O(s^{-1})\\
			&\lesssim   \sqrt{\frac{1}{L_n s}},
		\end{split}    
	\end{equation}
	where the last inequality above is due to $ \sum_{i = 1}^{\infty} \frac{1}{i (\log (i+1))^2} < \infty$, $p/p_0 \to 1$ and $ s^{-1} = o( 1/\sqrt{L_n s}) $. 
    Combining \eqref{eq_d_n} and \eqref{pf-le2-3} yields 
    \begin{equation}
        \mathbb{P} \bigg( \sup_{t \in (0,\, G^{-1} ( \frac{ q s }{2 p})] }  \bigg| \frac{\sum_{j \in \mathcal{H}_0} \mathbbm{1} (W_{j} \geq t) } {\sum_{j \in \mathcal{H}_0} \mathbb{P} (W_j \geq t)} - 1 \bigg| > \varepsilon \bigg)   \lesssim \varepsilon^{-2} \sqrt{\frac{L_n}{s}} \to 0. 
    \end{equation}
 By the same technique, we can show that 
    \begin{equation}
        \mathbb{P} \bigg( \sup_{t \in (0,\, G^{-1} ( \frac{ q s }{2 p})] }  \bigg| \frac{\sum_{j \in \mathcal{H}_0} \mathbbm{1} (W_{j} \leq -t) } {\sum_{j \in \mathcal{H}_0} \mathbb{P} (W_j \leq - t)} - 1 \bigg| > \varepsilon \bigg)   \lesssim \varepsilon^{-2} \sqrt{\frac{L_n}{s}} \to 0. 
    \end{equation}
This completes the proof of Lemma \ref{lemma-B_1-rate}.

\subsection{Proof of Lemma \ref{lemma-B_2-rate}} \label{Sec:pf-lemma-B_2-rate}
    The main idea is to apply the union bound and the assumed continuity of the conditional distribution of $W_k| W_j$ for any $k \neq j$. It follows by the union bound that 
    \begin{equation*}
        \begin{split}
            \mathbb{P} ( \mathcal{B}_2) & \leq \sum_{j \in [p]} \mathbb{P} \bigg( \min_{  k \neq j } \big|| W_j| - |W_k  | \big| \leq b_n \mbox{~and~}  |W_j| \geq G^{-1} \Big( \frac{3  m }{4 p }  \Big)    \bigg) \\
            & = \sum_{j \in [p]} \mathbb{P} \bigg(\bigcup_{k \neq j }   \Big\{ \big|| W_j| - |W_k  | \big| \leq b_n \Big\}  \cap  \Big\{ |W_j| \geq G^{-1} \Big( \frac{3  m }{4 p }  \Big) \Big\}   \bigg) \\
            & \leq  \sum_{j \in [p]}  \sum_{k \neq j } \mathbb{P} \Big(\Big\{ \big|| W_j| - |W_k  | \big| \leq b_n \Big\}  \cap  \Big\{ |W_j| \geq G^{-1} \Big( \frac{3  m }{4 p }  \Big) \Big\} \Big) \\
            & = \sum_{j \in [p]}  \sum_{k \neq j } \mathbb{P} \Big(\Big\{ \big|| W_j| - |W_k  | \big| \leq b_n \Big\}  \Bigm |   \Big\{ |W_j| \geq G^{-1} \Big( \frac{3  m }{4 p }  \Big) \Big\} \Big) \\
            & \quad \times  \mathbb{P} (|W_j| \geq G^{-1} \Big( \frac{3  m }{4 p }  \Big)) \\
            & = \sum_{j \in [p]}  \sum_{k \neq j } \mathbb{P} \Big(\Big\{ |W_j|- b_n \leq  |W_k| \leq  b_n + |W_j| \Big\}  \Bigm |   \Big\{ |W_j| \geq G^{-1} \Big( \frac{3  m }{4 p }  \Big) \Big\} \Big) \\
            & \quad \times  \mathbb{P} \Big(|W_j| \geq G^{-1} \Big( \frac{3  m }{4 p }  \Big) \Big) \\
            & \leq \sum_{j \in [p]}  \sum_{k \neq j } 4 C d_n b_n \mathbb{P} \Big(|W_j| \geq G^{-1} \Big( \frac{3  m }{4 p }  \Big) \Big) \\
            & \leq 4 C d_n b_n (p p_0 \frac{3  m }{4 p }  + p s)\lesssim m p d_n b_n  .
        \end{split}
    \end{equation*}
 This completes the proof of Lemma \ref{lemma-B_2-rate}.

\subsection{Proof of Lemma \ref{lemma-power-1st-error}} \label{Sec:pf-lemma-power-1st-error}
 
We observe that 
$$
\mathbb{P} (\breve{T} \leq b_n) =  \mathbb{P}(\cup_{j \in D_m} \{|W_{j}| \leq b_n, \breve{T} = |W_j|\}) \leq \sum_{j = 1}^p \mathbb{P} (|W_j| \leq b_n ).
$$
In addition, it holds that $ \mathbb{P} (|W_j| \leq b_n ) =   \mathbb{P} ( |d_n W_j| \leq d_n b_n )  \leq 2 C d_n b_n $ under assumption (i). 
Therefore, \eqref{eq-lemma3-result1} can be derived.  

Next we proceed to prove \eqref{power-le-result}. For simplicity, denote $M_n := G^{-1}(\frac{qs}{2p})$ and $v_n := \sup_{t \in (0, M_n + b_n)} \Var(\sum_{j \in \mathcal{S}_0} \mathbbm{1}(W_j > t) )$. It follows from Condition (ii) that $ d_n M_n  s^{-2} v_n \to 0$.   Recall that $\mathcal{S}_0$ is the relevant set and $s  = |\mathcal{S}_0|$. Define $\delta :=  [4 C s^2 v_n d_n(M_n + b_n)]^{1/4}$ and 
    \begin{equation}\label{eq:new-D}
         D_{n, \delta} :=\bigg \{ \sup_{t \in (0, M_n + b_n)} \bigg|  \sum_{j \in \mathcal{S}_0 } \mathbbm{1} (W_j > t) - \sum_{j \in \mathcal{S}_0 } \mathbb{P} (W_j > t)   \bigg|  > \delta \bigg\}, 
    \end{equation}
 First, we observe that naturally $s^{-1}   \sup_{t \in (0, M_n)}  \sum_{j \in \mathcal{S}_0} \mathbbm{1} (W_j \in (t, t+ b_n)) \leq 1 $ and 
   \begin{equation} \label{pf-power-le1-eq3}
   \begin{split}
        &  s^{-1}   \e \Big[ \sup_{t \in (0, M_n)}  \sum_{j \in \mathcal{S}_0 } \mathbbm{1} (W_j \in (t, t+ b_n)) \Big] \\
       & \leq \mathbb{P} (\mathcal{D}_{n, \delta}) + s^{-1} \e \bigg[ \sup_{t \in (0, M_n)}  \sum_{j \in \mathcal{S}_0  } \mathbbm{1} (W_j \in (t, t+ b_n)) \cdot \mathbbm{1} (D_{n, \delta}^c) \bigg]   \\
       & \leq   \mathbb{P} (\mathcal{D}_{n, \delta}) + s^{-1} \e \bigg[ \sup_{t \in (0, M_n)} \bigg| \sum_{j \in \mathcal{S}_0} \big( \mathbbm{1} (W_j > t)- \mathbb{P} (W_j > t  )  \big)\bigg|  \cdot \mathbbm{1} (D_{n, \delta}^c)   \bigg] \\
       & \quad +  s^{-1} \e \bigg[ \sup_{t \in (0, M_n)} \bigg| \sum_{j \in \mathcal{S}_0} \big( \mathbbm{1} (W_j > t + b_n ) - \mathbb{P} (W_j > t  + b_n )  \big) \bigg|  \cdot \mathbbm{1} (D_{n, \delta}^c)  \bigg] \\
       & \quad + s^{-1}  \sup_{t \in (0, M_n)} \mathbb{P} (W_j \in (t, t+ b_n))   \\
       & \leq   \mathbb{P} (\mathcal{D}_{n, \delta}) + 2 s^{-1} \delta + s^{-1} \sum_{j \in \mathcal{S}_0} \sup_{t \in (0, M_n)} \mathbb{P} (W_j \in (t, t+ b_n)).  
    \end{split}
   \end{equation}
 It follows by Conditions (i) and (iii) that the third term in the above upper bound tends to 0. Specifically, we have 
   \begin{equation}  \label{pf-power-le1-nnew-eq1}
        s^{-1} \sum_{j \in \mathcal{S}_0}  \sup_{t \in (0, M_n)} \mathbb{P} (W_j \in (t, t+ b_n))  \leq  C d_n b_n \to 0. 
   \end{equation}
Moreover, noting that naturally $v_n = \sup_{t \in (0, M_n + b_n)} \Var(\sum_{j \in \mathcal{S}_0} \mathbbm{1}(W_j > t) ) \leq s^2$,  we obtain under Conditions (ii) and (iii) that
   \begin{equation} \label{pf-lemma1-nnew-eq6}
       s^{-1} \delta =  [4 C s^{-2} v_n  d_n(M_n + b_n)]^{1/4}  \leq [4 C d_n b_n]^{1/4} + [4 C s^{-2} v_n d_n M_n]^{1/4}  \to 0. 
   \end{equation}

Now it remains to bound $ \mathbb{P} (D_{n, \delta})$. Let $l_n =  (s^2/v_n)^{1/4} (4 C d_n (M_n + b_n))^{3/4}$. Without loss of generality, let us assume that $l_n$ is an integer. We first divide the interval $(0, M_n + b_n] = \cup_{i \in [l_n] } I_i $ into $l_n$ disjoint sub-intervals, where $I_i := (t_{i-1}, t_{i}]$ with $t_i = i(M_n + b_n)/l_n$, for $i\in [l_n]$.  By definition of $D_{n, \delta}$ in \eqref{eq:new-D} and the monotonicity of functions $\sum_{j \in \mathcal{S}_0}\mathbbm{1} (W_j > t)$ and $\sum_{j \in \mathcal{S}_0} \mathbb{P} (W_j > t)$,  it follows that 
\begin{equation} \label{pf-lemma1-nnew-eq2}
\begin{split}
    \mathbb{P} (D_{n, \delta} ) & \leq  \mathbb{P} \Big( \sup_{i\in [l_n]} \sup_{t \in I_i} \Big|  \sum_{j \in \mathcal{S}_0}   \mathbbm{1} (W_j > t  ) - \sum_{j \in \mathcal{S}_0} \mathbb{P} (W_j > t )  \Big| > \delta  \Big ) \\
    & \leq \mathbb{P} \Big( \sup_{i\in [l_n]} \Big| \sum_{j \in \mathcal{S}_0}  \mathbbm{1} (W_j > t_{i-1} ) - \sum_{j \in \mathcal{S}_0} \mathbb{P} (W_j > t_{i})  \Big| > \delta/2 \Big) \\
    & \quad + \mathbb{P} \Big( \sup_{i\in [l_n]}\Big| \sum_{j \in \mathcal{S}_0} \mathbbm{1} (W_j > t_{i} ) - \sum_{j \in \mathcal{S}_0}\mathbb{P} (W_j > t_{i-1})  \Big| > \delta/2 \Big) \\
    & := P_1 + P_2.
\end{split}    
\end{equation}
We will only show the bound for $P_1$ since the same technique works for $P_2$. Observe that
\begin{equation*}
    \begin{split}
        P_1 & \leq \mathbb{P} \Big( \sup_{i\in [l_n]} \Big| \sum_{j \in \mathcal{S}_0}  \mathbbm{1} (W_j > t_{i-1} ) - \sum_{j \in \mathcal{S}_0} \mathbb{P} (W_j > t_{i-1})  \Big| \\
        & \qquad + \sup_{i \in [l_n]} \sum_{j \in \mathcal{S}_0} \mathbb{P} (W_j \in (t_{i-1}, t_i]) > \delta/2 \Big)  
    \end{split}
\end{equation*}
and by Condition (ii) and definitions of $l_n$ and $\delta$ that 
\begin{equation*}
\begin{split}
    \sup_{i \in [l_n] } \sum_{j \in \mathcal{S}_0} \mathbb{P} (W_j \in [t_{i-1}, t_{i}])  & = \sup_{i \in [l_n] } \sum_{j \in \mathcal{S}_0} \mathbb{P} (d_n W_j \in [d_n t_{i-1}, d_n t_{i}]) \\
    & \leq    C s d_n (M_n + b_n)/l_n =   \delta / 4. 
\end{split}
\end{equation*}
Then  we have by union bound inequality, Markov's inequality and Condition (i) that 
\begin{equation}  \label{pf-lemma1-nnew-eq4}
\begin{split}
    P_1 & \leq \mathbb{P}  \Big( \sup_{i\in [l_n]} \Big| \sum_{j \in \mathcal{S}_0}  \mathbbm{1} (W_j > t_{i-1} ) - \sum_{j \in \mathcal{S}_0} \mathbb{P} (W_j > t_{i-1})  \Big| > \delta / 4 \Big) \\
    & \leq l_n \sup_{i \in [l_n]} \mathbb{P} \Big(  \Big| \sum_{j \in \mathcal{S}_0}  \mathbbm{1} (W_j > t_{i-1} ) - \sum_{j \in \mathcal{S}_0} \mathbb{P} (W_j > t_{i-1})  \Big| > \delta / 4 \Big) \\ 
    & \leq 16 l_n \delta^{-2} \sup_{t \in (0, M_n + b_n) } \Var \Big(\sum_{j \in \mathcal{S}_0} \mathbbm{1} (W_j > t) \Big) \\
    & =  16   l_n \delta^{-2} v_n   =  16 [4 C s^{-2} v_n d_n (M_n + b_n)]^{1/4} \to 0,  
\end{split}
\end{equation}
where in the last step we have applied the same argument used in \eqref{pf-lemma1-nnew-eq6}. Combining \eqref{pf-lemma1-nnew-eq2} and \eqref{pf-lemma1-nnew-eq4} derives
\begin{equation}  \label{pf-lemma1-nnew-eq5}
    \mathbb{P} (D_{n, \delta}) \lesssim [   s^{-2} v_n d_n (M_n + b_n)]^{1/4} \to 0. 
\end{equation}
Finally, \eqref{pf-power-le1-eq3} together with \eqref{pf-power-le1-nnew-eq1}, \eqref{pf-lemma1-nnew-eq6} and \eqref{pf-lemma1-nnew-eq5}  leads to the desired result in \eqref{power-le-result}. This completes the proof of Lemma \ref{lemma-power-1st-error}. 

\subsection{Proof of Lemma \ref{lemma-cond(i)-marg}} \label{Sec:pf-density-marg}         
Recall that $ W_j := n^{-1} ( | \bX_j^{T} \by | - | \widetilde{\bX}_j^{T} \by | )$, where $\bX_j$ and $\widetilde{\bX}_j$ denote the $j$-th column of the data matrix $\bX$ and knockoff matrix $\widetilde{\bX}$, respectively.  Note that $ \widetilde{\bX} \independent \by  \,| \,\bX$ by the property of knockoff matrix. 
         Therefore, conditional on $(\bX, \by)$, we have $n^{-1} \widetilde{\bX}_j^{T} \by = n^{-1} \sum_{i = 1}^n \widetilde{\bX}_{i,j} y_i  $ is sum of independent variable with mean $ \mathbb{E}[\widetilde{\bX}_{i,j} y_i \,| \,(\bX, \by )] = y_i \mathbb{E}[\widetilde{\bX}_{i,j} \,| \, \bX_{i, \cdot}] $,  variance $ \Var(\widetilde{\bX}_{i,j} y_i \,| \, (\bX, \by )) = y_i ^2 \Var(\widetilde{\bX}_{i,j} \,| \,\bX_{i, \cdot}) $, and the third moment $ \mathbb{E}[|\widetilde{\bX}_{i,j} y_i |^3 \,| \,(\bX, \by )] = |y_i|^3 \mathbb{E}[|\widetilde{\bX}_{i,j}|^3 \,| \, \bX_{i, \cdot}]$. Under mild moment conditions,  it follows from the central limit theorem that conditional on $(\bX, \by)$, 
            $$
            n^{-1/2} \bigg(\widetilde{\bX}_j^{T} \by - \sum_{i = 1}^n y_i \mathbb{E}[\widetilde{\bX}_{i,j} \,| \, \bX_{i, \cdot}] \bigg)\stackrel{d}{\rightarrow} Z_j(\bX, \by):= N \Big(0, n^{-1}\sum_{i=1}^n y_i^2  \Var(\widetilde{\bX}_{i,j} \,| \, \bX_{i, \cdot}) \Big),
            $$
            and the following Berry-Esseen bound (\cite{berry1941accuracy}, \cite{esseen1942liapunov}) holds:
            \begin{equation*}
                \begin{split}
                      & \sup_{x \in \mathbb{R}} \Big| \mathbb{P} \Big( n^{-1/2} (\widetilde{\bX}_j^{T} \by - \sum_{i=1}^n y_i \mathbb{E}[\widetilde{\bX}_{i,j} \,| \, \bX_{i, \cdot}]) > x  \,\Big|\, (\bX, \by) \Big) - \mathbb{P} ( Z_j > x ) \Big|  \\
                      &  \lesssim \frac{   \sum_{i = 1}^n |y_i|^3  \mathbb{E}[|\widetilde{\bX}_{i,j}|^3 \,| \, \bX_{i, \cdot}] } { [\sum_{i = 1}^n  |y_i|^2\Var(\widetilde{\bX}_{i,j} \,| \, \bX_{i, \cdot})]^{3/2} } .
                \end{split}
            \end{equation*}

           To  lighten notation, denote by $ \tau_j := \mathbb{E}[Y^2 \Var(\widetilde{X}_j | X)], \sigma_j^2 := \Var (Y^2 \Var(\widetilde{X}_j | X))$, and $ \gamma_j:=\mathbb{E}[ |Y X_j|^3  $. Note that $\mathbb{E}[ |Y \widetilde{X}_j|^3  ] = \mathbb{E}[ |Y X_j|^3 = \gamma_j$ by the swapping property of knockoff variables. 
            Denote by $\mathscr{A}:= \{n^{-1} \sum_{i=1}^n y_i^2 \Var(\widetilde{\bX}_{i,j} \, | \, \bX_{i, \cdot}) \geq \tau_j /2\}$.
            It follows from the Chebyshev inequality that 
            $$
            \mathbb{P} ( \mathscr{A} ) \geq 1 - O(\frac{ \sigma_j^2 } { n \tau_j^2 }) = 1 -O(  n^{-1}),
            $$
            which implies that with high probability $1 -O(  n^{-1})$, the limit distribution $Z_j(\bX, \by)$ has   density function bounded by $ 1/\sqrt{\pi \tau_j} $ and the approximation error (i.e., Berry-Esseen bound) is of order $n^{-1/2}$. 
           Consequently, we can obtain that
            \begin{equation*}
                \begin{split}
                     & \mathbb{P} ( t \leq \sqrt{n} W_j \leq t + \sqrt{n} b_n ) 
                      = \mathbb{E} \big[\mathbb{P} ( t \leq \sqrt{n} W_j \leq t + \sqrt{n} b_n \, |\, \bX, \by ) \big] \\
                     & \leq  \mathbb{E} \big[\mathbb{P} ( -t - \sqrt{n} b_n + n^{-1/2} |\bX_j^{T} \by | \leq  n^{-1/2}  |\widetilde{\bX}_j^{T} \by | \leq -t  + n^{-1/2} |\bX_j^{T} \by | \, |\, \bX, \by ) \mathbbm{1} (\mathscr{A})\big] \\
                     & \quad + \mathbb{P}(\mathscr{A}^c) \\
                     & \lesssim \mathbb{E} \big[\mathbb{P} ( -t - \sqrt{n} b_n + n^{-1/2} |\bX_j^{T} \by | \leq   |Z_j| \leq -t  + n^{-1/2} |\bX_j^{T} \by | \, |\, \bX, \by )  \mathbbm{1} (\mathscr{A}) \big] \\
                     & \quad +   \mathbb{E} \Big[  \frac{   \sum_{i = 1}^n |y_i|^3  \mathbb{E}[|\widetilde{\bX}_{i,j}|^3 \,| \, \bX_{i, \cdot}] } { [\sum_{i = 1}^n  |y_i|^2\Var(\widetilde{\bX}_{i,j} \,| \, \bX_{i, \cdot})]^{3/2} }  \mathbbm{1} (\mathscr{A}) \Big] +  n^{-1} \\
                     & \lesssim \frac{1}{\sqrt{\pi \tau_j}} \cdot \sqrt n b_n +\frac{ \gamma_j }{\sqrt n \tau_j^{3/2}} + n^{-1}   \lesssim \sqrt n b_n  + n^{-1/2} .
                \end{split}
            \end{equation*}
In addition, recall that the sensitivity level $\Delta_n \asymp n^{-1}$ for marginal correlation knockoff statistics, and therefore, $\sqrt{n} b_n  \asymp   \frac{\sqrt{n}\Delta_n}{\mu} \sqrt{m } \log p \gtrsim n^{-1 } \sqrt{m }\log p \gg n^{-1/2}$. Finally, we obtain 
$$
\mathbb{P} ( t \leq \sqrt{n} W_j \leq t + \sqrt{n} b_n )  \lesssim \sqrt n b_n. 
$$
This completes the proof of Lemma \ref{lemma-cond(i)-marg}.

\subsection{Proof of Lemma \ref{lemma-cond(i)-ridge}} \label{Sec:pf-density-ridge}
Recall the knockoff statistic $W_j = | \widehat{\bbeta}_j | - | \widehat{\bbeta}_{j+p}|$, where $ \widehat{\bbeta} = (n^{-1} \breve{\bX}^T \breve{\bX} + \lambda I )^{-1} n^{-1} \breve{\bX}^T \by  $ with $\breve{\bX} = [\bX, \widetilde{\bX}]$ the augmented data matrix. 
Denote by $ \boldsymbol{\mu} := (n^{-1} \breve{\bX}^T \breve{\bX} + \lambda I )^{-1} n^{-1} \breve{\bX}^T \bX \bbeta \in \mathbb{R}^{2p}$ and $ \bSigma :=  \sigma^2 (n^{-1} \breve{\bX}^T \breve{\bX} + \lambda I )^{-1} n^{-1} \breve{\bX}^T \breve{\bX} (n^{-1} \breve{\bX}^T \breve{\bX} + \lambda I )^{-1} $. 
Note that conditional on $\breve{\bX}$, it follows from the central limit theorem that the joint distribution of $(\widehat{\bbeta}_j, \widehat{\bbeta}_{j+p})$ follows a normal distribution: 
\begin{equation}
 \begin{split}
     \Bigg(\begin{matrix}
\sqrt{n} (\widehat{\bbeta}_j  - \boldsymbol{\mu}_j) \\
 \sqrt{ n} (\widehat{\bbeta}_{j+p} - \boldsymbol{\mu}_{j+p})
\end{matrix} \Bigg)
\stackrel{d}{\sim}
 \Bigg(\begin{matrix}
Z_j   \\
  Z_{j+p} 
\end{matrix} \Bigg)  :=  N \Bigg( \mathbf{0}_2 , 
\bigg(\begin{matrix}
    \bSigma_{j, j}  & \bSigma_{j, j+p}  \\
    \bSigma_{j, j+p} &  \bSigma_{j+p, j+p}  \\
\end{matrix} \bigg) \Bigg)
 \end{split}
\end{equation}
Therefore, conditional on $\breve{\bX}$, the scaled knockoff statistic $\sqrt{n} W_j$ has the same distribution as $ | Z_j + \sqrt{n} \boldsymbol{\mu}_j |  - |  Z_{j+p} + \sqrt{n} \boldsymbol{\mu}_{j+p}  | $. Under event $\mathscr{A}:=\{ c_1 \leq \lambda_{\min} ( n^{-1} \breve{\bX}^T \breve{\bX}) \leq  \lambda_{\max} ( n^{-1} \breve{\bX}^T \breve{\bX}) \leq c_2\}$, we have that the conditional distribution $Z_{j+p}|Z_{j}$ is also normal with positive variance and hence has bounded density function. Therefore, we can obtain that 
\begin{equation}
    \begin{split}
         & \mathbb{P} ( t \leq \sqrt{n} W_j \leq t + \sqrt{n} b_n ) 
                      = \mathbb{E} \big[\mathbb{P} ( t \leq \sqrt{n} W_j \leq t + \sqrt{n} b_n \, |\, \breve{\bX}  ) \big] \\
         & \leq \mathbb{E} \big[\mathbb{P} ( t \leq \sqrt{n} W_j \leq t + \sqrt{n} b_n \, |\, \breve{\bX}  ) \mathbbm{1}(\mathscr{A}) \big] + O(n^{-1/2}) \\
         & \lesssim \sqrt n b_n +  n^{-1/2}.
    \end{split}
\end{equation}
Recalling that the sensitivity level $\Delta_n \gtrsim n^{-1} $ for ridge regression coefficient, we have $ n^{-1/2} \lesssim \sqrt{n} b_n $ and thus 
$$
 \mathbb{P} ( t \leq \sqrt{n} W_j \leq t + \sqrt{n} b_n )  \lesssim \sqrt n b_n .
$$
This concludes the proof of Lemma \ref{lemma-cond(i)-ridge}.

\subsection{Proof of Lemma \ref{le-decoder}} \label{Sec:pf-le-decoder}
Denote $T_u := |\widehat{S} \cap S_u|$ and $F_u := |\widehat{S} \cap S_u^c|$ and $\FDP_u := \frac{F_u}{ F_u + T_u }$. For notational simplicity, we suppress the subscript $u$ when there is no ambiguity. For any $w \neq u \in \mathcal{V}$, we have the set inclusion
\[
  \widehat S\cap S_w \subset (S_w\cap S_u)\ \cup\ (\widehat S\setminus S_u),
\]
hence
\begin{equation}\label{eq:overlap-upper}
  |\widehat S\cap S_w|\le |S_w\cap S_u| + |\widehat S\setminus S_u|
  \le  s/4 + F,
\end{equation}
where the last inequality uses the property $|S_w \cap S_u| \leq s/4$ for any $w \neq u$. Therefore, if
\begin{equation}\label{eq:strict-separation}
  T >  s/4 + F,
\end{equation}
then $|\widehat S\cap S_u|=T$ is strictly larger than $|\widehat S\cap S_w|$ for all $w\neq u$, so $u$ is the unique
maximizer in \eqref{eq:decoder-max-overlap} and thus $\widehat U=u$. This implies that $\{ T >  s/4 + F \} \subseteq  \{ \widehat{U} =  u \} $. 

Next, on the event $\{\FDP<  2/5\}$ and $\{|\widehat{S}|>0\}$, we have
\[
  \frac{F}{T+F} <   \frac{2}{5}\quad\Longrightarrow\quad F < \frac{2}{3}\,T.
\]
Hence on $\{T> 3s/4\}\cap\{\FDP< 2/5\}$,
\[
   s/4 + F \le  s/4 + 2 T/3 < T.
\]
 Therefore, it follows that under $\{|\widehat{S}| > 0\}$, 
\begin{equation}\label{eq:good-event-implies-success}
 \{T >   3s/4\} \cap \{\FDP < 2/5\} \subseteq  \{\widehat U= u\} .
\end{equation}
Taking complements, we obtain that 
\[
  \mathbb P_u(\widehat U\neq u, |\widehat{S}| > 0)
  \le \mathbb P_u(T\le 3 s/4, |\widehat{S}| > 0) + \mathbb P_u(\FDP\ge 2/5 ) .
\]
In addition, since $ T = 0 \leq 3s/4 $ when $|\widehat{S}| =0$, it holds that 
\[
  \mathbb P_u(\widehat U\neq u, |\widehat{S}| = 0)
  \le \mathbb P_u( |\widehat{S}| = 0) = \mathbb P_u( T\le 3 s/4, |\widehat{S}| = 0)  . 
\]
Combining the above two inequalities, we have 
\begin{equation} \label{eq-pf-le-decoder-1}
\mathbb P_u(\widehat U\neq u)
  \le \mathbb P_u(T\le 3s/4 ) + \mathbb P_u(\FDP\ge 2/5 )  .
\end{equation}
Observe that when $\mathrm{Pow}_u(\widehat{S}) = \mathbb{E}_u [T]/s \geq 1 - \eta$, 
$$
  s/4 \cdot \mathbb{P}_u(T\le  3 s/4)\le \E_u[s-T] = s-\E_u[T]\le \eta s, 
$$
where the first inequality is because on the event $\{T\le 3 s/4\}$ we have $s-T\ge s/4$. 
Hence 
\begin{equation} \label{eq-pf-le-decoder-2}
\mathbb{P}_u(T\le  3 s/4) \leq 4 \eta.
\end{equation}
In addition, we have 
\begin{equation}\label{eq-pf-le-decoder-3}
\mathbb P_u(\FDP\ge 2/5 ) \leq \frac{5}{2} \FDR = \frac{5}{2} q.  
\end{equation} 
A combination of \eqref{eq-pf-le-decoder-1} -- \eqref{eq-pf-le-decoder-3} yields the desired result in \eqref{eq-res-le-decoder}. This completes the proof of Lemma \ref{le-decoder}.

\subsection{Proof of Lemma \ref{le-bound-mutual}} \label{Sec:pf-le-bound-mutual}
Recall that $U $ follows a uniform distribution on $\mathcal{V}$ and $\widehat{U}$ is a function of the private output $\widehat{S}$. Denote $Q_u$ as the law of $\widehat{S}$ given $U = u$, and let $\bar{Q} := \frac{1}{|\mathcal{V}|} \sum_{u \in \mathcal{V}} Q_u$. Thus, the marginal distribution of $\widehat{S}$ is $\bar{Q}$. 
It follows by 
the chain rule of KL divergence, and the convexity of KL divergence that 
\begin{equation*}
\begin{split}
I(U; \widehat{S}) & = \frac{1}{|\mathcal{V}|} \sum_{u \in |\mathcal{V}|} \mathrm{KL} ( Q_u \, \| \, \bar{Q}) \\
& \leq \frac{1}{|\mathcal{V}|^2} \sum_{u \in \mathcal{V}} \sum_{u' \in \mathcal{V}} \mathrm{KL} ( Q_u \,\| \,Q_{u'} ).
\end{split}
\end{equation*}
Therefore, it suffices to derive the uniform upper bound of $\mathrm{KL} ( Q_u \,\| \,Q_{u'} )$  for any $ u \neq  u' $. We make two claims. First, 
\begin{equation} \label{eq-KL-claim1}
    \mathrm{KL} (Q_u \,\| \, Q_{u'}) \leq \frac{1}{2 \sigma^2} \mathbb{E}\big[\| \bX  (\bbeta^{(u)} - \bbeta^{(u')})  \|_2^2 \big].
\end{equation}
Second, for some coupling law $\gamma$ (to be introduced in the proof of \eqref{eq-KL-claim2}) of $(\bX, \by)$ and $(\bX, \by')$ such that $ \by|\bX \sim N(\bX \bbeta^{(u)} , \sigma^2 I_n) $ and $ \by'|\bX \sim N(\bX \bbeta^{(u')} , \sigma^2 I_n)  $, 
\begin{equation} \label{eq-KL-claim2}
    \mathrm{KL} (Q_u \,\| \, Q_{u'}) \leq \mu^2/2 \cdot \mathbb{E}_{\gamma} \Big[ \Big( \sum_{i = 1}^n \mathbbm{1}(y_i \neq y_i') \Big)^2 \Big], 
\end{equation}
where the expectation is taken wrt to $((\bX,\by), (\bX,\by'))\sim \gamma$.
The proof of \eqref{eq-KL-claim1} and \eqref{eq-KL-claim2} is postponed to the end of this subsection.  

Recall that $ \e[X X^{T}] = I_p $,  and $\bbeta^{(u)} = a b^{(u)}$ and $\bbeta^{(u')} = a b^{(u')}$ with $|\mathrm{supp}(\bbeta^{(u)})| = s  = |\mathrm{supp}(\bbeta^{(u')})| $. We obtain  
\begin{equation*}
\begin{split}
    \mathbb{E}\big[\| \bX  (\bbeta^{(u)} - \bbeta^{(u')})  \|_2^2 \big] = n \| (\bbeta^{(u)} - \bbeta^{(u')})  \|_2^2 \leq  4 n a^2 s
\end{split}
\end{equation*}
and therefore, using \eqref{eq-KL-claim1} we have
\begin{equation} \label{eq-pf-KL-conclusion1}
     \mathrm{KL} (Q_u \,\| \, Q_{u'}) \leq \frac{2 n a^2 s} {\sigma^2}. 
\end{equation}

In addition, we have 
\begin{equation} \label{eq-pf-KL-conclusion4}
    \begin{split}
        & \mathbb{E}_{\gamma} \Big[ \Big( \sum_{i = 1}^n \mathbbm{1}(y_i \neq y_i') \Big)^2 \Big]  = \mathbb{E}  \bigg\{\mathbb{E}  \Big[ \Big( \sum_{i = 1}^n \mathbbm{1}(y_i \neq y_i') \Big)^2 \Big| \bX \Big] \bigg\} \\
        & =  \mathbb{E}  \bigg[ \Var \Big\{   \sum_{i = 1}^n \mathbbm{1}(y_i \neq y_i')   \Big| \bX \Big\} + \Big\{\mathbb{E} \Big[   \sum_{i = 1}^n \mathbbm{1}(y_i \neq y_i')   \Big| \bX \Big]  \Big\}^2   \bigg] \\
        & \leq \mathbb{E}  \bigg[ \mathbb{E} \Big[  \sum_{i = 1}^n \mathbbm{1}(y_i \neq y_i')   \Big| \bX \Big]  + \Big\{\mathbb{E} \Big[ \sum_{i = 1}^n \mathbbm{1}(y_i \neq y_i')   \Big| \bX \Big]  \Big\}^2   \bigg] .
    \end{split}
\end{equation}
Observe that conditional on  $X$, the total variation between $ Y|X \sim N(X^{T} \bbeta^{(u)}, \sigma^2)$ and $Y'|X \sim N(X^{T} \bbeta^{(u')}, \sigma^2) $ satisfies 
\begin{equation*}
    \mathrm{TV}(N(X^{T} \bbeta^{(u)}, \sigma^2), N(X^{T} \bbeta^{(u')}, \sigma^2)) \leq \min \Big\{1, \frac{| X^{T}(\bbeta^{(u)} - \bbeta^{(u')}) |}{\sqrt{2 \pi }\sigma} \Big\}.
\end{equation*}
Since the coupling law $\gamma$ of $(\bX, \by)$ and $(\bX, \by')$ is  the maximal coupling  which achieves the total variation bound,  it holds that 
\begin{equation*}
\begin{split}
    \mathbb{P} (y_i \neq y_i' | \bX_i) = \mathrm{TV}
    & (N(\bX_i^{T} \bbeta^{(u)}, \sigma^2), N(\bX_i^{T} \bbeta^{(u')}, \sigma^2))  \\
    & \leq \min \Big\{1, \frac{| \bX_i^{T}(\bbeta^{(u)} - \bbeta^{(u')}) |}{\sqrt{2 \pi }\sigma} \Big\}.
\end{split}
\end{equation*}
Therefore, we have   
\begin{equation*}  
    \begin{split}
        \mathbb{E} \Big[ \sum_{i = 1}^n \mathbbm{1}(y_i \neq y_i')   \Big| \bX \Big]  & \leq \sum_{i = 1}^n \frac{ | \bX_i^{T}(\bbeta^{(u)} - \bbeta^{(u')}) | } { \sqrt{2 \pi }\sigma  }  \\
        & \leq \frac{\sqrt n }{\sqrt{2 \pi} \sigma} \| \bX (\bbeta^{(u)} - \bbeta^{(u')}) \|_2 .
    \end{split}
\end{equation*}
Hence 
\begin{equation} \label{eq-pf-KL-conclusion2}
\begin{split}
    \mathbb{E}  \bigg[ \mathbb{E} \Big[  \sum_{i = 1}^n \mathbbm{1}(y_i \neq y_i')   \Big| \bX \Big] \bigg] & \leq  \frac{\sqrt n }{\sqrt{2 \pi} \sigma} \mathbb{E} \big[\| \bX (\bbeta^{(u)} - \bbeta^{(u')}) \|_2 \big] \\
    & \leq  \frac{\sqrt n }{\sqrt{2 \pi} \sigma} \Big(\mathbb{E} \big[\| \bX (\bbeta^{(u)} - \bbeta^{(u')}) \|_2^2 \big] \Big)^{1/2} \\
    & \leq  \frac{\sqrt n }{\sqrt{2 \pi} \sigma} (4 n a^2 s)^{1/2} \lesssim \frac{  n a \sqrt{s} }{\sigma}
\end{split}
\end{equation}
and similarly, 
\begin{equation}  \label{eq-pf-KL-conclusion3}
    \Big\{\mathbb{E} \Big[ \sum_{i = 1}^n \mathbbm{1}(y_i \neq y_i')   \Big| \bX \Big]  \Big\}^2 \lesssim \frac{n^2 a^2 s}{\sigma^2}.
\end{equation}
Substituting  \eqref{eq-pf-KL-conclusion2}, \eqref{eq-pf-KL-conclusion3} and \eqref{eq-pf-KL-conclusion4} into \eqref{eq-KL-claim2} leads to 
\begin{equation}
     \mathrm{KL} (Q_u \,\| \, Q_{u'}) \lesssim \mu^2  \Big(\frac{n^2 a^2 s}{\sigma^2} +  \frac{  n a \sqrt{s} }{\sigma} \Big), 
\end{equation}
which together with \eqref{eq-pf-KL-conclusion1} derives the desired result in Lemma \ref{le-bound-mutual}.  

It remains to prove \eqref{eq-KL-claim1} and \eqref{eq-KL-claim2}. 

\textbf{Proof of \eqref{eq-KL-claim1}}. Let $\mathcal{L}(X)$ denote the law of a random variable $X$. 
Observe that we can write the selection outcome $\widehat{S}$ as a function of the input data $(\bX, \by)$ by $ \widehat{S} = \widehat{S}(\bX, \by) $. 
Conditional on a design matrix $\bX = \bx$, define $ \widehat{S}_{\bx} (\by) =  \widehat{S}(\bx, \by) $ and denote by $ Q_{u}^{\bx}:= \mathcal{L} (\widehat{S}_{\bx} (\by)) $ the distribution of $\widehat{S}_{\bx} (\by)$. Observe that 
\begin{equation} \label{eq-pf-KL-claim1-1}
     \mathrm{KL} ( Q_u \| Q_{u'} ) = \mathbb{E} [ \mathrm{KL} ( Q_u^{\bx} \| Q_{u'}^{\bx} ) ], 
\end{equation}
where the expectation is taken with respect to $\bX$. 
It follows from the data processing inequality of KL divergence that given $\bX = \bx$, 
\begin{equation} \label{eq-pf-KL-claim1-2}
\begin{split}
    \mathrm{KL} ( Q_u^{\bx} \, \| \, Q_{u'}^{\bx} ) & \leq  \mathrm{KL} ( N(\bX \bbeta^{(u)}, \sigma^2 I_n) \, \| \, N(\bX \bbeta^{(u')}, \sigma^2 I_n)) \\
    & = \frac{ \| \bx ( \bbeta^{(u)} - \bbeta^{(u')} ) \|_2^2 }{2\sigma^2 },
\end{split}
\end{equation}
which combined with \eqref{eq-pf-KL-claim1-1} leads to \eqref{eq-KL-claim1}.

\textbf{Proof of \eqref{eq-KL-claim2}}. Fix $u, u' \in \mathcal V$ and $u\neq u'$, first draw a common design matrix $\bX$ and then conditional on $\bX$, define a coupling law  that couples the responses row-by-row: for each $i\in [n]$, couple
$$
y_i \sim N(X_i^T\bbeta^{(u)}, \sigma^2), \qquad y_i \sim N(X_i^T\bbeta^{(u')}, \sigma^2)
$$
using a maximal coupling, independently across $i$. Denote the above coupling law of $(\bX, \by)$ and $(\bX, \by')$ by $\gamma$. Since $\bX$ is shared, the number of different rows in $(\bX, \by)$ and $(\bX, \by')$ is equal to $\sum_{i=1}^n{\mathbbm 1} \{y_i\neq y_i'\}$ under $\gamma$. 

Using the coupling law $\gamma$ defined above, further define an augmented pair of distributions $(\mathsf R, \mathsf S)$ on $((\bX, \by'),(\bX, \by'),\widehat S)$ as follows:\\
   Under $\mathsf R$: draw $((\bX, \by),(\bX, \by'))\sim \gamma$ and then output $\widehat S(\bX, \by)$\\
Under $\mathsf S$: draw $((\bX, \by),(\bX, \by'))\sim \gamma$ and then output $\widehat S(\bX, \by')$.

By construction, the marginal law of $\widehat S(\bX, \by)$ is $Q_u$, and the marginal law of $\widehat S(\bX, \by')$ is $Q_{u'}$. By the data processing inequality for KL divergence we have
$$
\mathrm{KL} ( Q_u  \| Q_{u'}  ) \leq \mathrm{KL} ( \mathsf R \, \|  \,\mathsf S  ). 
$$
By the definition of $\mathsf R$ and $\mathsf S$ above, we have
\begin{equation}
\begin{split}
     \mathrm{KL} ( \mathsf R \, \|  \,\mathsf S  ) = \mathbb{E}_{\gamma} \big[\mathrm{KL} ( \mathcal{L}(\widehat{S}(\bX, \by)) \, \|  \,\mathcal{L}(\widehat{S}(\bX, \by'))  )  \big],
\end{split}
\end{equation}
where the expectation is taken wrt to $((\bX, \by),(\bX, \by'))\sim \gamma$. 
In addition, since the selection outcome $ \widehat{S} $ guarantees $\mu$-GDP for neighboring datasets which differ in one row,  it follows from Theorem 3 in \cite{dong2019gaussian} that $\widehat{S} $ is $\mu t$-GDP for $t$-neighboring datasets, where at most $t$ rows are different. Corollary B.6 in \cite{dong2019gaussian} establishes that a $\mu$-GDP procedure is $(\alpha, (\mu^2/2) \alpha)$-R\'enyi differentially
private, where recall that R\'enyi divergence converges to the KL divergence when $\alpha \to 1$. Therefore, letting $\alpha \to 1$, we obtain that conditional on  $((\bX, \by),(\bX, \by'))$,
\begin{equation}
    \mathrm{KL}( \mathcal{L}(\widehat{S}(\bX, \by)') \, \|  \,\mathcal{L}(\widehat{S}(\bX, \by'))  )  \leq \mu^2/2 \cdot  \Big( \sum_{i = 1}^n \mathbbm{1}(y_i \neq y_i') \Big)^2 . 
\end{equation}
Consequently, the claim in \eqref{eq-KL-claim2} is derived by taking expectation with respect to the data $((\bX, \by),(\bX, \by'))\sim \gamma$. 

The proof of Lemma \ref{le-bound-mutual} is complete.

\subsection{Proof of Lemma \ref{lem: pure-to-mu}} \label{Sec: pf-lem-pure-to-mu}
Let $M$ be an $\varepsilon$-DP mechanism. By the connection between $f$-DP and $(\varepsilon, \delta)$-DP in \cite{dong2019gaussian}, we know that $\varepsilon$-DP implies $f_\varepsilon$-DP with 
$$f_\varepsilon(\alpha)=\max\left\{0,1-e^\varepsilon \alpha,e^{-\varepsilon}(1-\alpha)\right\}.$$
It remains to show that $f_\varepsilon\ge G_\mu(\alpha)$, $\forall \alpha\in[0,1]$,
where
$G_\mu(\alpha)=\Phi\left(\Phi^{-1}(1-\alpha)-\mu\right)$ is the Gaussian tradeoff function. For $\alpha_*=1/(1+e^\varepsilon)$, the two nonzero linear pieces of $f_\varepsilon$ meet at $\alpha_*$. More precisely,
$$
f_\varepsilon
=
\begin{cases}
1-e^\varepsilon\alpha, & 0\le \alpha\le \alpha_*,\\[4pt]
e^{-\varepsilon}(1-\alpha), & \alpha_*\le \alpha\le 1.
\end{cases}
$$

Now evaluate $G_\mu$ at three points. By calculation, we have
$$
G_\mu(0)=1,
\qquad
G_\mu(1)=0,
\qquad
G_\mu(\alpha_*)=\alpha_*
$$
Because $G_\mu$ is convex, its graph lies below the chord joining any two points on its graph. Thus,
$$
G_\mu(\alpha)\le f_\varepsilon(\alpha)
\qquad\forall \alpha\in[0,1].
$$
Thus, by Definition 2.6 in \cite{dong2019gaussian}, $M$ is $\mu$-GDP with
$\mu=2\Phi^{-1}\left(\frac{e^\varepsilon}{1+e^\varepsilon}\right).$

\subsection{Proof of Lemma \ref{le-1-pf-thm-power}}  \label{Sec.B.power.1}
For convenience of presentation, the following arguments are all under event $\mathcal{B}_{1, \varepsilon}^c$. We will prove the result by showing the upper bound and lower bound separately. 

\textit{Step 1}. Let us first consider the upper bound. By definition of $T^*$, it suffices to show that 
\begin{equation} \label{pf-lemm7-neww-eq0}
    1 + \sum_{j = 1}^p \mathbbm{1} \big( W_j \leq - G^{-1} ( \frac{ q s}{2p}) \big) \leq qs \leq q \sum_{j = 1}^p \mathbbm{1} \big( W_j \geq G^{-1} (\frac{ q s}{2p}) \big)
\end{equation}
holds with probability $1 - O(\kappa_n)$. 

First, it follows from the union bound that under Condition \ref{cond-power-2}, 
\begin{equation*}
    \begin{split}
        \mathbb{P} \big( \cup_{j \in \mathcal{S}_0}   \{ W_j <  \ell_n \} \big) & =  \sum_{j \in \mathcal{S}_0}  \mathbb{P} (  W_j - w_j + w_j  <  \ell_n ) \\
        & \leq \sum_{j \in \mathcal{S}_0}  \mathbb{P} (  W_j - w_j < - \ell_n )  \\
        & \leq \sum_{j = 1}^p  \mathbb{P} ( |W_j - w_j|  > \ell_n ) \leq \kappa_n \to 0. 
    \end{split}
\end{equation*}
Therefore, it holds that 
\begin{equation} \label{pf-lemm7-neww-eq1}
 \mathbb{P} \Big( \sum_{j = 1}^p \mathbbm{1} (W_j \geq  \ell_n) \geq s \Big) \geq   \mathbb{P} ( \cap_{j \in \mathcal{S}_0} \{ W_j \geq  \ell_n \} )  \geq 1 - \kappa_n. 
\end{equation}
In addition, since $w_j \geq 0$ for any $j \in [p]$ and $p_0 /p \to 1$ by assumption,  we have under Condition \ref{cond-power-2} that as $n$ is sufficiently large, 
\begin{equation*}
\begin{split}
    G(\ell_n) & = p_0 ^{-1} \sum_{j \in \mathcal{H}_0} \mathbb{P} (W_j \leq -  \ell_n) \leq p_0^{-1} \sum_{j = 1}^p \mathbb{P}( W_j - w_j \leq  -  \ell_n  )  \\
    & \leq p_{0}^{-1} \sum_{j = 1}^p \mathbb{P}( |W_j - w_j| \geq    \ell_n  ) \leq p_0^{-1} \kappa_n  < \frac{ q s}{2p}.
\end{split}
\end{equation*}
Therefore, it follows immediately from monotonicity of $G(\cdot)$ that 
$$
 G^{-1}( \frac{ q s}{ 2 p} ) \leq  \ell_n,
$$
which together with \eqref{pf-lemm7-neww-eq1} yields
\begin{equation} \label{pf-lemm7-neww-eq2}
\begin{split}
     \mathbb{P} \bigg( \sum_{j = 1}^p \mathbbm{1} \Big(W_j \geq   G^{-1} \big( \frac{  q s}{2p} \big) \Big) \geq s \bigg) &  \geq  \mathbb{P} \Big( \sum_{j = 1}^p \mathbbm{1} (W_j \geq  \ell_n) \geq s \Big) \geq 1 - \kappa_n. 
\end{split}
\end{equation}
This proves the second inequality in \eqref{pf-lemm7-neww-eq0}. It remains to prove the first inequality in \eqref{pf-lemm7-neww-eq0}. Under event $\mathcal{B}_{1, \varepsilon}^c$, we have
\begin{equation*}
    \begin{split}
        \frac{ q s }{ 2p}  & =   p_0^{-1} \sum_{j \in \mathcal{H}_0} \mathbb{P} (W_j \leq - G^{-1} (\frac{  q s}{2p}) ) \\
        & \geq   (1 - \varepsilon   )  p_0^{-1} \sum_{j \in \mathcal{H}_0} \mathbbm{1} (W_j \leq - G^{-1} (\frac{ q s}{2p} )),
    \end{split}
\end{equation*}
and thus when $\varepsilon < 1/4$, 
\begin{equation} \label{pf-lemm7-neww-eq3}
\begin{split}
    \sum_{j \in \mathcal{H}_0} \mathbbm{1} \Big(W_j \leq - G^{-1} (\frac{ q s}{2p} ) \Big) & \leq \frac{q s p_0}{2p(1 - \varepsilon)}  \leq \frac{q s  }{2(1 - \varepsilon)} < \frac{2 q s}{3}  . 
\end{split}
\end{equation}
In addition, we obtain by Markov's inequality and Condition \ref{cond-power-2} that 
\begin{equation} \label{pf-lemm7-neww-eq4}
\begin{split}
    & \mathbb{P} \bigg(\sum_{j \in \mathcal{S}_0}  \mathbbm{1} \Big(W_j \leq - G^{-1} (\frac{ q s}{2p} ) \Big) \geq  \frac{q s}{3} \bigg)   \\
    & \leq  \frac{3}{q s} \sum_{j \in \mathcal{S}_0} \mathbb{P} \Big(W_j - w_j \leq - G^{-1} (\frac{ q s}{2 p}) - w_j   \Big) \\
    & \leq  \frac{3}{q s} \sum_{j \in \mathcal{S}_0} \mathbb{P} \Big( |W_j - w_j| \geq    w_j  ) \Big)  \leq \frac{3}{q s} \sum_{j \in \mathcal{S}_0} \mathbb{P} \Big( |W_j - w_j| \geq   2 \ell_n  ) \Big) \leq \frac{3 \kappa_n }{q s} .
\end{split}   
\end{equation}
Combining \eqref{pf-lemm7-neww-eq3} and \eqref{pf-lemm7-neww-eq4} leads to 
\begin{equation*}
     \mathbb{P} \bigg(\sum_{j = 1}^p  \mathbbm{1} \Big(W_j \leq - G^{-1} (\frac{ q s}{ 2 p} ) \Big) <  q s \bigg)  \geq 1  -\frac{3 \kappa_n }{q s}, 
\end{equation*}
which together with \eqref{pf-lemm7-neww-eq2} derives \eqref{pf-lemm7-neww-eq0} with probability $ 1  -  O(\kappa_n) $. This results in the upper bound $T^* \leq G^{-1} ( \frac{q s}{2p} )$ with probability $1 - O(\kappa_n)$.

\noindent \textit{Step 2}. Now we proceed to prove the lower bound. It follows from the definition of $T^*$ in \eqref{knockoff-threshold} that 
    \begin{equation*}
        1 + \sum_{j = 1}^p \mathbbm{1} (W_j \leq - T^*) \leq q + q \sum_{j = 1}^p \mathbbm{1} (W_j \geq T^*).
    \end{equation*}
    Noting that $\sum_{j \in \mathcal{H}_0 } \mathbbm{1} (W_j \leq -  T^*) \leq \sum_{j=1}^p \mathbbm{1} (W_j \leq -  T^*) $ and $\sum_{j = 1}^p \mathbbm{1} (W_j \geq T^*) \leq s + \sum_{j \in \mathcal{H}_0 } \mathbbm{1} (W_j \geq T^*)  $, we have 
    \begin{equation} \label{pf-le-T-low-1}
        1 + \sum_{j \in \mathcal{H}_0 } \mathbbm{1} (W_j \leq -  T^*)  \leq q (1 + s) + q \sum_{j \in \mathcal{H}_0} \mathbbm{1} (W_j \geq T^*).
    \end{equation}
  Under event $ \mathcal B_{1,  \varepsilon}$ and noting that $T^* \leq G^{-1} (\frac{ q s}{2p})$, we obtain that 
    \begin{equation}   \label{pf-le-T-low-2}
       \frac{ \sum_{j \in \mathcal{H}_0} \mathbbm{1} (W_j \leq - T^*) } { p_0 G(T^*) } \geq 1 - \varepsilon
    \end{equation}
    and 
     \begin{equation}  \label{pf-le-T-low-3} 
       \frac{ \sum_{j \in \mathcal{H}_0} \mathbbm{1} (W_j \geq  T^*) } { p_0 G(T^*) } \leq 1 + \varepsilon .
    \end{equation}
    Combining \eqref{pf-le-T-low-1} -- \eqref{pf-le-T-low-3} yields that
    \begin{equation*}
        1 + (1 - \varepsilon) p_0 G(T^*) \leq q (1 + s) + q(1 + \varepsilon) p G(T^*)
    \end{equation*}
    and hence
    \begin{equation} \label{pf-le-T-low-4}
        G(T^*) \leq \frac{q (1 + s) -1 } {p_0 ( 1 - q - \varepsilon (1 + q) ) } \leq \frac{2 q s} { p_0 (1 - q) },
    \end{equation}
    where the last inequality follows by the range $\varepsilon \in (0, \frac{1 - q} {2 (1 + q)})$. By the continuity and monotonicity of $G(\cdot)$, we obtain from \eqref{pf-le-T-low-4} that $T^* \geq G^{-1} ( \frac{2 q s} {p_0 (1 - q)} ) $ under the event $ \mathcal{B}_{1, \varepsilon}^c$.
    
    Consequently, combining the results established for the upper bound and lower bound, we have with probability $ 1 - O(\kappa_n) - \mathbb{P} (\mathcal{B}_{1, \varepsilon})$, it holds that
    $$
     G^{-1} \Big( \frac{2 q s} {p_0 (1 - q)} \Big)  \leq  T^* \leq G^{-1} ( \frac{ q s }{ 2 p } ).
    $$
This completes the proof of Lemma \ref{le-1-pf-thm-power}. 

\subsection{Proof of Lemma \ref{le-2-pf-thm-power}} \label{Sec.B.power.2}
    At the first step, we prove that when $ \mathcal B_{1, \varepsilon}^c$ holds, there are at most $s \big( 1 + \frac{6 (1 + \varepsilon) q }{ 1 - q}  \big) $ knockoff statistics $W_j$'s satisfying $|W_j| \geq G^{-1} (\frac{2 q s} {p_0 (1 - q)})  - b_n  $. Indeed, under event $ \mathcal B_{1, \varepsilon}^c$, this can be proved by noting the following results 
    \begin{equation*}
    \begin{split}
        \sum_{j = 1}^p \mathbbm{1} \bigg( |W_j| \geq G^{-1} \Big(\frac{2 q s} {p_0 (1 - q)} \Big)  - b_n \bigg)  & 
        \leq s + \sum_{j \in \mathcal{H}_0} \mathbbm{1} \bigg( |W_j| \geq G^{-1} \Big(\frac{2 q s} {p_0 (1 - q)} \Big)  - b_n \bigg) \\
       & \leq s + 2 (1 + \varepsilon) p_0 G \Big(G^{-1} \Big(\frac{2 q s} {p_0 (1 - q)} \Big)  - b_n \Big) .
    \end{split}
    \end{equation*}
Moreover, it follows from Condition \ref{cond-power-1} that as $n $ is sufficiently large, 
\begin{equation*}
    \begin{split}
        p_0 G \Big(G^{-1} \Big(\frac{2 q s} {p_0 (1 - q)} \Big)  - b_n \Big) & = (1 + o(1)) p_0 G \Big(G^{-1} \Big(\frac{2 q s} {p_0 (1 - q)} \Big) \Big) \\
        & = (1 + o(1)) \frac{2 q s }{ 1 - q } < \frac{ 3 q s}{ 1 - q}, 
    \end{split}
\end{equation*}
which leads to 
\begin{equation}  \label{pf-le-filter-1}
    \sum_{j = 1}^p \mathbbm{1} \bigg( |W_j| \geq G^{-1} \Big(\frac{2 q s} {p_0 (1 - q)} \Big)  - b_n \bigg) <  s \Big( 1 + \frac{6 (1 + \varepsilon) q }{ 1 - q}  \Big).
\end{equation}
    
Next we proceed to show that $\mathbb{P} (\{j: | W_j  | \geq T^*\} \subset D_m) \geq  1- \mathbb{P} ( \mathcal B_{1,  \varepsilon}) - O( p^{-1})$. We have that $\varepsilon_{\mu, m} =  \log ( \Phi(\frac{\mu}{2 \sqrt{2m}}) /\Phi( -\frac{\mu}{2 \sqrt{2m}} )) \geq \frac{ \mu } {  \sqrt { \pi m } } $ based on the basic inequality $ \Phi(x)/\Phi(-x) \geq \frac{ 2 \sqrt{2} x} { \sqrt{  \pi}}  $ for $x > 0$.
Observe that it follows from the union bound inequality   that 
 \begin{equation}   \label{pf-le-filter-2}
    \begin{split}
        & \mathbb{P} \bigg(\max_{1 \leq k \leq p; 1 \leq j \leq m } |{Z}_{k, j}| \geq b_n/2 \bigg)  \\
        & \leq mp  \mathbb{P} \bigg(  |{Z}_{k, j}| \geq \frac{ 6 \Delta_n}{\mu} \sqrt{\pi m} \log p  \bigg) \\
        & \leq   m p  \exp \bigg\{ - \frac{ \frac{ 6 \Delta_n}{\mu} \sqrt{\pi m} \log p   } { 2 \Delta_n /\varepsilon_{\mu, m} }  \bigg\}  \leq   m p \exp \{-  3 \log p \} =   m / p^2 \leq p^{-1}.
    \end{split}
    \end{equation}
Under event $\mathcal{B}_{1, \varepsilon}^c$, in view of \eqref{pf-le-filter-1} and the fact that $|D_m| = m $, if $m >  s \big( 1 + \frac{6 (1 + \varepsilon) q }{ 1 - q}  \big)   $, we have that at least one index $i_j \in D_m$  such that $|W_{i_j}| <   G^{-1} (\frac{2 q s}{p_0 (1 - q)}) - b_n  $ (because otherwise it contradicts with \eqref{pf-le-filter-1}). 
Denote the event $ A_{1, n} := \{ \max_{1 \leq k \leq p; 1 \leq j \leq m } |{Z}_{k, j}| < b_n/2 \} $. We have established that $\mathbb{P}(A_{1, n}^c) \leq  p^{-1}$. 
Further under the event $A_{1,n} $, 
observe that for any $j\in \{1,\cdots, p\}$ such that $|W_j| \geq T^*$, it holds with probability $ 1 - O(\kappa_n) - \mathbb{P} (\mathcal{B}_{1, \varepsilon}) $ that $|W_j| \geq G^{-1} (\frac{2 q s} {p_0 (1 - q)} )$, and thus for any $k \in [m]$, 
\begin{equation*}
    (|W_j| + Z_{ j, k}) - (| W_{i_j}| + Z_{ i_j, k} ) \geq    |W_j | - |W_{i_j}|  -  |Z_{ j, k}| -  |Z_{ i_j, k}| >  b_n  - b_n/ 2 - b_n/2 = 0,
\end{equation*}
which implies that $S^* = \{ j: |W_j| \geq T^*\}$ will pass through the mirror peeling algorithm. Therefore, it follows from \eqref{pf-le-filter-2} and Lemma \ref{le-1-pf-thm-power} that
\begin{equation*}
\begin{split}
       \mathbb{P} ( S^* \subset D_m ) 
       & \geq 1  - O(\kappa_n) - \mathbb{P} (\mathcal{B}_{1, \varepsilon}) - \mathbb{P} (A_{1,n}^c)  \\
       &\geq 1 -  \mathbb{P} ( \mathcal{B}_{1, \varepsilon}) - O( p^{-1} + \kappa_n )  . 
\end{split}
\end{equation*}
This completes the proof of Lemma \ref{le-2-pf-thm-power}.

\subsection{Proof of Lemma \ref{le-3-pf-thm-power}} \label{Sec.B.power.3}

It is easy to see that when $ S^* = \{j: |W_j| \geq T^* \} \subset D_m$, we have 
$$
\sum_{j \in D_m } \mathbbm{1} \{ W_{j} \geq T^* \} = \sum_{j = 1}^p \mathbbm{1} \{ W_{j} \geq T^* \}
$$
and 
$$
\sum_{j\in D_m} \mathbbm{1} \{ W_{j} \leq -T^* \} = \sum_{j = 1}^p \mathbbm{1} \{ W_{j} \leq - T^* \}.
$$
Therefore, it holds that $1 + \sum_{j \in D_m} \mathbbm{1} \{ W_{j} \leq -T^* \}  \leq q \sum_{j \in D_m } \mathbbm{1} \{ W_{j} \geq T^* \}$ by the definition of $T^*$. 
Recalling the definition of $\breve{T}$, we obtain
\begin{equation}
    \breve{T} \leq T^*,
\end{equation}
which immediately implies that $ S^* \subset  \breve{S}   $ when $S^* \subset D_m$. This completes the proof of Lemma \ref{le-3-pf-thm-power}.

\subsection{Proof of Lemma \ref{le-4-pf-thm-power}} \label{Sec.B.power.4}
First, since $\widetilde{Z}_{j} \sim N(0, 2m \Delta_n^2/\mu^2)$,  it can be shown that 
\begin{equation} \label{pf-le-4-new1}
    \max_{1 \leq j \leq m} |\widetilde{W}_{i_j} - W_{i_j} | = \max_{1 \leq j \leq m} |\widetilde{Z}_{j}| \leq b_n / 2
\end{equation}
holds with probability at least $1 - O( p^{-1})$. 

Recall the definition of event $ A_{1, n} := \{ \max_{1 \leq k \leq p; 1 \leq j \leq m } |{Z}_{k, j}| < b_n/2 \} $ and we have shown in the proof of Lemma \ref{le-2-pf-thm-power} that $\mathbb{P} (A_{1, n}^c) \leq    p^{-1}$. 
Under event $A_{1, n} \cap \mathcal{B}_{1,  \varepsilon}^c$, we claim that  $  {D}_m  \subset \{j:  | {W}_j | > G^{-1} ( \frac{ 3 m}{4p} )\}$.  Indeed, under  $ A_{1, n} \cap \mathcal{B}_{1,  \varepsilon}^c$ and Condition \ref{cond-power-1},  it holds that for large $n$ and $\varepsilon < 1/4$,  
\begin{equation} \label{pf-lemma10-neww-eq1}
\begin{split}
     \sum_{j \in \mathcal{H}_0} \mathbbm{1} \bigg( | W_j | > G^{-1} \Big( \frac{3m}{4 p } \Big) + b_n \bigg) 
    & \geq 2 (1 - \varepsilon) p_0 G \bigg( G^{-1} \Big( \frac{3m}{4 p } \Big) + b_n\bigg) \\
    & \geq 2 (1 - \varepsilon)p_0  G\bigg( G^{-1} \Big( \frac{3m}{4 p } \Big)  \bigg) (1 + o(1)) \\
    & =  3 m  (1 + o(1))(1 - \varepsilon)  / 2 >  m. 
\end{split}
\end{equation}
Therefore, there are at least $m + 1$ knockoff statistics whose magnitude exceeds $ G^{-1}  ( \frac{3m}{4 p } ) + b_n$. We now show by contradiction that if $ |{W}_{k_0}| \leq G^{-1}  (\frac{3m}{4p}   )$ for some $k_0 \in \mathcal{H}_0$, then $k_0 \notin {D}_m$ must hold.  Observe that if $ k_0 \in {D}_m $, then there exists some $j_0 \in [m]$ and a set $\mathcal{S}_{j_0}$ with $|\mathcal{S}_{j_0} | = p-j_0+1$ such that 
$$
G^{-1}  \Big(\frac{3m}{4p}   \Big) + b_n/2 > | {W}_{k_0}| +  {Z}_{k_0, j_0} \geq | {W}_{k }| + Z_{k, j_0}, \quad \mbox{for}~ k \in \mathcal{S}_{j_0}, 
$$
which means
$$
| {W}_{k}| < G^{-1}  \Big(\frac{3m}{4p}   \Big) + b_n, \quad \mbox{for}~ k \in \mathcal{S}_{j_0}, 
$$
This implies there are at most $j_0$ ($j_0 \in[m]$) knockoff statistics whose magnitude exceeds $G^{-1}   (\frac{3m}{4p}) + b_n$, which contradicts with \eqref{pf-lemma10-neww-eq1}. Consequently, we obtain $  \{j:  | {W}_j | \leq G^{-1} ( \frac{3m}{4p} )\} \subset \{ j: j \notin {D}_m \} $ and hence $ |{W}_j| > G^{-1} ( \frac{ 3 m}{4p} ) $ for any $j \in D_m$. 

Next we prove that $ \widetilde{T} < \breve{T} + b_n/2 $ with high probability. Denote $A_{3,n} = \{ \max_{1 \leq j\leq m}  | \widetilde{Z}_{j} | \leq b_n /2 \}$ and it has been shown that $\mathbb{P}( A_{3,n}^c ) \leq O( p^{-1})$. 
Note that $\breve{T} \in \{ | W_j | \}_{j \in D_m} $ by definition, then we have 
\begin{equation}  \label{pf-lemma10-neww-eq0}
    \begin{split}
         &   \big\{ \widetilde{T} \leq \breve{T} + b_n /2, A_{1, n} \cap A_{3, n}  \cap \mathcal{B}_{1, \varepsilon}^c \big\} \\
         & = \cup_{j^* \in D_m}   \big\{ \breve{T} = |W_{j^*}|,  \widetilde{T} \leq \breve{T} + b_n /2 , A_{1, n} \cap A_{3, n} \cap \mathcal{B}_{1, \varepsilon}^c \big\} \\
         & \supset \cup_{j^* \in D_m} \big\{  \breve{T} = |W_{j^*}|,  \widetilde{T} \leq   | \widetilde{W}_{j^*} |,   A_{1, n} \cap A_{3, n} \cap \mathcal{B}_{1, \varepsilon}^c \big\} \\
         & \supset \cup_{j^* \in D_m} \big\{  \breve{T} = |W_{j^*}|,  \widetilde{T} \leq   | \widetilde{W}_{j^*} |,   A_{1, n} \cap A_{3, n} \cap \mathcal{B}_{1, \varepsilon}^c ,  |W_{j^*}| > b_n  \big\}
    \end{split}
\end{equation}
where the second-to-last inequality is due to $| \widetilde{W}_{j^*} |  \leq  |W_{j^*}| + b_n /2 $   under event $  A_{3, n}  $,  for $j^* \in D_m$. 

In addition, under event $A_{1, n} \cap A_{3, n} \cap \mathcal{B}_{1, \varepsilon}^c  \cap \{ \breve{T} = |W_{j^*}| \}$, we claim that $ \{ \cap_{j \neq j^*; j \in [p]} \big| |W_{j^*}| - |W_j| \big| > b_n  \} \subset \{ \widetilde{T} \leq |\widetilde{W}_{j^*}|  \} $. 
In fact, $  \cap_{j \neq j^*; j \in [p]}\big\{ \big|   W_j - W_{j^*} \big| \land \big|   W_j + W_{j^*} \big| \geq b_n \big\} $ indicates that $ \sum_{j \neq j^*} \mathbbm{1} (  |W_{j^*}| -b_n \leq | W_j| \leq |W_{j^*}| + b_n)  = 0 $ and $ \sum_{j \neq j^*} \mathbbm{1} (  -|W_{j^*}| -b_n \leq | W_j| \leq -|W_{j^*}| + b_n)  = 0 $. Therefore, under event 
$A_{1, n} \cap A_{3, n} \cap \mathcal{B}_{1, \varepsilon}^c \cap \{ \breve{T} = |W_{j^*}| \}  \cap \{ |W_{j^*}| > b_n \}   $,  
\begin{equation} \label{pf-lemma10-neww-eq1-nn}
    \begin{split}
        & 1 + \sum_{j \in D_{m} } \mathbbm{1} (\widetilde{W}_{j} \leq - |\widetilde{W}_{j^*}|) \\
        & = 1 + \mathbbm{1} (\widetilde{W}_{j^*} \leq - |\widetilde{W}_{j^*}|) + \sum_{j \neq j^*}  \mathbbm{1} (\widetilde{W}_{j} \leq - |\widetilde{W}_{j^*}|) \\
        & \leq 1 + \mathbbm{1} (\widetilde{W}_{j^*} \leq - |\widetilde{W}_{j^*}|) + \sum_{j \neq j^*}  \mathbbm{1} ( {W}_{j} \leq - | {W}_{j^*}| + b_n) \\
        &  = 1 + \mathbbm{1} (\widetilde{W}_{j^*} \leq - |\widetilde{W}_{j^*}|) + \sum_{j \neq j^*}  \mathbbm{1} ( {W}_{j} \leq - | {W}_{j^*}| ) \\
        & = 1 + \sum_{j \in D_m} \mathbbm{1} ( {W}_{j} \leq - | {W}_{j^*}|) + \mathbbm{1} (\widetilde{W}_{j^*} \leq - |\widetilde{W}_{j^*}|) - \mathbbm{1} ( {W}_{j^*} \leq - | {W}_{j^*}|) \\
        & =  1 + \sum_{j \in D_m} \mathbbm{1} ( {W}_{j} \leq - | {W}_{j^*}|), 
    \end{split}
\end{equation}
where  the last equality holds since  the sign of $ \widetilde{W}_{j^*} $ and the sign of $ {W}_{j^*} $ are identical under event $ \{ |W_{j^*}| > b_n \}  $. 
By the same technique for proving  \eqref{pf-lemma10-neww-eq1-nn}, we can show that 
\begin{equation}  \label{pf-lemma10-neww-eq3}
    \sum_{j \in D_{m} } \mathbbm{1} (\widetilde{W}_{j} \geq |\widetilde{W}_{j^*}|) \geq \sum_{j \in D_m}  \mathbbm{1} ( {W}_{j} \geq  | {W}_{j^*}|).  
\end{equation} \label{pf-lemma10-neww-eq4-0}
In addition, $\breve{T} = |W_{j^*}|$ implies 
\begin{equation} \label{pf-lemma10-neww-eq4}
    1 + \sum_{j \in D_{m} } \mathbbm{1} (W_{j} \leq - |W_{j^*}|) \leq q \sum_{j \in D_m} \mathbbm{1} (W_{j} \geq  |W_{j^*}|). 
\end{equation}

Combining \eqref{pf-lemma10-neww-eq1-nn}--\eqref{pf-lemma10-neww-eq4} derives that under event $A_{1, n} \cap A_{3, n} \cap \mathcal{B}_{1, \varepsilon}^c  \cap \{ \breve{T} = |W_{j^*}| \} \cap \{ |W_{j^*}| > b_n \}$, 
\begin{equation*}
     1 + \sum_{j \in D_{m} } \mathbbm{1} (\widetilde{W}_{j} \leq - |\widetilde{W}_{j^*}|) \leq  q \sum_{j \in D_{m} } \mathbbm{1} (\widetilde{W}_{j} \geq |\widetilde{W}_{j^*}|) 
\end{equation*}
and hence 
\begin{equation*}
    \widetilde{T} \leq |\widetilde{W}_{j^*}|.
\end{equation*}
This proves $  \cap_{j \neq j^*; j \in [p]}\big\{ \big|   W_j - W_{j^*} \big| \land \big|   W_j + W_{j^*} \big| \geq b_n \big\}  \subset \{ \widetilde{T} \leq |\widetilde{W}_{j^*}|  \} $. Therefore, it follows from \eqref{pf-lemma10-neww-eq0} that 
\begin{equation*}  
    \begin{split}
         &  \big\{ \widetilde{T} \leq \breve{T} + b_n /2, A_{1, n} \cap A_{3, n} \cap \mathcal{B}_{1, \varepsilon}^c \big\} \\
         & \supset  \cup_{j^* \in D_m}  \Big\{ \breve{T} = |W_{j^*}|,    \cap_{j \neq j^*; j \in [p]}\big\{ \big|   W_j - W_{j^*} \big| \land \big|   W_j + W_{j^*} \big| \geq b_n \big\}, \\
         & \quad A_{1, n} \cap A_{3, n} \cap \mathcal{B}_{1, \varepsilon}^c, |W_{j^*}| > b_n \Big\},
    \end{split}
\end{equation*}
and hence 
\begin{equation}  \label{pf-lemma10-neww-eq5}
    \begin{split}
        &   \big\{ \widetilde{T}  >  \breve{T} + b_n /2 , A_{1, n} \cap A_{3, n} \cap \mathcal{B}_{1, \varepsilon}^c \big\} \\
        &\subset  \big\{ A_{1, n} \cap A_{3, n} \cap \mathcal{B}_{1, \varepsilon}^c \big\}\\
        & \quad -   \cup_{j^* \in D_m}  \Big\{ \breve{T} = |W_{j^*}|,    \cap_{j \neq j^*; j \in [p]}\big\{ \big|   W_j - W_{j^*} \big| \land \big|   W_j + W_{j^*} \big| \geq b_n \big\}, \\
         & \quad A_{1, n} \cap A_{3, n} \cap \mathcal{B}_{1, \varepsilon}^c, |W_{j^*}| > b_n \Big\} \\
        & =\cup_{j^* \in D_m}   \big\{\breve{T} = |W_{j^*}|, A_{1, n} \cap A_{3, n} \cap \mathcal{B}_{1, \varepsilon}^c \big\}\\
        & \quad -  \cup_{j^* \in D_m}  \Big\{ \breve{T} = |W_{j^*}|,    \cap_{j \neq j^*; j \in [p]}\big\{ \big|   W_j - W_{j^*} \big| \land \big|   W_j + W_{j^*} \big| \geq b_n \big\}, \\
         & \quad A_{1, n} \cap A_{3, n} \cap \mathcal{B}_{1, \varepsilon}^c, |W_{j^*}| > b_n \Big\}\\
        & \subset \cup_{j^* \in D_m} \bigg\{ \Big\{ \breve{T} = |W_{j^*}|,    \cup_{j \neq j^*; j \in [p]}\big\{ \big|   W_j - W_{j^*} \big| \land \big|   W_j + W_{j^*} \big| \leq b_n \big\}, \\
        & \quad A_{1, n}  \cap A_{3, n}  \cap \mathcal{B}_{1, \varepsilon}^c \Big\}   
       \cup \Big\{\breve{T} = |W_{j^*}| , A_{1, n}  \cap A_{3, n}  \cap \mathcal{B}_{1, \varepsilon}^c , |W_{j^*}| \leq b_n \Big\} \bigg\} \\
        & \subset \cup_{j^* \in D_m} \bigg\{ \Big\{ \breve{T} = |W_{j^*}|,  \cup_{j \neq j^*; j \in [p]}\big\{ \big|   W_j - W_{j^*} \big| \land \big|   W_j + W_{j^*} \big| \leq b_n\big\}, \\
        & \quad |W_{j^*}| > G^{-1} (\frac{3m}{4p})  \Big\}  
  \cup \Big\{\breve{T} = |W_{j^*}|,   |W_{j^*}| \leq b_n  \Big\} \bigg\},
    \end{split}
\end{equation}
where the last inequality holds since $ |{W}_j| > G^{-1} ( \frac{ 3 m}{4p} ) $ for any $j \in D_m$, under event $ A_{1, n}   \cap \mathcal{B}_{1, \varepsilon}$. Therefore, we can obtain from \eqref{pf-lemma10-neww-eq5} that 
\begin{equation}   \label{pf-lemma10-neww-eq6}
    \begin{split}
        &  \mathbb{P} (\widetilde{T}  >  \breve{T} + b_n /2 )  \\
        & \leq  \mathbb{P}( \widetilde{T}  >  \breve{T} + b_n /2,  A_{1, n} \cap A_{3, n} \cap \mathcal{B}_{1, \varepsilon}^c  ) + \mathbb{P}(A_{1, n}^c) + \mathbb{P}(A_{3, n}^c) + \mathbb{P}(\mathcal{B}_{1, \varepsilon})\\
        & \leq   \mathbb{P} \bigg( \cup_{j^* \in D_m} \Big\{ \breve{T} = |W_{j^*}|,   \cup_{j \neq j^*; j \in [p]}\big\{ \big|   W_j - W_{j^*} \big| \land \big|   W_j + W_{j^*} \big| \leq b_n\big\}, \\
        & \quad | W_{j^*} |> G^{-1} ( \frac{ 3 m}{4p} ) \Big\}    \bigg) \\
        & \quad + \mathbb{P} \big(\cup_{j^* \in D_m}  \big\{\breve{T} = |W_{j^*}|,   |W_{j^*}| \leq b_n \} \big)   + \mathbb{P}(A_{1, n}^c) + \mathbb{P}(A_{3, n}^c) + \mathbb{P}(\mathcal{B}_{1, \varepsilon})\\
        & \leq   \mathbb{P} \bigg( \cup_{j^* \in [p]} \Big\{  \cup_{j \neq j^*; j \in [p]}\big\{ \big|   W_j - W_{j^*} \big| \land \big|   W_j + W_{j^*} \big| \leq b_n\big\}, \\
        & \quad | W_{j ^*}|> G^{-1} ( \frac{ 3 m}{4p} ) \Big\}    \bigg)    +    \mathbb{P} (  \breve{T} \leq b_n )  + \mathbb{P}(A_{1, n}^c) + \mathbb{P}(A_{3, n}^c) + \mathbb{P}(\mathcal{B}_{1, \varepsilon})\\
        & \leq \mathbb{P} (\mathcal{B}_2)+  \mathbb{P} (  \breve{T} \leq b_n ) + \mathbb{P}(\mathcal{B}_{1, \varepsilon} )   + O( p^{-1}). \\
    \end{split}
\end{equation}

  Finally, note that under the event $\{\widetilde{T}  \leq  \breve{T} + b_n /2\} \cap A_{3, n}$, 
    \begin{equation*}
    \begin{split}
        \breve{S}  \cap \widetilde{S}^c \cap \mathcal{S}_0 & = \{j \in \mathcal{S}_{0} \cap D_m:  \widetilde{W}_j < \widetilde{T}~   \mbox{and}~ W_{j} \geq \breve{T}\} \\
        & \subset\{ j \in \mathcal{S}_0 \cap D_m: \breve{T} \leq  W_j \leq \breve{T} + b_n \}. 
    \end{split} 
    \end{equation*}
   This together with \eqref{pf-lemma10-neww-eq6} completes the proof of Lemma \ref{le-4-pf-thm-power}.

\subsection{Proof of Lemma \ref{pf-thm8-lemma1}} \label{Sec:pf-newle6}
The main idea of the proof is to approximate the sum of indicator functions by the corresponding sum of probabilities, and then apply the definition of $\wh{D}_m$. For a small number $\varepsilon > 0$, under event $\mathscr{A} = \mathscr{A}_{1} \cap \mathscr{A}_{2} \cap \mathscr{A}_3 \cap \mathscr{A}_{4, \varepsilon}$, we obtain that 
\begin{equation} \label{pf-newle6-eq1}
\begin{split}
    & \sum_{j \in \mathcal{H}_0} \mathbbm{1} \left(|\wh{W}_j| > G^{-1} \Big(\frac{(1 + c_1)m}{2p_0}  \Big) + b_n \right)  \\
    & \geq \sum_{j \in \mathcal{H}_0} \mathbbm{1} \left(| {W}_j  | > G^{-1} \Big(\frac{(1 + c_1)m}{2p_0} \Big) + b_n + \rho_n  \right) \\
    & \geq 2(1 - \varepsilon) \sum_{j \in \mathcal{H}_0} \mathbb{P} \left( {W}_j > G^{-1} \Big(\frac{(1 + c_1)m}{2p_0} \Big) + b_n  + \rho_n \right)  \\
    &  = 2(1 - \varepsilon) p_0 G \left(  G^{-1} \Big(\frac{ (1 + c_1) m}{2 p_0} \Big) + b_n + \rho_n \right). 
\end{split}
\end{equation}
Moreover, it follows from Condition \ref{cond:rho-n} that 
    $$
       G \left(  G^{-1} \Big(\frac{ (1 + c_1) m}{2 p_0} \Big) + b_n + \rho_n \right) =  \frac{(1 + c_1) m }{2p_0}  \cdot (1 + o(1))    \geq \frac{(1 + c_1/2 ) m}{2 p_0},
    $$
which combines with \eqref{pf-newle6-eq1} yields that when $ \varepsilon < c_1/4 $ and $n $ is sufficiently large, 
\begin{equation} \label{pf-newle6-eq2}
\begin{split}
    & \sum_{j \in \mathcal{H}_0} \mathbbm{1} \left(|\wh{W}_j| > G^{-1} \Big(\frac{(1 + c_1)m}{2p_0}  \Big) + b_n \right)  \\
    & \geq  (1 - \varepsilon)(1 + c_1/2) (1 + o(1)) m > m.
\end{split}
\end{equation}
Therefore, there are at least $m+1$  knockoff statistics whose magnitude exceeds $ G^{-1}  (\frac{(1 + c_1)m}{2p_0}   ) + b_n $. 

We now show by contradiction that if $ |\wh{W}_{k_0}| \leq G^{-1}  (\frac{(1 + c_1)m}{2p_0}   )$ for some $k_0 \in \mathcal{H}_0$, then $k_0 \notin \wh{D}_m$. 
Observe that if $ k_0 \in \wh{D}_m $, then there exists some $j_0 \in [m]$ and a set $\mathcal{S}_{j_0}$ with $|\mathcal{S}_{j_0} | = p-j_0+1$ such that 
$$
G^{-1}  \Big(\frac{(1 + c_1)m}{2p_0}   \Big) + b_n/2 > |\wh{W}_{k_0}| +  {Z}_{k_0, j_0} \geq |\wh{W}_{k, j_0}| + Z_{k, j_0}, \quad \mbox{for}~ k \in \mathcal{S}_{j_0}, 
$$
which means
$$
|\wh{W}_{k, j_0}| < G^{-1}  \Big(\frac{(1 + c_1)m}{2p_0}   \Big) + b_n, \quad \mbox{for}~ k \in \mathcal{S}_{j_0}, 
$$
This implies there are at most $j_0$ ($j_0 \in[m]$) knockoff statistics whose magnitude exceeds $G^{-1}  \Big(\frac{(1 + c_1)m}{2p_0}   \Big) + b_n$, which contradicts with \eqref{pf-newle6-eq2}. Consequently, we obtain that $ \wh{D}_m  \subset \{ j: \wh{W}_j > G^{-1} ( \frac{(1 +c_1)m}{2p_0} )\}$ for a small constant $c_1 >0$. 

Next we proceed to prove the second part of Lemma \ref{pf-thm8-lemma1}. Using the same proof technique for proving \eqref{pf-newle6-eq2}, we can obtain that under event $\mathscr{A}$ and when $\varepsilon < c_1/4$ and $n$ is sufficiently large, 
\begin{equation}
    \begin{split}
        & \sum_{j \in \mathcal{H}_0} \mathbbm{1} \left( |\wh{W}_j| > G^{-1} \Big( \frac{(1 - c_1) (m-s)}{2 p_0} \Big) - b_n   \right) \\
        & \leq \sum_{j \in \mathcal{H}_0} \mathbbm{1} \left( | {W}_j| > G^{-1} \Big( \frac{(1 - c_1) (m-s)}{2 p_0} \Big) - b_n -\rho_n 
       \right) \\
       & \leq 2(1 + \varepsilon)  p_0 G \left(  G^{-1} \Big(\frac{ (1 - c_1) m}{2 p_0} \Big) + b_n + \rho_n \right) \\
       & \leq (1 + \varepsilon) (1 - c_1/2) (1 + o(1)) (m-s)< m-s. 
    \end{split}
\end{equation}
There are at most $m-s-1$ null knockoff statistics with magnitude exceeding $G^{-1}  ( \frac{(1 - c_1) (m-s)}{2 p_0}  ) - b_n $. Since $|\wh{D}_m| = m$ and $s$ is the number of nonnull variables,   there must exist a $j^* \in \wh{D}_m \cap \mathcal{H}_0$ such that $|\wh{W}_{j^*}| \leq G^{-1}  ( \frac{(1 - c_1) (m-s)}{2 p_0}  ) - b_n$. If $|\wh{W}_{k_0}| > G^{-1}  ( \frac{(1 - c_1) (m-s)}{2 p_0}  )$ for some $k_0 \in \mathcal{H}_0$, then  
$$
|\wh{W}_{k_0}| + \wt{Z}_{k_0} > |\wh{W}_{j^*}| + \wt{Z}_{j^*},
$$
which implies that $k_0 \in \wh{D}_m$. This proves that $\big\{j: |\wh{W}_{j}| > G^{-1}  ( \frac{(1 - c_1) (m-s)}{2 p_0}  ) \big\} \subset  \wh{D}_m $. The proof of Lemma \ref{pf-thm8-lemma1} is completed.

\subsection{Proof of Lemma \ref{pf-thm8-lemma2}} \label{Sec:pf-newle7}
The main idea of proof is to apply the definition of the threshold $\wh{T}$ and the approximation for indicator functions. Note that by the definition of $\wh{T}$, we have
\begin{equation} \label{pf-newle7-eq1}
    \begin{split}
        1 + \sum_{j \in \wh{D}_m } \mathbbm{1} (\wh{W}_j + \wt{Z}_j \leq - \wh{T}) \leq q + q \sum_{j \in \wh{D}_m} \mathbbm{1} (\wh{W}_j +  \wt{Z}_j  \geq \wh{T}) .
    \end{split}
\end{equation}
For a small constant $c_1 > 0$, under event $ \mathscr{A} $ with $\varepsilon$ sufficiently small, it follows from Lemma \ref{pf-thm8-lemma1} that for a small constant $c_1 > 0$
\begin{equation}  \label{pf-newle7-eq2}
\begin{split}
    & \sum_{j \in \wh{D}_m} \mathbbm{1} (\wh{W}_j + \wt{Z}_j \leq - \wh{T}) \\
    & = \sum_{j \in [p]} \mathbbm{1} (\wh{W}_j + \wt{Z}_j \leq - \wh{T}, j \in \wh{D}_m) \\
    & \geq \sum_{j \in \mathcal{H}_0} \mathbbm{1} \left( \wh{W}_j \leq - \wh{T}  - b_n/2, | \wh{W}_j | \geq G^{-1}\Big(\frac{(1 - c_1) (m-s)}{2 p_0} \Big)  \right) \\
    & =  \sum_{j \in \mathcal{H}_0} \mathbbm{1} \left( \wh{W}_j \leq - \wh{T}  - b_n/2,  \wh{W}_j  \leq -G^{-1}\Big(\frac{(1 - c_1) (m-s)}{2 p_0} \Big)  \right) \\
    & \geq \sum_{j \in \mathcal{H}_0} \mathbbm{1} \left( {W}_j \leq - \wh{T}  - b_n/2 - \rho_n,   {W}_j  \leq -G^{-1}\Big(\frac{(1 - c_1) (m-s)}{2 p_0} \Big) - \rho_n \right) \\
    & \geq  (1 - \varepsilon)    p_0 (1 + o(1))G\left( \wh{T}  \vee G^{-1}\Big(\frac{(1 - c_1) (m-s)}{2 p_0}\Big) + b_n/2 + \rho_n \right) \\
    & \geq (1 - \varepsilon)    p_0 (1 + o(1))G\left( \wh{T}  \vee G^{-1}\Big(\frac{(1 - c_1) (m-s)}{2 p_0}\Big)  \right),
\end{split}
\end{equation}
where the last step is due to Condition \ref{cond:rho-n}.
Similarly, under event $ \mathscr{A} $, we have 
\begin{equation}  \label{pf-newle7-eq3}
    \begin{split}
        & \sum_{j \in \wh{D}_m} \mathbbm{1} (\wh{W}_j +  \wt{Z}_j  \geq \wh{T}) \\
        & \leq  s + \sum_{j \in \mathcal{H}_0\cap  \wh{D}_m} \mathbbm{1} (\wh{W}_j +  \wt{Z}_j  \geq \wh{T}) \\
        & \leq s + \sum_{j \in \mathcal{H}_0} \mathbbm{1} \left( W_j \geq \wh{T} - b_n/2 - \rho_n, W_j \geq G^{-1} \Big( \frac{(1 + c_1) m}{2 p_0}\Big)  - \rho_n \right)\\
        & \leq s + (1 + \varepsilon) p_0 (1 + o(1))G\left( \wh{T}   \vee G^{-1}\Big(\frac{(1 + c_1) m}{2 p_0}\Big) - b_n/2 - \rho_n \right) \\
        & \leq s + (1 + \varepsilon) p_0 (1 + o(1))G\left( \wh{T}   \vee G^{-1}\Big(\frac{(1 + c_1) m}{2 p_0}\Big)  \right).
    \end{split}
\end{equation}
Combining \eqref{pf-newle7-eq1}--\eqref{pf-newle7-eq3} leads to the inequality
\begin{equation} \label{pf-newle7-eq4}
    \begin{split}
        & (1 - \varepsilon)    p_0 (1 + o(1))G\left( \wh{T}  \vee G^{-1}\Big(\frac{(1 - c_1) (m-s)}{2 p_0}\Big)  \right) \\
        & \leq q s + q (1 + \varepsilon) p_0 (1 + o(1))G\left( \wh{T}   \vee G^{-1}\Big(\frac{(1 + c_1) m}{2 p_0}\Big)  \right).
    \end{split}
\end{equation}
If $n$ is sufficiently large and $\varepsilon < c_1/4$,  it is seen that
\begin{equation} \label{pf-newle7-eq5}
\begin{split}
    & (1 - \varepsilon) p_0 (1 + o(1))G\left( G^{-1}\Big(\frac{(1 - c_1) (m-s)}{2 p_0}\Big)  \right) \\
    & = (1 - \varepsilon) p_0 (1 + o(1)) \frac{(1 - c_1) (m-s)}{2 p_0} \\
    & > (1 -3c_1/2) (m-s) /2 
\end{split}
\end{equation}
and 
\begin{equation} \label{pf-newle7-eq6}
    \begin{split}
        & q s + q (1 + \varepsilon) p_0 (1 + o(1))G\left( G^{-1}\Big(\frac{(1 + c_1) m}{2 p_0}\Big)  \right) \\
        & \leq q s + q (1 + 3 c_1/2) m /2.
    \end{split}
\end{equation}
Moreover, when $m > \frac{(2q + 1 + c)s}{1 - q}$ for some $c >0$ and $c_1 < \frac{2c(1-q)}{3(1+q)(1 + 2q) - 3(1-q) + 3c(1+q)}$,  we have
\begin{equation}  \label{pf-newle7-eq7}
    (1 -3c_1/2) (m-s)   > 2 q s + q(1 + 3c_1/2)m. 
\end{equation}
When $m > \frac{(2q + 1 + c)s}{1 - q}$ for some $c >0$, $n$ is sufficiently large, and $c_1$ and $\varepsilon$ are sufficiently small,  a combination of \eqref{pf-newle7-eq5}--\eqref{pf-newle7-eq7} yields that 
\begin{equation}  \label{pf-newle7-eq8}
\begin{split}
    & (1 - \varepsilon) p_0 (1 + o(1))G\left( G^{-1}\Big(\frac{(1 - c_1) (m-s)}{2 p_0}\Big)  \right)  \\
    & > q s + q (1 + \varepsilon) p_0 (1 + o(1))G\left(  G^{-1}\Big(\frac{(1 + c_1) m}{2 p_0}\Big)  \right). 
\end{split}
\end{equation}
Since $G(\cdot)$ is a non-increasing function, \eqref{pf-newle7-eq4} entails that 
\begin{equation}  \label{pf-newle7-eq9}
      \begin{split}
        & (1 - \varepsilon)    p_0 (1 + o(1))G\left( \wh{T}  \vee G^{-1}\Big(\frac{(1 - c_1) (m-s)}{2 p_0}\Big)  \right) \\
        & \leq q s + q (1 + \varepsilon) p_0 (1 + o(1))G\left(  G^{-1}\Big(\frac{(1 + c_1) m}{2 p_0}\Big)  \right),
    \end{split}
\end{equation}
which together with \eqref{pf-newle7-eq8} implies that $\wh{T} > G^{-1} ( \frac{(1 - c_1)(m-s)}{2 p_0} )$ must hold. This completes the proof of Lemma \ref{pf-thm8-lemma2}.

\subsection{Lemma \ref{lemma: knockoff-couling-ridge} and Its Proof} \label{Sec: Lemma-knock-accu-ridge}
In this subsection, we verify the coupling accuracy Condition \ref{cond:gene-W-cou-accu} for ridge regression coefficient difference knockoff statistics. We consider a linear model $ Y = X ^T \bbeta + \varepsilon$, where $ X \sim N({\bf 0}_p, \bOmega^{-1} ) $ and $\varepsilon \sim N(0, \sigma^2)$. 
The approximate knockoff matrix $\widehat{\bX}$ and the coupled ideal knockoff matrix $\widetilde{\bX}$ are defined in \eqref{Gaussian-knockoff-matrix-est} and \eqref{Gaussian-knockoff-matrix}, respectively. Denote $ \widehat{\bf{D}}  = [\bX, \widehat{\bX}] $ and $ \widetilde{\bf{D}}  = [\bX, \widetilde{\bX}] $. The ridge regression coefficient difference knockoff statistics $\{ \widehat{W}_j \}_{j \in[p]}$ and the coupled ideal counterpart $\{ \widetilde{W}_j \}_{j \in[p]}$ are defined according to \eqref{eq:ridge} and \eqref{def:ridge-coeff}, based on the augmented design matrix $\widehat{\bf{D}}$ and $\widetilde{\bf{D}} $, respectively.
 
\begin{lemma} \label{lemma: knockoff-couling-ridge}
    Suppose that $\widehat{\bOmega}$ is an estimator of $\bOmega$ such that  
    \begin{equation} \label{eq: lemma19-cond0}
        \mathbb{P} (\|\wh{\bOmega} - \bOmega \|_2 \leq u_n ) \to 1 .
    \end{equation}
    In addition, assume that $p = o(n)$ and with probability tending to one, 
    \begin{equation} \label{eq: lemma19-cond1}
        \begin{split}
        & c_1\leq  \sigma_{\min}( n^{-1} \widetilde{\bf{D}} ^T \widetilde{\bf{D}} ) \leq \sigma_{\max}( n^{-1} \widetilde{\bf{D}} ^T \widetilde{\bf{D}} )    \leq c_2,  \\        
        & c_1\leq  \sigma_{\min}( n^{-1} \widehat{\bf{D}} ^T \widehat{\bf{D}} ) \leq \sigma_{\max}( n^{-1} \widehat{\bf{D}} ^T \widehat{\bf{D}} )  \leq c_2 
        \end{split}
    \end{equation}
    for positive constants $c_1$ and $c_2$.
    Then we have 
    \begin{equation}
         \mathbb{P} \left( \max_{1 \leq j \leq p} | \wh{W}_j -  \wt{W}_j | \geq  \lambda u_n \| \bbeta \|_2 +  \sigma u_n \sqrt{ \frac{\log n}{ n }} \right) \to 0.
    \end{equation}
\end{lemma}

\begin{proof}[Proof of Lemma \ref{lemma: knockoff-couling-ridge}]
Denote $ \widehat{\bf R} = ( n^{-1} \widehat{\bf{D}} ^T  \widehat{\bf{D}} + \lambda I_{2p} )^{-1} $, $ \widetilde{\bf R} = ( n^{-1} \widetilde{\bf{D}} ^T  \widetilde{\bf{D}} + \lambda I_{2p} )^{-1} $, and $\bf H = \widehat{\bf D} - \widetilde{\bf D} = [\bf 0, \widehat{\bX } - \widetilde{\bX}]$. By assumption in \eqref{eq: lemma19-cond1}, we have 
\[
\| \widehat{\bf R} \|_2 \leq c_1^{-1}, \quad \| \widetilde{\bf R} \|_2 \leq c_1^{-1}. 
\]
Noting that 
\[
\widehat{\bX} - \widetilde{\bX} = \bX( \bOmega - \widehat{\bOmega} ) \diag(\boldsymbol{r}) + \bZ {\bf B}, 
\]
where $ {\bf B} = (2 \diag(\boldsymbol{r}) - \diag(\boldsymbol{r})  \wh\bOmega \diag(\boldsymbol{r})  )^{1/2} - (2 \diag(\boldsymbol{r}) - \diag(\boldsymbol{r})  \bOmega \diag(\boldsymbol{r})  )^{1/2} $, and it has been shown in (A.51) in \cite{fan2023ark} that 
\[
\| {\bf B }\|_2 \leq C \| \widehat{\bOmega} - \bOmega \|_2, 
\]
we obtain that with probability tending to one, 
\begin{align*}
\| {\bf H} \|_2 &  = \|  \widehat{\bX} - \widetilde{\bX} \|_2 \leq   \| \diag(\boldsymbol{r}) \|_2 \| \|\bX\|_2 \|  \bOmega - \widehat{\bOmega}\|_2 + \| \bZ \|_2 \| {\bf B }\|_2   \\
&\lesssim \sqrt{n} u_n.
\end{align*}

In addition, observe that 
\begin{align*}
        \widehat{\bf R} - \widetilde{\bf R} & = \widehat{\bf R} ( \widetilde{\bf R}^{-1} - \widehat{\bf R}^{-1}  ) \widetilde{\bf R}  \\
        & = \widehat{\bf R} ( n^{-1} \widetilde{\bf{D}} ^T  \widetilde{\bf{D}}  - n^{-1} \widehat{\bf{D}} ^T  \widehat{\bf{D}}  ) \widetilde{\bf R}  
\end{align*}
and 
\begin{align*}
     n^{-1} \widehat{\bf{D}} ^T  \widehat{\bf{D}} - n^{-1} \widetilde{\bf{D}} ^T  \widetilde{\bf{D}}  = n^{-1} ( {\widetilde{\bf D}} ^{T} {\bf H} +  {\bf H} ^T {\widetilde{\bf D}} + {\bf H}^T {\bf H} ).
\end{align*}
It follows from the Cauchy-Schwarz inequality that 
\[
\| n^{-1} \widehat{\bf{D}} ^T  \widehat{\bf{D}} - n^{-1} \widetilde{\bf{D}} ^T  \widetilde{\bf{D}} \|_2 \leq 2 n^{-1} \| {\widetilde{\bf D}} \|_2 \| {\bf H} \|_2 + n^{-1}\| {\bf H}^T {\bf H}\|_2\lesssim u_n + u_n^2 \lesssim u_n.
\]
Therefore, it holds that 
\begin{align*}
    \| \widehat{\bf R} - \widetilde{\bf R}  \|_2 \lesssim u_n. 
\end{align*}

Recall that the ridge regression coefficients are defined by 
\[
\widehat{\bbeta}  = \widehat{\bf R}  n^{-1} {\widehat{\bf D}}^T \by, \quad \widetilde{\bbeta} = \widetilde{\bf R}  n^{-1} {\widetilde{\bf D}}^T \by.
\]
Next we proceed to bound $\| \widehat{\bbeta} - \widetilde{\bbeta}\|_{\infty}$. Denote by $\bbeta^* = (\bbeta^T, {\bf 0}_p^T)^T $ the augmented true parameter. It holds that $\by = \bX {\bbeta} + \beps = \widehat{\bf D} \bbeta^* +  \beps = \widetilde{\bf D} \bbeta^* +  \beps$. We obtain that 
\begin{align*}
     \|\widehat{\bbeta} - \widetilde{\bbeta}\|_{\infty}  \leq \| \lambda (\widetilde{\bf R}  - \widehat{\bf R} ) \bbeta^* \|_{\infty} + \|\widehat{\bf R}  n^{-1} {\bf H}^T \beps \|_{\infty}+ \| (\widetilde{\bf R}  - \widehat{\bf R} ) n^{-1} \widetilde{\bf D}^T \beps \|_{\infty}.
\end{align*}

For the first error term, we have with probability tending to one, 
\begin{align*}
   \| \lambda (\widetilde{\bf R}  - \widehat{\bf R} ) \bbeta^* \|_{\infty} & \leq  \| \lambda (\widetilde{\bf R}  - \widehat{\bf R} ) \bbeta^* \|_{2} \\
   & \leq \lambda \|   (\widetilde{\bf R}  - \widehat{\bf R} ) \|_{2} \| \bbeta \|_2 \lesssim \lambda u_n \| \bbeta \|_2.
\end{align*}

For the second error term, we denote $ {\bf A} =  n^{-1} {\bf H} \widehat{\bf R}  $.
Noting that $ {\bf A}_j^T  \beps \sim N( 0,   \sigma^2 \| {\bf A}_j\|_2^2   ) $, we have 
\begin{align*}
    \mathbb{P} ( \| {\bf A}^{T} \beps  \|_{\infty} \geq 2  \sigma \max_{j \in [2p] } \| {\bf A}_j \|_2  \sqrt{\log n}) & \leq 
    \sum_{j = 1}^{2p}  \mathbb{P} ( | {\bf A}_j^{T} \beps  |  \geq  2\sigma  \| {\bf A}_j \|_2 \sqrt{\log n} ) \\
    & \lesssim n^{-1} \to 0
\end{align*}
In addition, $ \max_{j \in [2p]} \| {\bf A}_j \|_2 \leq n^{-1} \| {\bf H} \|_2  \| \widehat{\bf R}  \|_2 \lesssim u_n / \sqrt{n} $, we obtain that with probability tending to one, 
\[
\|n^{-1}\widehat{\bf R}  {\bf H}^T \beps \|_{\infty} \lesssim \sigma u_n \sqrt{ \frac{\log n}{ n }}.
\]

For the third error term, denote $ {\bf G} =  n^{-1} \widetilde{\bf D} (\widetilde{\bf R}  - \widehat{\bf R} )  $. Since $\max_{j \in [2p]} \| {\bf G}_j  \|_{2} \leq n^{-1} \|\widetilde{\bf D} \|_2 \| \widetilde{\bf R} - \widehat{\bf R} \|_{2}  \lesssim u_n /\sqrt{n}$, then it can be shown  using the same technique for addressing the second error term that with asymptotic probability one
\[
\| (\widetilde{\bf R} - \widehat{\bf R} ) n^{-1} \widetilde{\bf D}^T \beps \|_{\infty} \lesssim  \sigma u_n \sqrt{ \frac{\log n}{ n }}.
\]
Combining the above results, we obtain that $ \| \widehat{\bbeta} - \widetilde{\bbeta} \|_{\infty} \lesssim  \lambda u_n \| \bbeta \|_2 +  \sigma u_n \sqrt{ \frac{\log n}{ n }}$ with probability tending to one. The same bound holds for the knockoff statistics according to the triangle inequality. This completes the proof of Lemma \ref{lemma: knockoff-couling-ridge}. 
\end{proof}

\section{Additional Simulation}\label{Sec:add-simu}

\subsection{Comparison among different privacy budget}\label{sec:simu_dp_budget}

\begin{figure}[!h]
		\centering
		\begin{subfigure}[b]{0.48\textwidth}
			\centering
			\includegraphics[width=\textwidth]{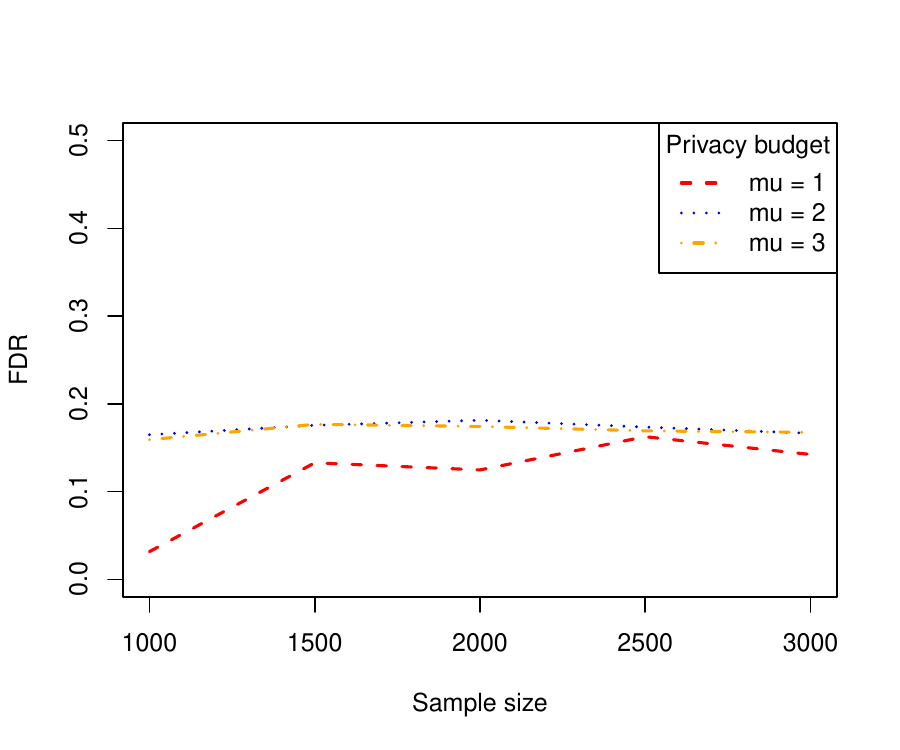}
		\end{subfigure}
		\hfill
		\begin{subfigure}[b]{0.48\textwidth}
			\centering
			\includegraphics[width=\textwidth]{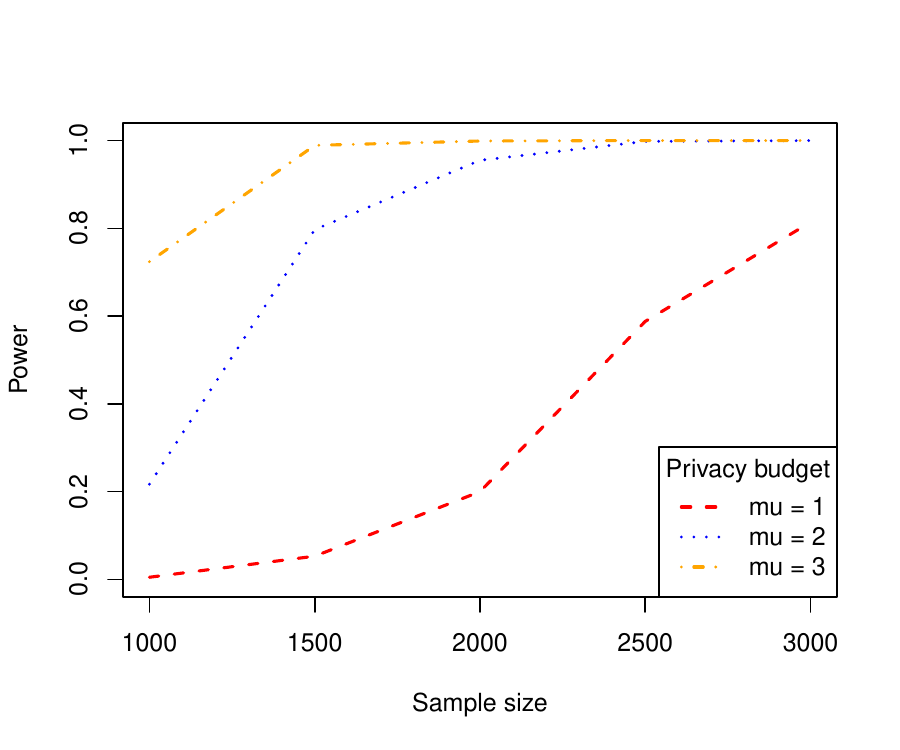}
		\end{subfigure}
		\caption{The FDR and Power of DP-knockoff among different privacy budgets. $n$ increases from 1000 to 3000, and $p$ is fixed at $1000$. $\beta = 0.6$.}
		\label{fig:budget_N}
	\end{figure}

\begin{figure}[!h]
		\centering
		\begin{subfigure}[b]{0.48\textwidth}
			\centering
			\includegraphics[width=\textwidth]{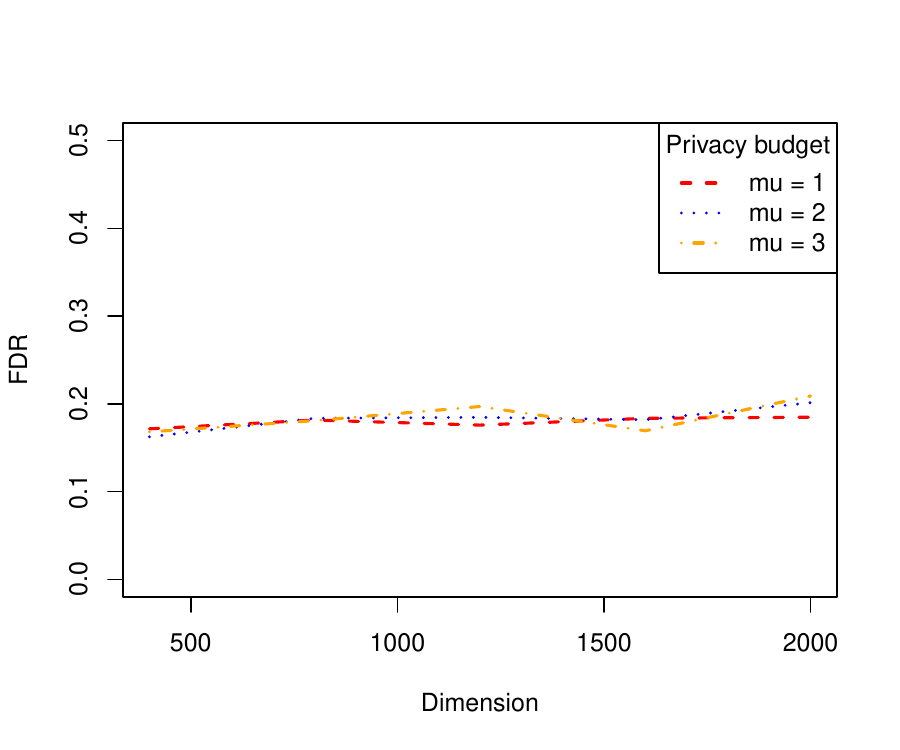}
		\end{subfigure}
		\hfill
		\begin{subfigure}[b]{0.48\textwidth}
			\centering
			\includegraphics[width=\textwidth]{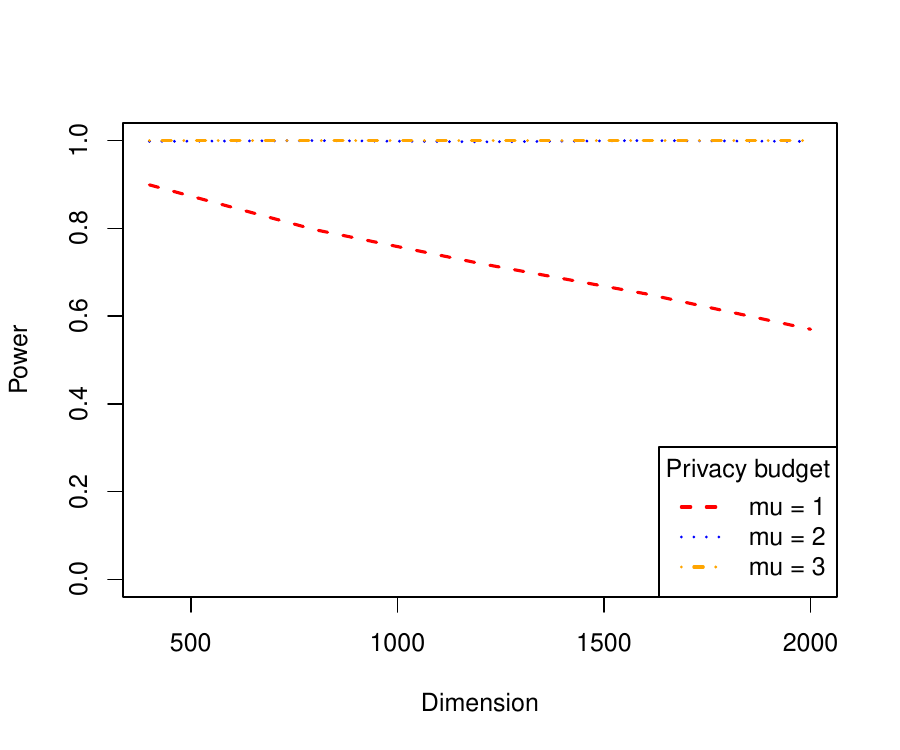}
		\end{subfigure}
		\caption{The FDR and Power of the DP-knockoff among different privacy budgets. $p$ increases from 400 to 2000, and $n$ is fixed at $2000$.  $\beta = 1$.}
		\label{fig:budget_P}
	\end{figure}

    In this section, we examine the role of the privacy budget $\mu$ in the performance of the DP-knockoff method. Figure \ref{fig:budget_N} shows that while FDR control is maintained across all settings, a larger privacy budget $\mu$ leads to improved power. Figure \ref{fig:budget_P} further investigates the effect of varying $\mu$ in high-dimensional settings, where $p$ increases from 400 to 2000. Consistent with Figure \ref{fig:budget_N}, a larger $\mu$ enhances the power. Similar to Figure \ref{fig:P_increase}, the dimensionality $p$ has minimal impact on power, which can be attributed to sample splitting and feature screening.
    
\subsection{Comparison between marginal correlation and HSIC}\label{sec:MCDC}

\begin{figure}[!h]
    \centering
    \begin{subfigure}[b]{0.48\textwidth}
        \centering
        \includegraphics[width=\textwidth]{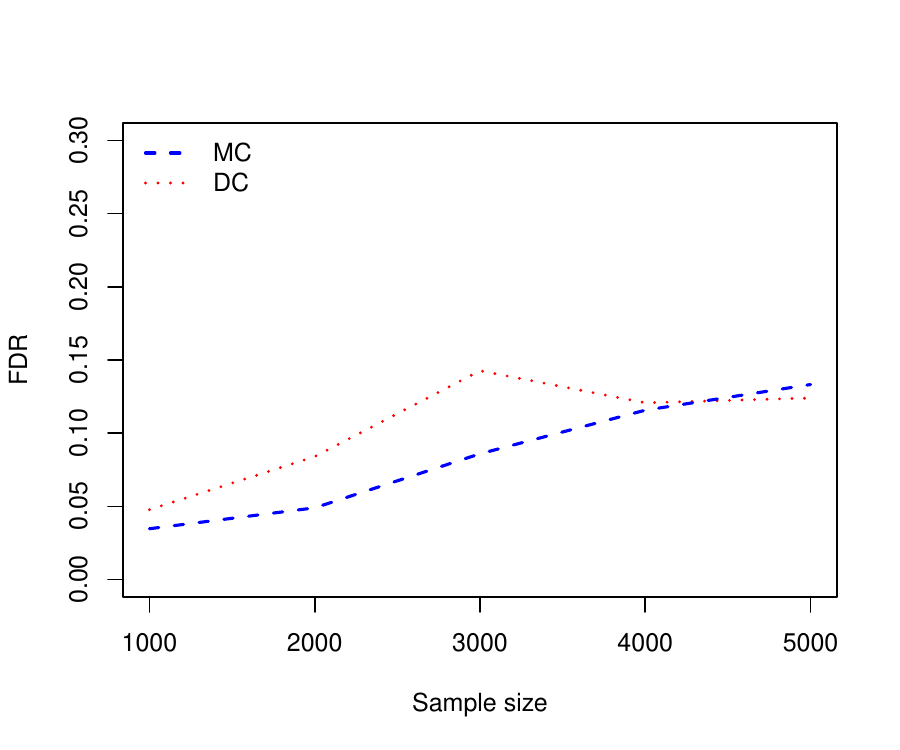}
    \end{subfigure}
    \hfill
    \begin{subfigure}[b]{0.48\textwidth}
        \centering
        \includegraphics[width=\textwidth]{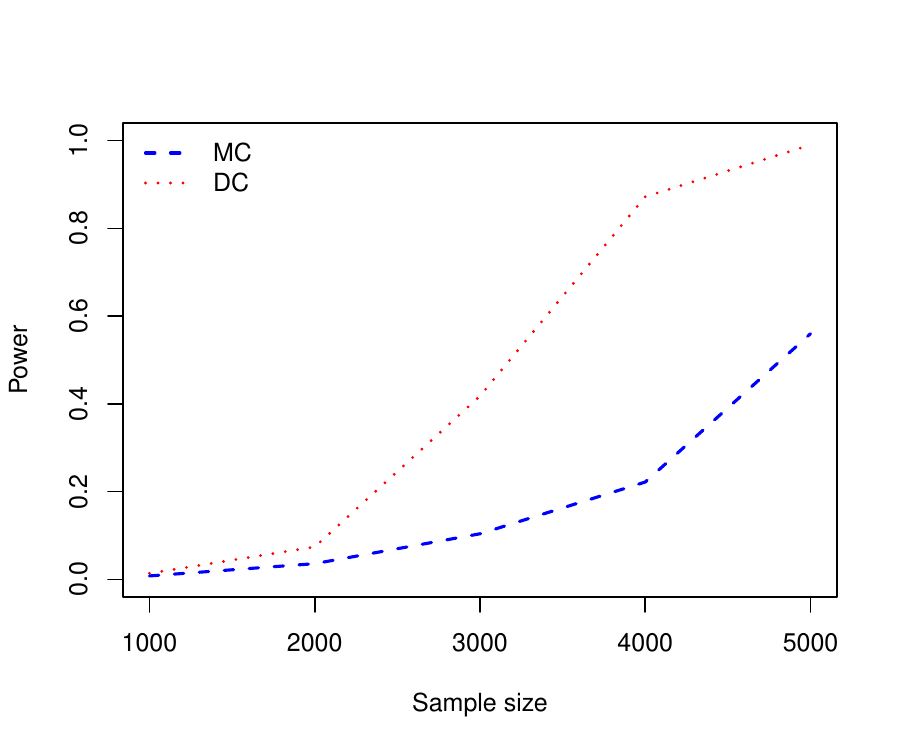}
    \end{subfigure}
    \caption{FDR and power of the DP-knockoff based on marginal correlation (MC) and distance covariance (DC) under the linear model.}
    \label{fig:MD-linear}
\end{figure}

We evaluate the numerical performance of the two screening knockoff statistics introduced in Sections \ref{Sec:sensi-mar-corr} and \ref{Sec:sensi-hsic}. For the HSIC-based method, we employ the distance kernel to implement distance covariance (DC) and use it as the screening statistic. We first consider a linear model $Y = X^T \beta + \varepsilon$ with $\varepsilon \sim N(0,1)$. The covariate vector $X$ is drawn from a $p$-dimensional independent uniform distribution on $[0,1]$. The response $Y$ depends only on the first five covariates, with the corresponding coefficients in $\beta$ set to 1.5 and all others set to 0. The peeling size is fixed at $m=10$. To ensure comparability, we scale $Y$ to the unit interval as $Y = (Y - \min(Y))/(\max(Y)-\min(Y))$. Under this setup, the sensitivity of marginal correlation (MC) is $4/n$, and the sensitivity of DC is $8/n$. Throughout, we set $p=200$, and the nominal FDR level $q=0.2$. The averaged FDR and power for MC and DC under the linear model are reported in Figure \ref{fig:MD-linear}. Both methods achieve finite-sample FDR control, while DC consistently exhibits higher power than MC, despite having twice the sensitivity.  

\begin{figure}[!h]
    \centering
    \begin{subfigure}[b]{0.48\textwidth}
        \centering
        \includegraphics[width=\textwidth]{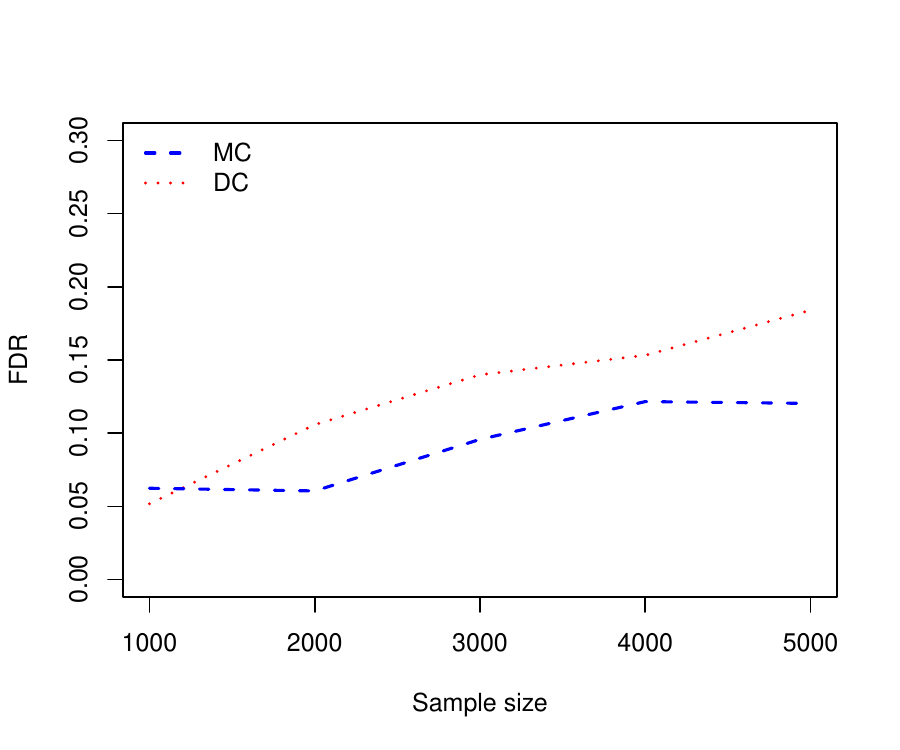}
    \end{subfigure}
    \hfill
    \begin{subfigure}[b]{0.48\textwidth}
        \centering
        \includegraphics[width=\textwidth]{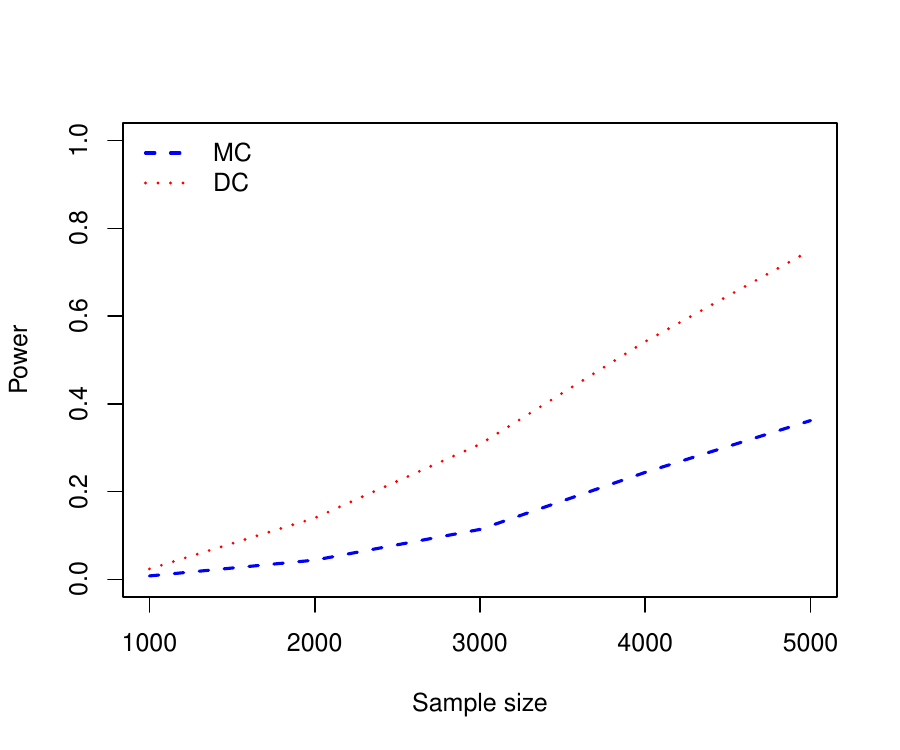}
    \end{subfigure}
    \caption{FDR and power of the DP-knockoff based on marginal correlation (MC) and distance covariance (DC) under the nonlinear model.}
    \label{fig:MD-non-linear}
\end{figure}

We next consider a nonlinear model where the response still depends only on the first five covariates:  
\[
Y = 3\sin(X_1) + 2X_2 + X_3^3 + 2X_4^2 + 1.5X_5 + \varepsilon, \quad \varepsilon \sim N(0,1).
\]  
All other settings remain the same as in the linear case. The results, shown in Figure \ref{fig:MD-non-linear}, indicate that both MC and DC maintain FDR control, while DC again achieves higher power.  

Although DC outperforms MC in terms of power for both linear and nonlinear models, it incurs substantially higher computational cost: DC requires at least $O(n^2)$ operations \citep{Szekely2007}, whereas MC requires only $O(n)$. This highlights a trade-off between statistical efficiency and computational efficiency.  

\subsection{DP-knockoff based on stochastic gradient descent}\label{sec:SGDsimu}

\begin{figure}[!h]
		\centering
		\begin{subfigure}[b]{0.48\textwidth}
			\centering
			\includegraphics[width=\textwidth]{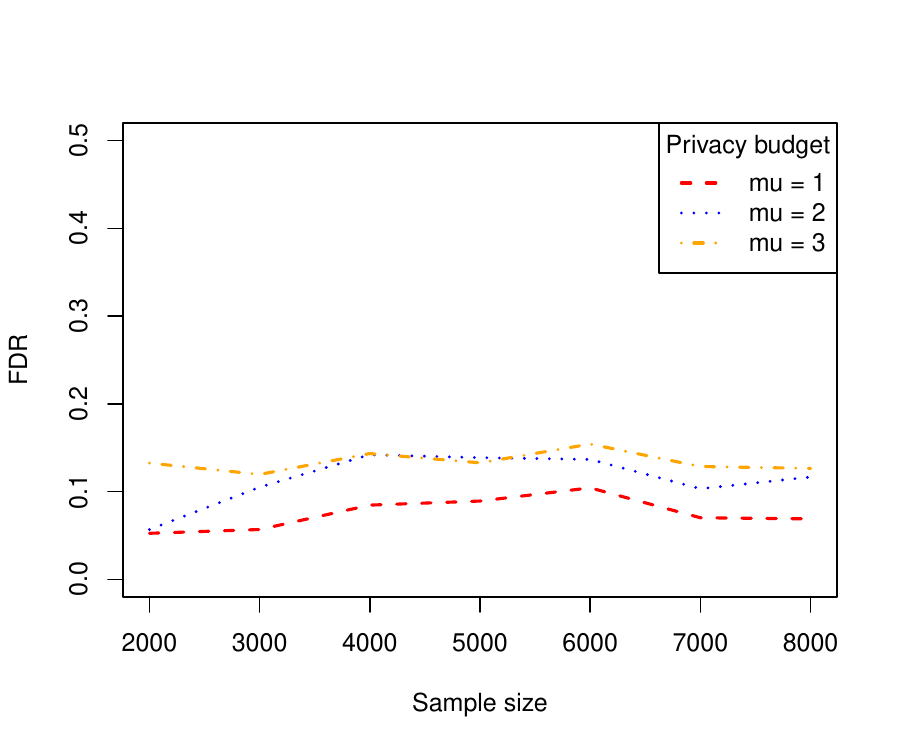}
		\end{subfigure}
		\hfill
		\begin{subfigure}[b]{0.48\textwidth}
			\centering
			\includegraphics[width=\textwidth]{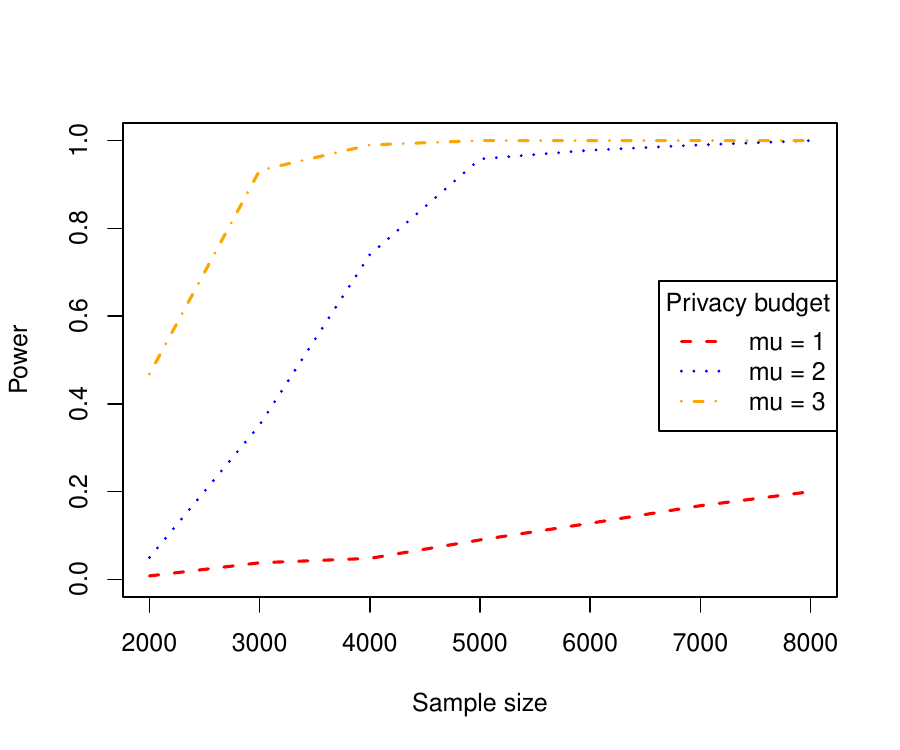}
		\end{subfigure}
		\caption{FDR and power of the DP-knockoff with SGD-based logistic regression, under privacy budgets $\mu \in \{1,2,3\}$. $p=400$. }
		\label{fig:SGD-log}
	\end{figure}

We evaluate the numerical performance of the stochastic gradient descent (SGD) proposed in section \ref{Sec:sensi-SGD}. Here we consider a logistic regression model, and use SGD as the estimation algorithm.  The covariate vector $X$ is drawn from a $p$-dimensional independent uniform distribution on $[-1,1]$ with $p=400$. The response $Y$ follows a logistic regression model that depends only on the first five covariates, with the corresponding coefficients set to 1.5 and the remaining coefficients set to 0. The results are shown in Figure \ref{fig:SGD-log}. As the sample size increases, the DP-knockoff procedure maintains FDR control, while the power steadily increases. This trend is consistent across all three privacy budgets $\mu \in \{1,2,3\}$.

\subsection{Choice of $m$ and $K_n$}\label{sec:Kn}

\begin{figure}[!h]
		\centering
		\begin{subfigure}[b]{0.48\textwidth}
			\centering
			\includegraphics[width=\textwidth]{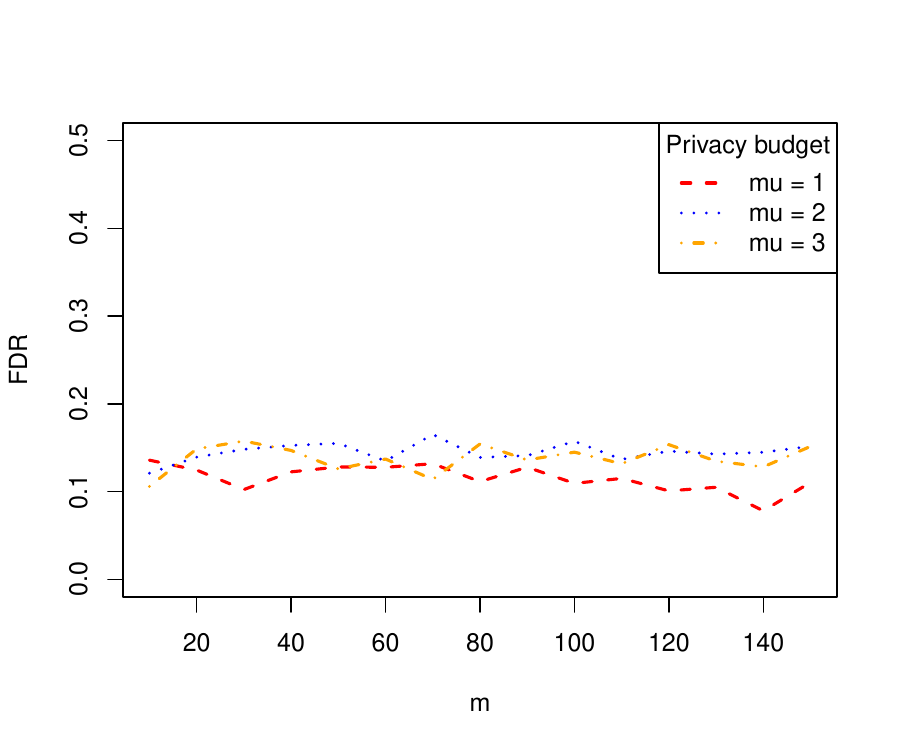}
		\end{subfigure}
		\hfill
		\begin{subfigure}[b]{0.48\textwidth}
			\centering
			\includegraphics[width=\textwidth]{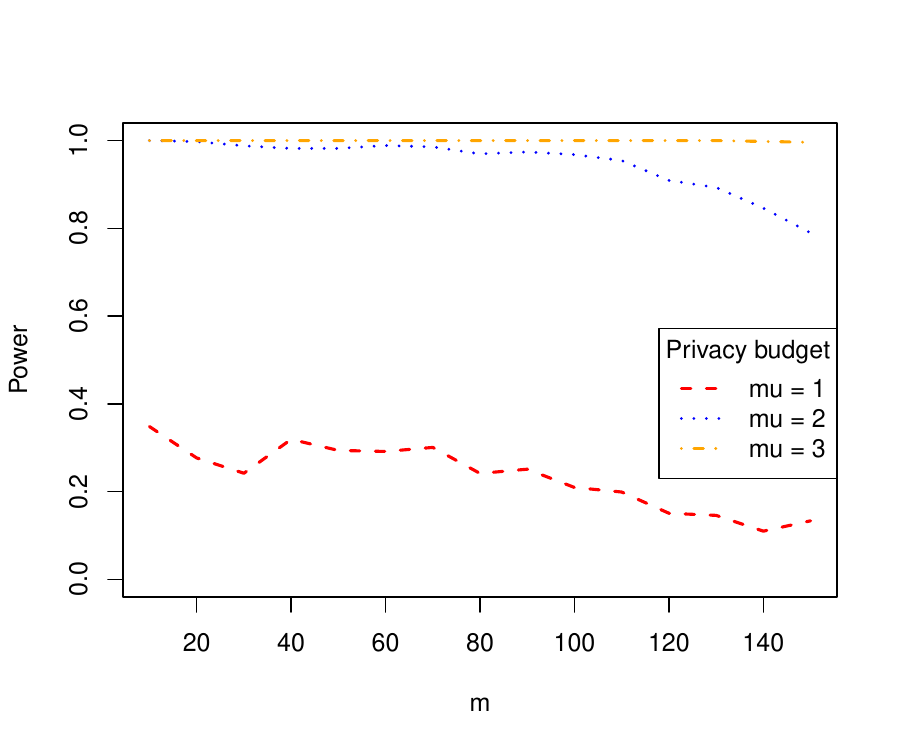}
		\end{subfigure}
		\caption{FDR and power of the DP-knockoff when $m$ increases, under privacy budgets $\mu \in \{1,2,3\}$. }
		\label{fig:mm}
	\end{figure}

\begin{figure}[!h]
		\centering
		\begin{subfigure}[b]{0.48\textwidth}
			\centering
			\includegraphics[width=\textwidth]{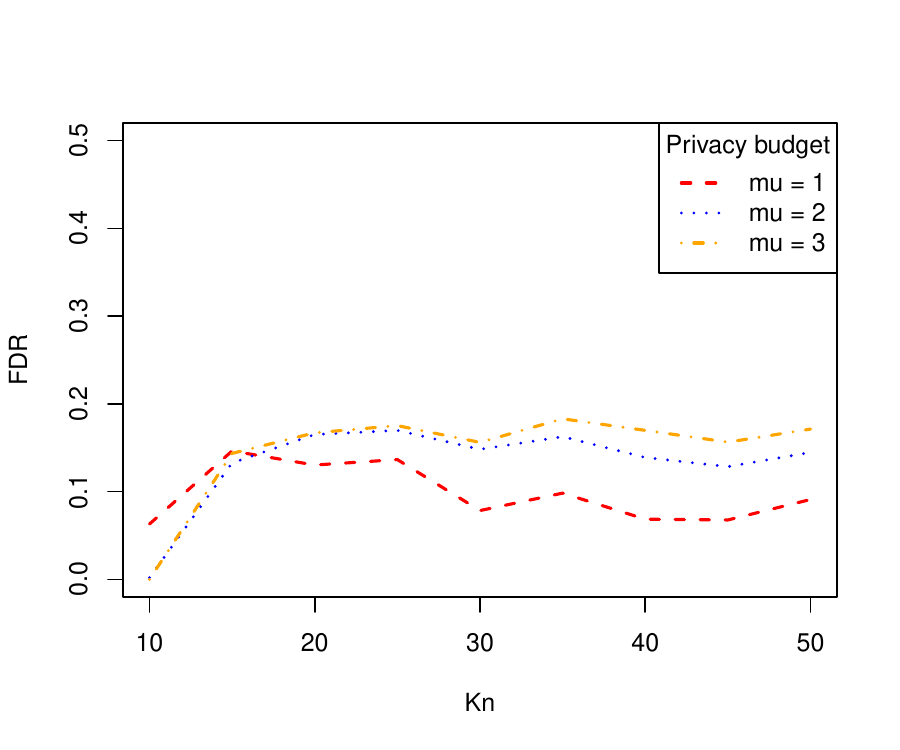}
		\end{subfigure}
		\hfill
		\begin{subfigure}[b]{0.48\textwidth}
			\centering
			\includegraphics[width=\textwidth]{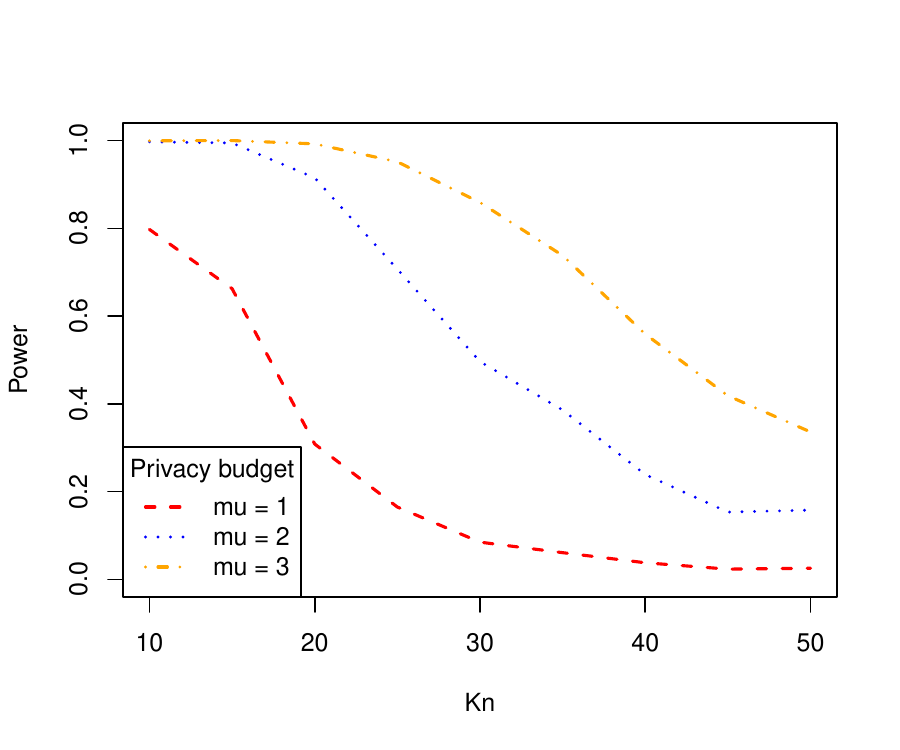}
		\end{subfigure}
		\caption{FDR and power of the DP-knockoff when $K_n$ increases, under privacy budgets $\mu \in \{1,2,3\}$. }
		\label{fig:Kn}
	\end{figure}

In this subsection, we examine the impact of the tuning parameters \( m \) and \( K_n \) in Algorithms~1 and~3, respectively. To study the effect of \( m \) in low-dimensional settings, we adopt the same linear model configuration as in Section~\ref{sec:MCDC}, with non-zero $\beta = 1$, \( n = 2000 \) and \( p = 200 \). Figure~\ref{fig:mm} reports the false discovery rate (FDR) and power obtained using the marginal correlation statistic introduced in Section~4.1 as the knockoff statistic. To investigate the influence of \( K_n \) in high-dimensional regimes, we consider the same linear model setting as in Section~6, with \( n = 2000 \) and \( p = 1000 \). Figure~\ref{fig:Kn} presents the FDR and power as \( K_n \) increases from 10 to 50 under three different privacy budgets.

The two figures convey a consistent message. Varying \( m \) and \( K_n \) does not affect FDR control: for all considered values, the FDR remains below the nominal level. In contrast, increasing either \( m \) or \( K_n \) leads to a reduction in power. This behavior aligns with the design of the proposed algorithms, in which the variance of the added noise grows proportionally with \( m \) and \( K_n \).

\end{appendices}

\end{document}